\documentclass[12pt]{unibz_ase_thesis}
\usepackage[hyphens]{url}
\usepackage{lipsum}
\usepackage{graphicx}
\usepackage{pdfpages}
\usepackage{enumitem}
\usepackage{hyperref}
\usepackage{changepage} 
\usepackage{graphicx}   
\usepackage{subcaption} 
\usepackage{amsmath}
\usepackage{todonotes}
\usepackage[utf8]{inputenc}
\usepackage[T1]{fontenc}
\usepackage{newtxtext}
\usepackage{tcolorbox}
\usepackage{titlesec}
\usepackage{hyperref}
\usepackage{caption} 
\usepackage[a-2b]{pdfx}

\usepackage{tabularx}
\usepackage{array}

\newenvironment{pubinfo}[1]{
\begin{center}
\setlength{\fboxsep}{10pt}
\noindent\fbox{%
\begin{minipage}{0.9\textwidth}
\textbf{Publication information} \\[0.8em]
#1
\end{minipage}}
\end{center}
}{}

\usepackage{blkarray}
\usepackage{longtable}
\usepackage{rotating}
\usepackage{algorithm}
\usepackage{algpseudocode}
\usepackage{textcomp}
\usepackage[ngerman,italian,english]{babel}
\usepackage{csquotes}
\usepackage{tikz}
\usepackage{enumitem} 
\usetikzlibrary{trees}
\renewcommand{\footnotesize}{\scriptsize}
\usepackage{amsmath, amsthm, amssymb}
\usepackage[backend=biber,style=ieee,defernumbers=true]{biblatex}
\usepackage{microtype}
\usepackage{xurl}
\usepackage{pdfpages}

\usepackage{chancery} 
\usepackage[T1]{fontenc} 

\renewcommand{\vec}[1]{\boldsymbol{#1}}

\usepackage{hyperref}
\usepackage{multirow}
\usepackage{array} 
\newcounter{paper}
\usepackage[a-2b]{pdfx}
\renewcommand{\thepaper}{\Alph{paper}} 

\makeatletter
\newcommand{\myPaper}[1]{
  \cleardoublepage
  \refstepcounter{paper}%
  \begingroup
    \let\orig@addcontentsline\addcontentsline
    \edef\orig@thechapter{\thechapter}%
    \@ifundefined{theHchapter}{}{\edef\orig@theHchapter{\theHchapter}}%
    \edef\orig@chaptername{\chaptername}%
    \@ifundefined{chapapp}{}{\edef\orig@chapapp{\chapapp}}%
    \renewcommand{\thechapter}{\thepaper}%
    \@ifundefined{theHchapter}{}{%
      \renewcommand{\theHchapter}{paper.\thepaper}}%
    \renewcommand{\chaptername}{Paper}%
    \@ifundefined{chapapp}{}{\renewcommand{\chapapp}{Paper}}%
    \renewcommand{\addcontentsline}[3]{}%
    \chapter{#1}%
    \let\addcontentsline\orig@addcontentsline
    \phantomsection
    \addcontentsline{toc}{chapter}{Paper \thepaper:\hspace{0.5em}#1}
    \renewcommand{\chaptername}{\orig@chaptername}%
    \@ifundefined{chapapp}{}{\renewcommand{\chapapp}{\orig@chapapp}}%
    \renewcommand{\thechapter}{\orig@thechapter}%
    \@ifundefined{theHchapter}{}{%
      \renewcommand{\theHchapter}{\orig@theHchapter}}%
  \endgroup
  \addtocounter{chapter}{-1}%
  \setcounter{figure}{0}\setcounter{table}{0}\setcounter{equation}{0}%
}
\makeatother

\makeatletter
\newcommand{\letterChapter}[2][]{%
  \cleardoublepage
  \refstepcounter{chapter}%
  \begingroup
    \renewcommand{\chaptername}{Chapter}%
    \def\chapletter{\Alph{chapter}}%
    \if\relax\detokenize{#1}\relax\else
      \edef\chapletter{#1}%
    \fi
    \renewcommand{\thechapter}{\chapletter}%
    \addcontentsline{toc}{chapter}{Chapter \chapletter: #2}%
    \chapter*{Chapter \chapletter\\#2}%
  \endgroup
}
\makeatother

\algnewcommand\algorithmicswitch{\textbf{switch}}
\algnewcommand\algorithmiccase{\textbf{case}}
\algdef{SE}[SWITCH]{Switch}{EndSwitch}[1]{\algorithmicswitch\ #1\ \algorithmicdo}{\algorithmicend\ \algorithmicswitch}%
\algdef{SE}[CASE]{Case}{EndCase}[1]{\algorithmiccase\ #1}{\algorithmicend\ \algorithmiccase}%
\algtext*{EndSwitch}%
\algtext*{EndCase}%

\begin{document}
\pagenumbering{gobble}

\title{
Predicting Known Vulnerabilities from Attack Descriptions Using Sentence Transformers}

\author{Refat Othman} 

\degreeaward{Ph.D.\ in Advanced-Systems Engineering}       
\PhDcycle{$38^\mathrm{th}$ Cycle}          
\copyyear{2026}  
\defenddate{}          

\birthday{[02.10.1991]}				
\birthplace{[Nablus, Palestine]}			
\defenddate{27.01.2026}				
\signplace{[Bolzano, Italy]}			

\supervisor{Professor Barbara Russo}					
\secondsupervisor{ Professor Bruno Rossi  \\ & Professor Mengyuan Zhang}
\orcid{0000-0003-3791-399X}

\rightsstatement{All rights reserved}

\maketitle[logo]

\begin{acknowledgements} 
 
\noindent

My deepest and most sincere gratitude goes to my supervisor, Professor Barbara Russo. Her unwavering support, constant encouragement, and invaluable guidance have been at the heart of this journey. She provided not only the intellectual direction needed to shape this thesis but also the patience and understanding that sustained me through its most challenging phases. Professor Russo gave me the freedom to explore my ideas while always being there to provide clarity, constructive feedback, and motivation when I needed it most. Her dedication to excellence, combined with her kindness and humanity, made her an exceptional mentor and role model. I will always remain grateful for the opportunity to learn from her and for the trust she placed in me throughout these years. I am truly honored to have had the privilege of working under her supervision.

I am also sincerely grateful to Professor Bruno Rossi and Professor Andrea Janes for their supervision and valuable suggestions during my period abroad. Their guidance, constructive feedback, and encouragement greatly enriched my research experience and provided me with new perspectives that strengthened the quality of this thesis. I deeply appreciate the time and effort they devoted to supporting me throughout this important phase of my doctoral journey.

I sincerely thank the members of my thesis examination committee, Professor Bruno Rossi and Professor Mengyuan Zhang, for serving as examiners of this dissertation and for the time and care they dedicated to the evaluation of my work. I am deeply grateful for their insightful comments and constructive feedback.



This research was supported by the European Social Fund Plus (ESF+), Project ESF2f30005, CUP: B56F24000100001. It also gratefully acknowledges the support of the Cybersecurity Laboratory (CSLab) at the Free University of Bozen-Bolzano, funded by the EFRE-FESR 2021–2027 program, project EFRE1039, CUP: I53C23001690009.

\newpage
\noindent
\begin{minipage}{\linewidth}
    \raggedleft
    {\fontfamily{pzc}\selectfont \LARGE To My Family}  
\end{minipage}

\vspace{1cm} 

To my father, Tahseen, who is no longer with us,  I dedicate this work with profound love, respect, and sorrow. You left this world before you could witness this milestone. You taught me integrity, patience, and perseverance, and those lessons have guided me through every challenge of this journey.  Though I could not share this moment with you in person, I feel your presence in every success I achieve. This work is a testament to the seeds you planted with your devotion and hard work.

To my mother, Amneh, and my wife, Saja, I owe a debt of gratitude that words can never fully express. Your unwavering love, patience, and strength have been the pillars of my life and the foundation upon which this journey was built. You have both sacrificed so much so that I could pursue my dreams, my mother, through her endless prayers, resilience, and quiet strength that guided me through every challenge; and my wife, through her boundless patience, encouragement, and steadfast belief in me during the most demanding moments. You are my inspiration and my safe haven, and every achievement I make is a reflection of your love and support. To my beloved sons, Majdaldeen and Mohammad, your joy, laughter, and light remind me every day of why every effort was worth it.


To my brother, Abed, I will always remain deeply grateful. From the moment I began my academic journey, from my bachelor’s studies, through my master’s, into my professional life, and now this doctorate, you have been by my side, carrying the role of our father with love, sacrifice, and responsibility. You filled a place that no one else could, guiding me with wisdom, supporting me with strength, and standing by me every step of the way. I will forever remain indebted to you, and your presence will always have a special place in my heart.
I am also sincerely grateful to my brothers and sisters, Tareq, Malek, Yusra, Mohammad, Rafe, and Sewar. Thank you for your encouragement, care, and love.

Last but not least, to my friends who became like family, Dr. Diaeddin Rimawi and Dr. Usman Rafiq, thank you for your companionship, encouragement, and the countless shared moments that made this journey brighter, more meaningful, and far less lonely. Your friendship has been a source of strength and joy throughout this path.

\end{acknowledgements}

\begin{abstract}


Modern infrastructures and services rely heavily on software systems, which remain persistently vulnerable to cyberattacks. These attacks frequently exploit vulnerabilities documented in public repositories such as MITRE’s Common Vulnerabilities and Exposures (CVE). To achieve timely and proactive defense, it is essential to anticipate which vulnerabilities might be exploited as soon as new attack information becomes available. However, current Cyber Threat Intelligence (CTI) resources, such as MITRE’s Adversarial Tactics, Techniques, and Common Knowledge (ATT\&CK) and CVE repositories, provide only partial and disconnected coverage of the attack–vulnerability relationships. They often describe adversarial behaviors and vulnerabilities independently, leaving many of the relationships between them undocumented.
This thesis addresses the problem of forecasting and recommending likely vulnerabilities from textual descriptions of cyberattacks. The motivation stems from the fact that attack information usually appears first, often before the corresponding vulnerability is officially disclosed. Thus, being able to infer vulnerabilities directly from attack descriptions would enable countermeasures to prevent and mitigate cyberattacks. The key challenge lies in automatically linking natural-language attacks to known vulnerability descriptions, a task that is infeasible to perform manually at scale. Additional challenges include the heterogeneous abstraction levels of attack information, the selection of appropriate text representation models, inconsistencies in existing repositories, and the generalization of models trained on curated data to unseen, real-world reports.
To overcome these challenges, this research develops automated methods grounded in advanced Natural Language Processing (NLP), specifically transformer-based sentence embedding models that transform attack and vulnerability descriptions into vector representations capturing their semantic similarity. 
These semantic representations enable the system to infer and prioritize vulnerabilities that are most likely associated with attacks.

Our methodology follows a systematic and iterative process: First, annotated datasets were constructed using existing links within MITRE repositories to map attacks to vulnerabilities.
Second, fourteen state-of-the-art (SOTA) transformer models were compared across four types of attack descriptions (Tactic, Technique, Procedure, and Attack Pattern) to determine which type provides the most informative signal for vulnerability prediction.
The results show that Technique descriptions in the MITRE ATT\&CK repository, which describe attack methods and tools, are the most informative for predicting vulnerabilities. Among all evaluated models, the \texttt{multi-qa-mpnet-base-dot-v1 (MMPNet)} transformer achieved the highest performance, because its hybrid pre-training strategy integrates masked and permuted language modeling, enabling it to capture long-range contextual dependencies and nuanced semantic relationships between attack and vulnerability descriptions. Additionally, its fine-tuning on large-scale question–answer datasets optimizes it for semantic similarity tasks, enabling it to capture nuanced contextual dependencies and semantic relations within short, technical texts that are particularly beneficial for vulnerability inference.
The proposed methodology was implemented in the \texttt{VULDAT} tool, which automates the process of linking attack descriptions to known vulnerabilities. Manual validation of its predictions revealed numerous previously undocumented relationships in MITRE, confirming the incompleteness of existing repositories and highlighting the value of automation for knowledge enrichment. 
Finally, the approach was applied to unseen, real-world cyberattack reports, demonstrating that the proposed models can generalize beyond curated datasets and forecast relevant vulnerabilities from unstructured narratives and provide timely, automated support for vulnerability awareness and mitigation.
In conclusion, this thesis advances the state of vulnerability detection by introducing a data-driven, scalable framework that bridges structured and unstructured sources of cybersecurity knowledge. It reduces reliance on manual curation, accelerates the identification of vulnerabilities, and enhances the completeness of existing threat intelligence repositories. By combining methodological rigor with practical validation, the thesis provides both researchers and practitioners with scalable, automated strategies to strengthen the security of systems in the face of evolving cyberthreats.

\end{abstract}

\chapter*{List of Publications}
\addcontentsline{toc}{chapter}{List of Publications}

\begin{refsection}[publications.bib]







  \bigskip
  \textbf{Papers included in the thesis:}\par
  \textbf{Paper A:} 
  \AtNextCite{\defcounter{maxnames}{4}}\fullcite{othman2025attack}.
  \\ \hspace*{2em}Link: \url{https://doi.org/10.1016/j.jss.2025.112615}

  \par
  \textbf{Paper B:} \fullcite{othman2024comparison}.
  \\ \hspace*{2em}Link: \url{https://ieeexplore.ieee.org/abstract/document/10803510/}
  \par
  \textbf{Paper C:} \fullcite{othman2024cybersecurity}.
  \\ \hspace*{2em}Link: \url{https://ieeexplore.ieee.org/abstract/document/10803317/}
  
  \par
  \textbf{Paper D:} \fullcite{othman2023vuldat}.
  \\ \hspace*{2em}Link: \url{https://dl.acm.org/doi/abs/10.1007/978-3-031-46077-7_36}
  \par
  \textbf{Paper E:} \fullcite{othman2024vulnerability}.
  \\ \hspace*{2em}Link: \url{https://dl.acm.org/doi/abs/10.1145/3661167.3661170}
  \par
  \textbf{Chapter F:} \AtNextCite{\defcounter{maxnames}{4}}\fullcite{othman2026predictingknownvulnerabilitiesattack}.\par

\end{refsection}










\tableofcontents
\listoffigures
\listoftables


\mainmatter

\nomenclature{AUC}{\textbf{A}rea \textbf{U}nder the \textbf{C}urve}
\nomenclature{ATT\&CK}{\textbf{A}dversarial \textbf{T}actics, \textbf{T}echniques, and \textbf{C}ommon \textbf{K}nowledge}
\nomenclature{CAPEC}{\textbf{C}ommon \textbf{A}ttack \textbf{P}attern \textbf{E}numeration and \textbf{C}lassification}
\nomenclature{CVE}{\textbf{C}ommon \textbf{V}ulnerabilities and \textbf{E}xposures}
\nomenclature{CWE}{\textbf{C}ommon \textbf{W}eakness \textbf{E}numeration}
\nomenclature{CVSS}{\textbf{C}ommon \textbf{V}ulnerability \textbf{S}coring \textbf{S}ystem}
\nomenclature{CTI}{\textbf{C}yber \textbf{T}hreat \textbf{I}ntelligence}
\nomenclature{EER}{\textbf{E}qual \textbf{E}rror \textbf{R}ate}
\nomenclature{FPR}{\textbf{F}alse \textbf{P}ositive \textbf{R}ate}
\nomenclature{LSI}{\textbf{L}atent \textbf{S}emantic \textbf{I}ndexing}
\nomenclature{ML}{\textbf{M}achine \textbf{L}earning}
\nomenclature{NLP}{\textbf{N}atural \textbf{L}anguage \textbf{P}rocessing}
\nomenclature{ROC}{\textbf{R}eceiver \textbf{O}perating \textbf{C}haracteristic}
\nomenclature{SOTA}{\textbf{S}tate of the \textbf{A}rt}
\nomenclature{TF-IDF}{\textbf{T}erm \textbf{F}requency--\textbf{I}nverse \textbf{D}ocument \textbf{F}requency}
\nomenclature{TTPs}{\textbf{T}actics, \textbf{T}echniques, and \textbf{P}rocedures}
\nomenclature{TPR}{\textbf{T}rue \textbf{P}ositive \textbf{R}ate}
\nomenclature{VULDAT}{\textbf{V}ulnerability \textbf{D}etection from \textbf{A}ttack \textbf{T}ext}
\nomenclature{XSS}{\textbf{S}tored \textbf{C}ross-\textbf{S}ite \textbf{S}cripting}

\printabbreviations



\newcommand{\rqone}{\textbf{RQ1:} \textit{ Which sentence transformer models and attack information are most effective in predicting known vulnerability types?}}
\newcommand{\rqtwo}{\textbf{RQ2:} \textit{Can our approach recommend new semantic links that are not explicitly listed in MITRE repositories? }}
 \newcommand{\rqthree}{\textbf{RQ3:} \textit{How well do transformer models generalize to real-world data when predicting known vulnerability types from attack news reports?} }
\chapter{Introduction and Research Overview}
\label{ch:intro}

\huge{S}\normalsize{oftware systems have become the foundation of modern infrastructures and services. Critical domains such as finance, healthcare, transportation, energy, and communication depend on complex, interconnected systems to function reliably and efficiently. This increasing reliance on digital infrastructures has been accompanied by a significant escalation in cyber threats. Cyberattacks today are not isolated incidents but persistent, global challenges with devastating impacts on organizations, economies, and societies at large~\cite{admass2024cyber,rahman2023attackers,reis2021fixing}.
The cost of cybercrime illustrates the magnitude of the problem. According to recent estimates, cyberattacks caused damages worth approximately 3 trillion USD in 2015, doubled to 6 trillion USD in 2021, and are projected to surpass 10.5 trillion USD annually by 2025~\cite{Cybercrime}. These numbers place cybersecurity among the most urgent technological challenges of our time. Beyond financial loss, cyberattacks lead to disruptions in services, breaches of sensitive personal data, erosion of trust in digital systems, and in some cases, direct risks to human safety~\cite{pelletreau2024cybersecurity}~\cite{seh2020healthcare}.}

The scale and frequency of attacks further highlight the seriousness of the problem. Reports indicate that on average, an organization experiences over a thousand attempted attacks per week~\cite{CResearch}. Attackers exploit a broad spectrum of techniques, from social engineering and phishing campaigns to advanced persistent threats targeting critical software weaknesses. These attacks are carried out by a wide range of adversaries, including cybercriminal groups, state-sponsored actors, and hacktivist collectives, all of whom continuously evolve their methods. 
Furthermore, the number of reported vulnerabilities in the MITRE Common Vulnerabilities and Exposures (CVE)~\cite{CVEdataset} repository has surpassed 295{,}000. 
Analysts attempting to manually link these attacks with relevant vulnerabilities must often read lengthy and inconsistent documents, interpret unstructured narratives, and cross-reference multiple repositories. 
This process is time-consuming, demands high expertise, and is prone to omissions and human error. 
Consequently, manual interpretation delays detection and response, leaving systems exposed to further exploitation. 
Hence, there is a strong need for automation to improve both efficiency and reliability.
In this context, timely detection of vulnerabilities is of paramount importance. A vulnerability refers to a flaw in software, hardware, or system design that can be exploited by attackers to compromise confidentiality, integrity, or availability~\cite{CWE}. Once exploited, a vulnerability can allow adversaries to gain unauthorized access, disrupt services, steal sensitive information, or install malicious payloads. The faster vulnerabilities are identified and mitigated, the lower the risk of successful exploitation. However, delays in detection and remediation extend the window of opportunity for attackers, thereby amplifying the potential damage.

To support proactive defense, cybersecurity professionals rely on Cyber Threat Intelligence (CTI). CTI refers to the systematic collection, processing, and analysis of knowledge about adversaries, their behaviors, and the vulnerabilities they exploit~\cite{rahman2023attackers}. By providing evidence-based insights, CTI helps organizations anticipate threats, strengthen defenses, and respond effectively to incidents. At the core of CTI are structured knowledge bases maintained by the MITRE Corporation. These repositories have become standard references in cybersecurity research and practice: (1) MITRE Adversarial Tactics, Techniques, and Common Knowledge (ATT\&CK) provides a comprehensive framework of adversary tactics, techniques, and procedures (TTPs) observed in real-world attacks~\cite{ATTACK}.
(2) Common Attack Pattern Enumeration and Classification (CAPEC) offers a catalog of recurring attack strategies, including detailed information on execution, potential consequences, and mitigations~\cite{CAPEC}.
(3) Common Weakness Enumeration (CWE) documents common software and hardware weaknesses, such as coding errors and design flaws, that can lead to vulnerabilities~\cite{CWE}.
(4) CVE serves as the global dictionary of publicly disclosed vulnerabilities, providing unique identifiers and basic descriptions~\cite{CVEdataset}.
Collectively, these repositories constitute a powerful source of knowledge for defenders. They enable analysts to trace connections between adversary behavior and system vulnerabilities, prioritize patching, and design effective countermeasures. Furthermore, they have become critical in academic research, security product development, and industry standards.

However, these repositories are curated manually and evolve independently. This manual process leads to fragmentation and missing links between related entities. For example, ATT\&CK techniques often describe adversary behavior in detail but rarely specify the CVEs that can be exploited to realize those behaviors. Conversely, CVE entries describe vulnerabilities in affected products but usually lack contextual information about the attack techniques that might exploit them~\cite{sonmez2021classifying,elder2022really}. 
The CAPEC and CWE repositories partially help bridge these gaps by providing intermediate layers of abstraction: CAPEC links attack patterns to the techniques used by adversaries, while CWE defines the underlying software weaknesses that give rise to specific vulnerabilities.
Nevertheless, both resources remain incomplete and inconsistent due to the complexity of the domain and the rapid pace of cyber threat evolution.
The scale of these repositories further complicates matters. 
ATT\&CK includes 14 tactics, 201 techniques, and more than 400 sub-techniques, while the CVE database contains over 295,000 entries~\cite{CVEdataset}. The size and heterogeneity of these resources make manual mapping across repositories infeasible. Analysts seeking to trace a link from an observed attack technique to relevant vulnerabilities must engage in time-consuming cross-referencing, which is prone to omissions. This limitation hinders comprehensive threat analysis and timely vulnerability detection.

The central problem addressed in this research is the lack of automated methods to establish links between attack descriptions and known vulnerabilities. Despite the richness of CTI repositories, the absence of explicit cross-references between ATT\&CK, CAPEC, CWE, and CVE creates a fragmented landscape. Security analysts are left to manually search, interpret, and connect entries across multiple repositories. Consequently, when a cyberattack occurs, defenders may not know which vulnerabilities were exploited, or conversely, when a new vulnerability is disclosed, they may not know which attack techniques could leverage it. These blind spots weaken defensive strategies, delay remediation, and increase the risk of successful exploitation.

This research presents the existing research objectives and the motivations that guided this PhD work. 
The main goal is to \textit{infer vulnerabilities from cyberattack information by establishing links between attack descriptions and vulnerability descriptions}.
The motivation for this research stems from both practical and scientific needs. On the practical side, organizations require faster and more reliable ways to detect vulnerabilities and respond to cyberattacks. These incidents have been shown to cause wide-ranging consequences, including service disruptions, breaches of sensitive personal data, erosion of trust in digital infrastructures, and even direct risks to human safety, particularly in domains such as healthcare and critical infrastructures~\cite{andersson2024classification}.
On the scientific side, there is still limited understanding of how to automatically connect heterogeneous cybersecurity knowledge sources, such as ATT\&CK, CAPEC, CWE, and CVE, that are expressed in natural language and evolve independently. Existing research mainly focuses on isolated repositories or relies on manual expert analysis, leaving open questions about how modern Natural Language Processing (NLP) and representation learning techniques can be leveraged to model the semantic relationships between adversary behaviors and known software vulnerabilities. Addressing this gap contributes to advancing automated vulnerability intelligence and the broader field of empirical cybersecurity analytics~\cite{seh2020healthcare,li2021comprehensive}.

\section{Research Challenges}
\label{sec:intro:reseachchallenges}

This research identifies four main challenges in detecting vulnerabilities from cyberattack descriptions. 
Traditional threat intelligence systems are designed to work with structured repositories, such as MITRE ATT\&CK and CVE, which are well-established and widely adopted knowledge bases where attacks and vulnerabilities are explicitly documented.
However, these repositories evolve independently and often lack timely or detailed cross-links between related entities. Complementary information about emerging threats is frequently reported in unstructured sources such as news articles, security blogs, and incident reports.
Therefore, integrating insights from both structured and unstructured data presents four key challenges addressed in this research.
First, attack descriptions are expressed at multiple levels of abstraction, such as Tactics, Techniques, Procedures, and Attack Patterns in the MITRE ATT\&CK repository, and it is not clear which type provides the most informative signal for linking to vulnerabilities.
Second, selecting effective representation models is non-trivial: traditional NLP approaches fail to capture semantic meaning, while transformer-based models differ significantly in how they embed textual information and compute similarity between attack and vulnerability descriptions.
Third, constructing annotated datasets is difficult because repositories evolve independently, differ in structure, and often lack complete or consistent cross-references, making reproducibility and coverage a challenge. 
Finally, the variability and noise in real-world cyberattack reports complicate generalization, as models trained on curated datasets may not transfer effectively to unstructured sources. 
To address these challenges, this thesis develops automated methods that leverage advanced NLP techniques, particularly sentence transformers, to infer CVEs and predict links from cyberattack descriptions across both structured repositories and real-world reports.

Thus, summarizing, this PhD research addresses the following research challenges.
\begin{itemize}
\item \textbf{RC1}. \textit{Selecting informative attack descriptions:}
MITRE repositories capture adversarial behavior at multiple levels of abstraction, such as Tactics, Techniques, Procedures, and Attack Patterns. Each level varies in its degree of detail and vocabulary, which directly influences the effectiveness of predicting vulnerabilities. High-level Tactics tend to be too abstract to provide strong predictive signals, while detailed Procedures may include tool names or commands that add noise. 
The challenge is to determine which type of attack description provides the most reliable semantic signal for linking to CVEs, and to evaluate how models perform across these heterogeneous inputs.

\item \textbf{RC2}. \textit{Choosing effective representation models:}
 Another key challenge is selecting the most effective representation model to capture the semantic relationship between attack descriptions and vulnerabilities. 
    Traditional NLP techniques such as Term Frequency–Inverse Document Frequency (TF–IDF) and Latent Semantic Indexing (LSI) often fail to capture contextual and semantic relationships in cybersecurity text because they rely primarily on word co-occurrence and frequency statistics rather than understanding word meaning or sentence context. As a result, they treat words as independent tokens and cannot recognize that semantically related expressions.
    Transformer-based models, although more powerful, differ widely in architecture, embedding strategies, and computational requirements. For example, transformers use different pre-training tasks and tokenization methods, which influence how they encode semantic relationships and contextual meaning. Consequently, their performance and resource demands vary significantly across downstream tasks such as vulnerability detection.    
    The challenge is to systematically compare state-of-the-art (SOTA) models, identify their strengths and weaknesses, and select those that provide the best predictive performance across different types of attack information.

\item \textbf{RC3}. \textit{Constructing vulnerability datasets:}
Constructing annotated datasets that connect attack descriptions with CVE descriptions requires integrating multiple fragmented repositories, such as ATT\&CK, CAPEC, CWE, and CVE. 
    These repositories evolve independently, follow different structures, and often contain incomplete or inconsistent links. 
    Many CVEs lack explicit references to the techniques that exploit them, and many attack techniques remain without associated vulnerabilities. 
    The challenge is to design a reproducible and representative dataset that ensures sufficient coverage, consistency, and balance between positive and negative samples, thereby enabling robust model training and evaluation.

\item \textbf{RC4}. \textit{Generalization to unseen real-world data:}  
Cyberattack information is not limited to structured repositories such as CVE and ATT\&CK, but also appears in unstructured real-world sources, including news articles and blogs.
These sources often employ inconsistent terminology and informal language, making it more challenging to interpret and extract reliable threat intelligence. 
Therefore, the key challenge is to develop an approach that can link attack descriptions from such unstructured reports to relevant vulnerabilities, even when the text is noisy, ambiguous, or lacks technical detail.
Addressing this challenge requires evaluating whether similarity-based approaches, methods that use text embeddings to measure the closeness between attack and vulnerability descriptions, can generalize beyond curated repositories and still identify semantic links between attack descriptions and CVE reports.

\end{itemize}
\section{Research Goals and Questions}
\label{sec:intro:researchgoals}

\begin{figure*}[htb!]
\centering
\includegraphics[scale=1.4]{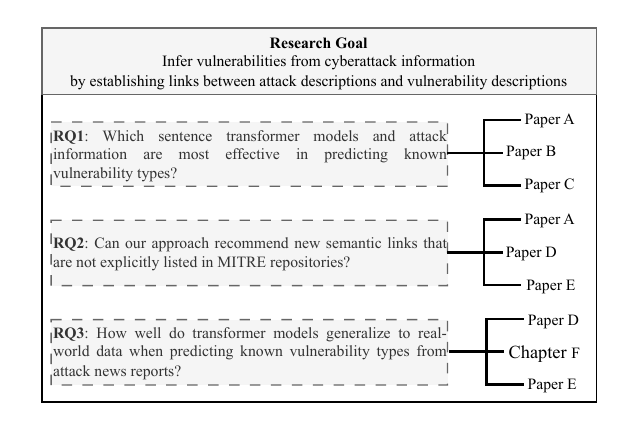}
\caption{Mapping of RQs to thesis papers.}
\label{fig:researchQuestion}
\end{figure*}

This PhD research aims to automatically detect and recommend vulnerabilities based on cyberattack descriptions using NLP techniques. 
From these challenges, we derive three research goals (RG1–RG3), each linked to a corresponding research question (RQ1–RQ3). The following subsections describe each goal and research question in detail. In addition, Figure~\ref{fig:researchQuestion} provides an overview of the three research questions and their correspondence to the thesis papers and chapters that address them.

\begin{itemize}
    \item \textbf{RG1}. \textit{Identify effective sentence transformer models and attack information types for vulnerability prediction.}  
The first goal is to evaluate which sentence transformer architectures (e.g., MPNet, MiniLM, RoBERTa) and types of attack information (e.g., Tactic, Technique, Procedure, Attack Pattern) yield the most accurate vulnerability predictions, since model performance may depend on the granularity and abstraction level of the input descriptions. Evaluating both dimensions together allows assessing how different model architectures interact with various types of attack information to influence predictive effectiveness.
The evaluation is based on standard performance metrics such as precision, recall, and F1-score.
This research goal aims to address the research challenges RC1 and RC2 by answering the following research question: 

\begin{itemize}
    \item \rqone
    
    To answer this question, we embed both attack and CVE descriptions into the same vector space using a sentence transformer model and then compute the similarity between their vectors. CVE reports are ranked according to their similarity scores to the attack description.
    Multiple types of attack information and several sentence transformer models are evaluated to identify the best-performing combination. The detailed answer to this question is presented in our papers: \textbf{Paper A~\cite{othman2025attack}, Paper B~\cite{othman2024comparison}, and Paper C~\cite{othman2024cybersecurity}}.
    %
\end{itemize}

\item \textbf{RG2}. \textit{Recommend semantic CVE links that are not explicitly listed in MITRE repositories.}  
The second goal is to uncover implicit but relevant CVE links that are not currently recorded in MITRE repositories. These missing links could significantly enhance vulnerability knowledge bases and strengthen threat intelligence systems. This goal explores whether NLP models can propose valid links by capturing the semantics between attack descriptions and CVE reports. This research goal aims to address the research challenge RC3 by answering the following research question:
\begin{itemize}
    \item \rqtwo 
    
To answer this question, we developed a semantic recommendation approach that detects undocumented links between attack descriptions and CVE reports.
By analyzing the semantic similarity between the two, the approach identifies semantic links that are absent from existing repositories. To validate these links, we manually examined the predicted links to determine whether the attack and CVE descriptions were truly related.
This process confirmed that many of the suggested links represent realistic and relevant relationships between vulnerabilities and attack behaviors. The detailed answer to this question is discussed in our papers: \textbf{Paper A, Paper D~\cite{othman2023vuldat}, and Paper E~\cite{othman2024vulnerability}}.

\end{itemize}
\item \textbf{RG3}. \textit{ Evaluate model generalization to unseen real-world data.}  
The third goal is to evaluate whether the proposed approach can accurately predict vulnerabilities from unseen, unstructured sources such as cyberattack news articles, thereby assessing its ability to generalize beyond the structured repositories used for training.
This evaluation combines quantitative measures (e.g., semantic similarity thresholds, top-K accuracy) with qualitative assessment through manual validation. This research goal aims to address the research challenge RC4 by answering the following research question:

\begin{itemize}
    \item \rqthree 
    
    To answer this question, we evaluated the effectiveness of the approach on a dataset of real-world cybersecurity news reports. Unlike structured repositories, these reports often contain noisy, informal, and diverse language. We applied the similarity-based method to infer links between news descriptions of attacks and CVE reports. To validate the predictions, we manually assessed whether the news descriptions and CVE reports were related. This evaluation provided insights into the model’s ability to generalize to heterogeneous, real-world scenarios. The detailed answer to this question is discussed in our papers: \textbf{Paper D, Paper E, and Chapter F~\cite{othman2026predictingknownvulnerabilitiesattack}}.
\end{itemize}

Together, these goals advance the automation and accuracy of vulnerability detection from cyberattack descriptions. They contribute to broader objectives of enhancing threat intelligence coverage, enabling semantic reasoning across cybersecurity knowledge bases, and improving the robustness of NLP models for practical security applications.

\end{itemize}

\section{Our Contributions}
We initiated this PhD research to address the challenge of detecting and recommending vulnerabilities from cyberattack descriptions. 
Our early investigations introduced the research methodology for linking attack descriptions to vulnerabilities using the TF-IDF technique~\textbf{(Paper D, Paper E)}. Then, we conducted an empirical comparison between traditional NLP approaches, specifically TF-IDF and LSI combined with multiple classifiers, and modern sentence transformer models. Using CAPEC attack pattern descriptions as input to predict CVE reports~\textbf{(Paper B)}.
We subsequently designed and developed VULDAT, a proof-of-concept tool for automated vulnerability detection that links ATT\&CK Technique descriptions to CVEs and evaluates the effectiveness of nine transformer models~\textbf{(Paper C)}. This work was followed by a comprehensive evaluation of 14 state-of-the-art sentence transformer models across four types of attack information (Tactic, Technique, Procedure, and Attack Pattern). The study systematically analyzed model performance, identified the most effective model–description combinations, and uncovered 275 previously undocumented Technique–CVE links, many of which were later incorporated into MITRE repositories~\textbf{(Paper A)}. We later extended this approach beyond structured repositories by investigating the generalization of transformer models to unstructured sources, evaluating their ability to infer vulnerabilities from noisy and heterogeneous cyberattack news reports~\textbf{(Chapter F)}.
Altogether, these papers and chapters are the cornerstone of this PhD research, which consolidates our findings and orchestrates the following key contributions:
\begin{enumerate}
    \item \textit{Automated Attack-to-Vulnerability Linking:} We developed a proof-of-concept approach that leverages sentence transformer models to automatically infer links between attack descriptions and CVE reports. This work reduced reliance on manual mapping and enabled faster incident response.
    \item \textit{Comprehensive Transformer Model Evaluation:} We conducted a large-scale comparative study of 14 sentence transformer models across different types of attack descriptions, demonstrating that Technique descriptions are the most informative, with the multi-qa-mpnet-base-dot-v1 (MMPNet) model achieving an F1-score of 89\%.
    \item \textit{Annotated Datasets:} We constructed a novel annotated mapping dataset that explicitly links attack descriptions with vulnerabilities documented in MITRE repositories (CVE, CWE, CAPEC, and ATT\&CK)~\cite{VULDATDataSet}. This dataset provides a reproducible benchmark for evaluating automated vulnerability detection methods and supports further research on enriching cyber threat intelligence knowledge bases.

    \item \textit{Discovery of Undocumented Links:} Through empirical analysis and manual validation, we uncovered 275 previously undocumented Technique–CVE associations, many of which were later added to the MITRE repositories. This contribution demonstrates that current repositories are incomplete and highlights the potential of automated methods to enrich existing vulnerability knowledge bases.
    \item \textit{Generalization to Unseen Real-World Data:}
    We provided a proof-of-concept tool that supports CTI research by enabling automated detection of vulnerabilities from cyberattack news reports. Key
findings include:
    \begin{itemize}
    \item  A tool implementation based on the SOTA sentence transformer model MPNet, capable of predicting vulnerabilities from attack reports in real-time.
 \item 
A systematic evaluation on a dataset of 100 real-world SecurityWeek news articles, combining expert manual review with three oracle-based validation methods to assess prediction accuracy and robustness.
    \end{itemize}
    \item \textit{Empirical Validation and Findings:}
    Across multiple empirical studies and experiments, we validated the effectiveness of our approaches. The key findings can be summarized as follows:

    \begin{itemize}
        \item Transformer models outperform traditional NLP methods in detecting vulnerabilities from attack descriptions.
        \item Technique descriptions are the most informative attack type for vulnerability prediction.
        \item Sentence transformers can recommend valid, undocumented CVE links, enriching repositories.
        \item Models can generalize to real-world cyberattack news, supporting proactive vulnerability detection.
    \end{itemize}
\end{enumerate}

\section{Outline}
\label{sec:intro:outline}
This thesis is organized into seven chapters, each addressing a specific aspect of the research. The following overview outlines the chapters that collectively address the problem of automated vulnerability detection from cyberattack descriptions.
Chapter~\ref{ch:intro} introduces the research problem, outlines the main challenges, and motivates the need for automated approaches to link cyberattacks with vulnerabilities. It also presents the research goals, questions, and contributions of the study. 
Chapter~\ref{ch:background} provides the necessary background by discussing vulnerabilities and weaknesses, the MITRE family of knowledge bases (ATT\&CK, CAPEC, CWE, CVE), and the principles of sentence transformer models. 
Chapter~\ref{ch:lr} presents related work, situating the contributions of this thesis within the existing literature and identifying the research gaps it seeks to address.
Chapter~\ref{ch:methods} presents the research methodology, covering data collection and annotation, preprocessing of attack and vulnerability descriptions, embedding strategies, similarity analysis, and evaluation procedures. 
Chapter~\ref{ch:results} reports the experimental results in relation to the three research goals: identifying effective models and attack information types, recommending missing links between attacks and vulnerabilities, and evaluating model generalization on real-world news.
Finally, Chapter~\ref{ch:threats} discusses the limitations and threats to validity. Chapter~\ref{ch:conclusion} concludes the thesis by summarizing its main contributions and outlining future directions for advancing automated, scalable vulnerability detection.

\chapter{Background}
\label{ch:background}

\huge{I}\normalsize{n this chapter, we provide an overview of the key concepts relevant to our study, including software vulnerabilities, weaknesses, and attacks, as well as the knowledge bases used in this study. We also describe the sentence transformer models as they have been used extensively in the current work.
}
\section{Vulnerability knowledge bases}
\label{sec:vulnerability}
A \textbf{vulnerability} is defined as a flaw in software, firmware, hardware, or a service component resulting from a weakness that can be exploited, causing a negative impact on the confidentiality, integrity, or availability of an impacted component or components~\cite{dong2023dekedver,elder2024survey}. According to NIST, a \textbf{software vulnerability} is "a security flaw, glitch, or weakness found in software code that could be exploited by an attacker (threat source)"~\cite{dempsey2017automation}. By contrast,  a \textbf{ weakness} is a condition in the software, firmware, hardware, or service components that, under certain circumstances, could contribute to introducing vulnerabilities~\cite{CWE, esposito2024validate}. Thus, weaknesses denote potential security problems even in the absence of a known exploit, whereas a weakness becomes a vulnerability once an attacker identifies a viable method of exploitation.

A weakness in a system can enable an attacker to gain elevated privileges, such as assuming the identity of a superuser or system administrator, thereby achieving unauthorized access and compromising the system’s security~\cite{ alevizopoulou2021social, gasmi2019information,vuldatPaper}. For example, improper sanitization of user input in a web application constitutes a weakness (e.g., CWE-79: Improper Neutralization of Input During Web Page Generation). If exploited through script injection, then it becomes a vulnerability that allows the execution of a Stored Cross-Site Scripting (XSS) attack. A specific case is documented in CVE-2022-4826, affecting the Simple Tooltips WordPress plugin prior to version 2.1.4. The vulnerability enabled contributors to inject malicious JavaScript, leading to persistent XSS. In addition, such vulnerabilities are recorded in the CVE repository, and often mapped to CAPEC attack patterns (e.g., CAPEC-38: Command Injection), which illustrate how adversaries can exploit the underlying weakness.  Since it is impossible to predict whether a weakness exists when a vulnerability will be exploited or what impact an attack will have on a system, it is of foremost importance to provide stakeholders instruments to detect and, possibly, remove vulnerabilities~\cite{iorga2021yggdrasil, queiroz2019eavesdropping,baccar2021automated, dionisio2019cyberthreat,tang2023csgvd}.

When a vulnerability is suspected in a piece of software, a CVE identifier can be reserved through the CVE program~\cite{CVEdataset}. The CVE dictionary consists of independent entries, each of which contains (1) a unique identification number (CVE-ID), (2) a concise description of the issue, and (3) at least one public reference to external sources that provide additional details. A CVE identifier follows the syntax \texttt{CVE-YEAR-5digits}, where \emph{YEAR} denotes the year of reservation and \emph{5digits} is a sequence number for that year. Importantly, identifiers may be reserved before public disclosure: MITRE assigns the identifier but does not publish the entry until the vulnerability is confirmed. Thus, in some cases, reserved identifiers are never confirmed and hence never published. Once a vulnerability is confirmed and the CVE record is published, the responsible vendor typically begins remediation, such as developing and releasing a patch, to eliminate the flaw before attackers can exploit it~\cite{dempsey2017automation}. Moreover, the CVE repository is publicly accessible and aggregates vulnerability reports from diverse systems and applications~\cite{CVEdataset, othman2024vulnerability, WhatCVE}. For every publicly known vulnerability, a CVE record includes the CVE-ID and description, references to advisories or vendor resources, when available evidence of observation or exploitation~\cite{sun2023automatic}, known fixes and mitigations, and severity scores based on the Common Vulnerability Scoring System (CVSS)~\cite{CVSS}. Accordingly, Table~\ref{tab:cveDataset} shows an example of an attack and description for CWE, and CVE records. We can see the common vulnerability and exposure CVE-2022-4826 affecting the Simple Tooltips WordPress plugin prior to version 2.1.4. This vulnerability is related to improper validation and escaping of shortcode attributes before rendering them on a page or post. As a result, users with contributor roles and above could inject malicious scripts, leading to Stored Cross-Site Scripting (XSS) attacks. This vulnerability can be linked to the pattern of attack CAPEC-38 with an attacker attempting to load a malicious resource into the program so that it will be executed.

\begin{figure}[t]
\centering
\includegraphics[width=\linewidth]{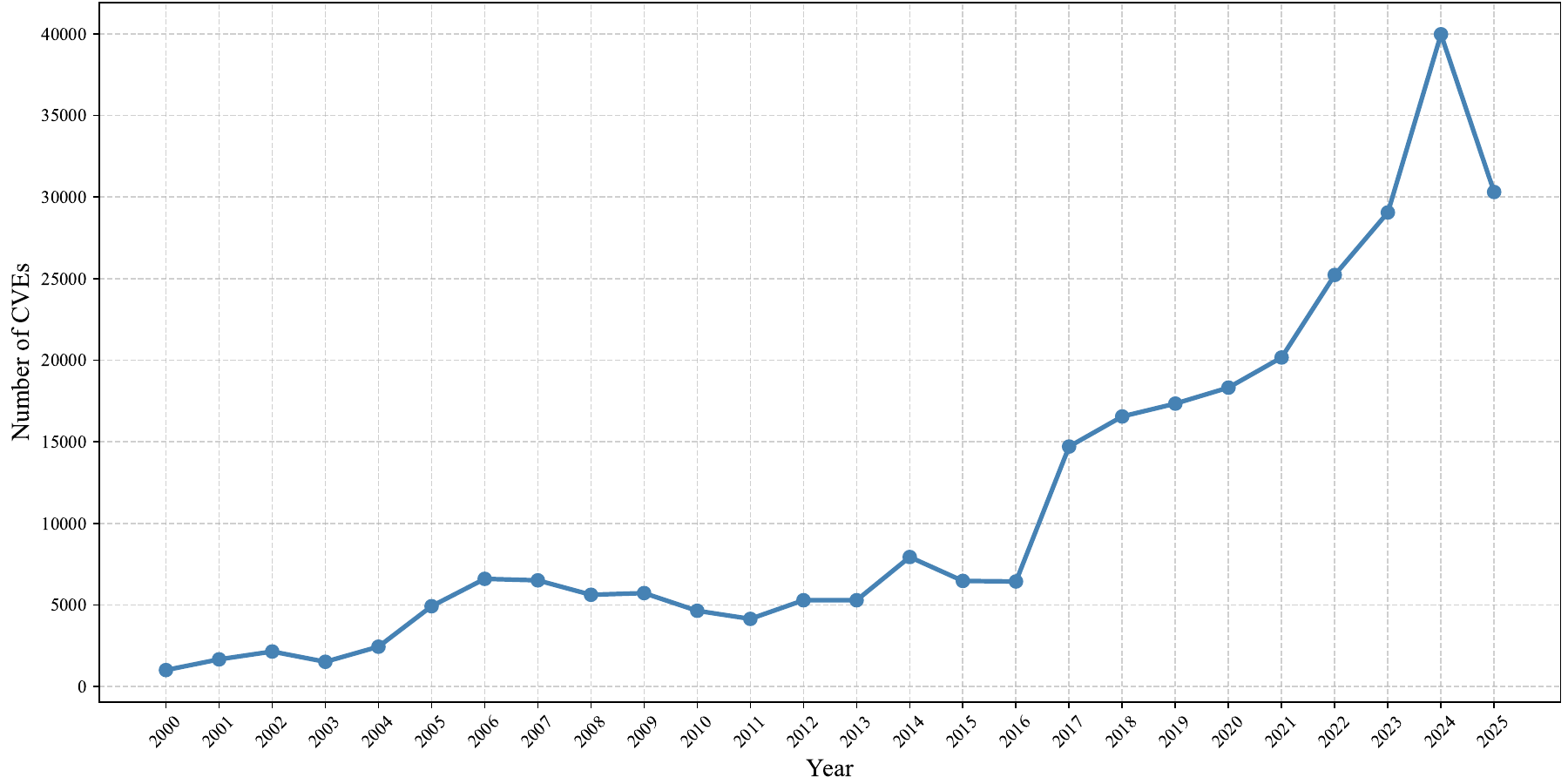}
\caption{Annual number of published CVEs.}
\label{fig:cve_trend}
\end{figure}

CWE is a community-developed collection of typical weaknesses in software, coding errors, and security flaws. A CWE report includes various information such as a name, an identification number (CWE-ID), a description, an observed example containing CVE reports ID related to CWE, and related attack patterns highlighting the relationships between specific attack patterns and CWE. CWE and CVE reports are connected through the CVE-ID. 
Figure~\ref{fig:cve_trend} summarizes the annual number of published CVE records since 2000. As visible in the figure, disclosure volumes remained in the low thousands until the mid‑2010s, then accelerated sharply beginning in 2017 and continued to rise. Thus, the sustained growth underscores the expanding scale of the vulnerability landscape and motivates the need for automated, semantically informed methods for linking and understanding how adversaries operate, which requires attack‑oriented repositories.
 
\section{Attack knowledge bases}
\label{sec:TTP}
While CWE and CVE offer a defender‑centric catalogue of weaknesses and concrete vulnerabilities, understanding how adversaries operate requires attack‑oriented repositories. Accordingly, this thesis leverages the information on attacks contained in the two complementary sources: MITRE ATT\&CK and CAPEC. Together, they provide an attacker-centric perspective that characterizes tactics, techniques, procedures, and attack patterns; thus, they complement vulnerability repositories and enable end-to-end reasoning across the vulnerability–attack landscape.

\par
\begin{figure*}[htb!]
\centering
\includegraphics[scale=0.9]{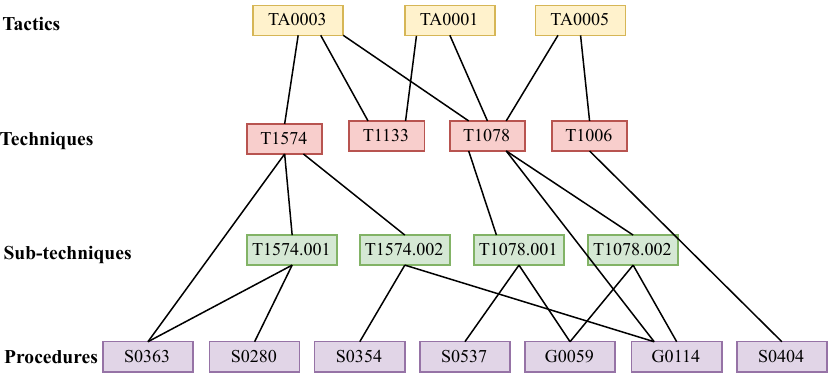}
\caption{An example of the ATT\&CK relations.}
\label{fig:mitrerelationship}
\end{figure*}

 \noindent \textbf{MITRE ATT\&CK}.
ATT\&CK~\cite{ATTACK} is a framework that provides a comprehensive inventory of \textbf{Tactics}, \textbf{Techniques}, and \textbf{Procedures} (TTPs) adopted by attackers during various stages of a cyberattack~\cite{rahman2024attackers,son2023introduction, irshad2023cyber}. TTPs help security experts understand adversary behavior, guide threat detection and response efforts, and enhance organizational defenses against cyberthreats. According to the ATT\&CK model, tactics represent the goal of an adversary's attack, like ways to get access to a secured network, whereas techniques define how to carry out an attack in the context of a certain Tactic, like using phishing attempts~\cite{MITRE, satvat2021extractor}. Techniques can be further detailed into subtechniques. Procedures provide detailed insights into how adversaries effectively use the techniques in the actual attacks. For example, a procedure could involve sending emails camouflaged as legitimate communications from a trusted source to trick users into clicking malicious links.
Researchers typically use TTPs to profile or examine the attack life cycle on a specific system. 
Our study used Version 14 of the ATT\&CK framework, which contains 14 Tactics, 201 Techniques, 424 Subtechniques, and 809 procedures. The framework also provides the ATT\&CK Matrix, which breaks each attack down into tactics, techniques, and procedures. 
An example of the relations among Tactics, Techniques, Subtechniques, and Procedures are shown in Figure~\ref{fig:mitrerelationship}. All relations are many-to-many. For instance,  the Technique \textit{``event triggered execution"} has 16 Subtechniques, whereas, \textit{``account access removal, acquire access, and audio capture"} has no Subtechnique. 

\par \noindent \textbf{Attack Pattern.} The Common Attack Pattern Enumeration and Classification (CAPEC)~\cite{CAPEC} is a catalog of common Attack Patterns. CAPEC is linked to CWE using the CWE-ID, creating an association between specific attack techniques and the underlying weaknesses they exploit~\cite{ATTACK}. Attack Patterns allow us to see what techniques in the ATT\&CK repository an adversary could use to target the weaknesses and vulnerabilities.
For instance, the Pattern\textit{ ``privilege escalation (CAPEC-233)"} elevates attackers' privilege to perform an action they are not supposed to be authorized to perform. Related to this Pattern, one can link the Technique \textit{``Abuse Elevation Control Mechanism (T1548)"} that describes the techniques that attackers use to gain higher-level permissions on a system or network. Among those, there is the Technique \textit{``Access token manipulation (T1134)"}, which, can be implemented in Metasploit~\cite{Metasploit, Metasploitprocedure} with the procedure  \textit{``named-pipe impersonation"}~\cite{MITRE, wu2021price, noor2019machine}.
We will connect ATT\&CK with CAPEC in our dataset, matching the techniques' IDs.

ATT\&CK focuses on how attackers behave TTPs at tactical and technical levels, whereas Attack Patter characterizes what reusable patterns they employ to exploit weaknesses. For example, ATT\&CK Tactics are abstract and short, while Attack Patterns are typically longer and more technically detailed. Table~\ref{tab:AttacsDescriptions} summarizes the typical characteristics of the four types of attack descriptions. 
In addition, CAPEC provides rich textual descriptions and explicit cross-references to ATT\&CK techniques and to CWE; via CVE tags present in CWE records, this creates an indirect bridge to CVE.
Consequently, integrating ATT\&CK and CAPEC with CWE and CVE yields a unified perspective that links attacks to vulnerabilities. This connection is essential, as it enables analysts to trace adversarial behaviors to the specific weaknesses and CVEs they exploit, thereby improving mitigation and defense strategies.
However, manually linking 625 ATT\&CK techniques to over 295,000 CVEs is impractical. To address this challenge, we employ sentence transformers to achieve a fully automated, semantically driven integration across ATT\&CK, CAPEC, CWE, and CVE repositories.

\begin{table}[htb]
 \caption{Typical characteristics of the attacks types.}
    \label{tab:AttacsDescriptions}
    \centering
    \begin{tabular}{@{\extracolsep{\fill}}p{1.7cm}p{3cm}p{3cm}p{3cm}p{3cm}}
    \hline
&\textbf{Tactic}&\textbf{Technique}&\textbf{Procedure}         & \textbf{Pattern} \\\hline
\textbf{Definition}&               	 High-level goal of an adversary  &	 General method of how the adversary achieves a goal 	 &Specific instance of an attacker using a technique 	 &Structured description of a recurrent method of attack \\          \textbf{Vocabulary} &      	 Abstract, it uses mostly generic terms    &                  	 It is more specific, but it still uses reusable terms        	& It includes tool names, scripts, commands         &  	It uses  technical terms, exploit-specific language      \\                
     \hline
    \end{tabular}
   
\end{table}

\section{Sentence transformers}\label{sec:transformer}
Sentence transformers are pre-trained, Transformer-based encoders that map sentences into dense, fixed-length vectors that preserve semantic meaning. These embeddings enable similarity-based operations such as semantic search, paraphrase identification, clustering, and large-scale retrieval using simple similarity functions (e.g., cosine similarity) in the embedding space~\cite{nie2022fine}. However, these models often contain millions of parameters, making fine-tuning and deployment challenging due to latency and capacity constraints. 

Most sentence-transformer pipelines comprise four standard components: (i) a Transformer encoder (e.g., BERT, RoBERTa, MPNet, T5) that produces contextual token representations; (ii) subword tokenization to handle open-vocabulary text; (iii) a pooling layer (commonly mean pooling over the last hidden layer,  occasionally CLS pooling or max pooling) to obtain a single fixed-size vector per sentence; and (iv) a similarity head that scores pairs via cosine similarity. In evaluation and downstream usage, vectors are typically L2-normalized, which stabilizes cosine similarity across lengths~\cite{reimers2019sentence}.

\begin{figure}[h!]
  \centering
  \begin{subfigure}[b]{0.45\textwidth}
    \centering
    \hspace*{3em} 
    \includegraphics[width=\textwidth]{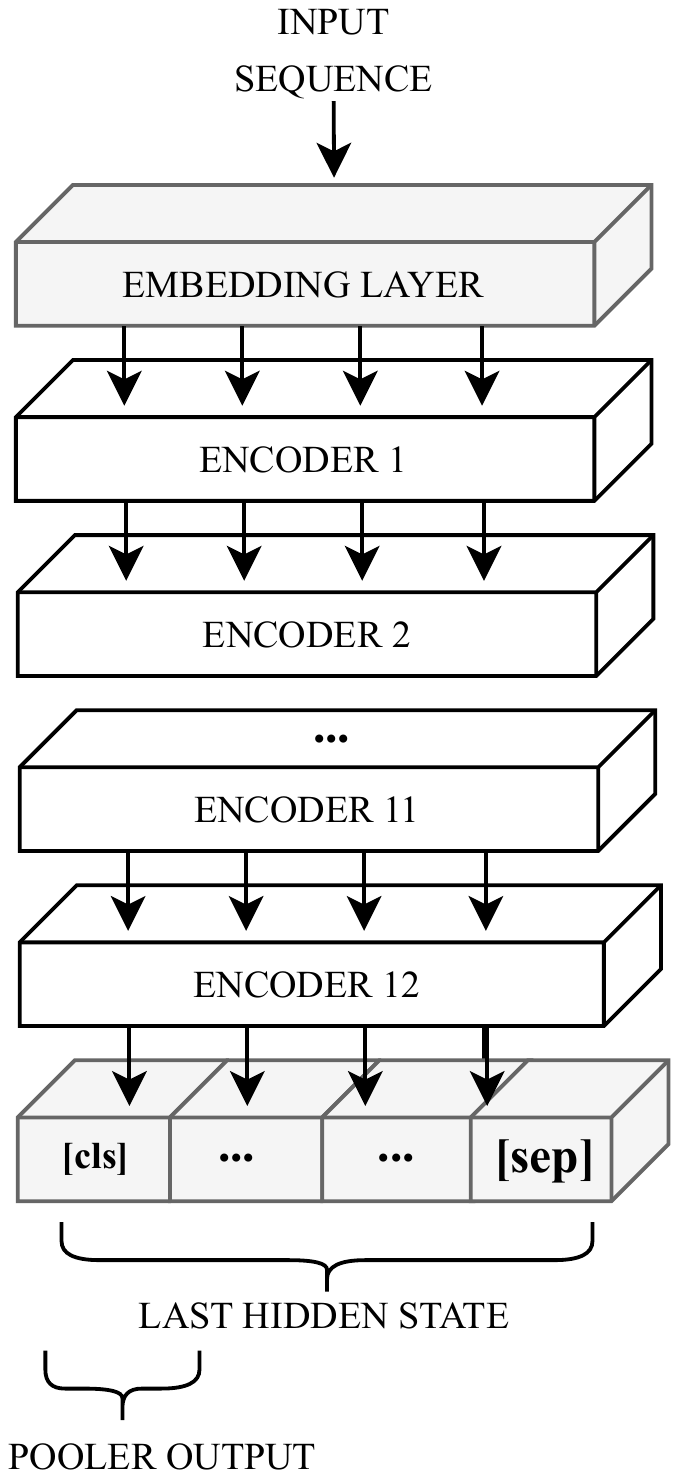}
    \caption{Transformer-based encoder architecture~\cite{reimers2019sentence}.}
    \label{fig:encoder}
  \end{subfigure}
  \hfill
  \begin{subfigure}[b]{0.45\textwidth}
    \centering
    \includegraphics[width=0.95\textwidth]{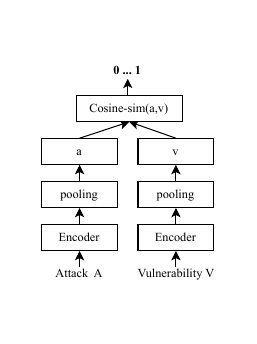}
    \caption{Cosine similarity computation between attack and vulnerability embeddings~\cite{reimers2019sentence}.}
    \label{fig:cosine}
  \end{subfigure}

  \caption{Overview of the embedding-based vulnerability detection pipeline.}
  \label{fig:embedding-pipeline}
\end{figure}

As shown in Figure~\ref{fig:embedding-pipeline}, the transformer encoder architecture (Figure~\ref{fig:encoder}) processes the input sequence through stacked self-attention layers to produce contextualized embeddings. A pooling operation then aggregates token-level representations into a single sentence embedding. Finally, in the embedding-based vulnerability detection pipeline (Figure~\ref{fig:cosine}), attack and vulnerability descriptions are encoded separately, pooled, and compared using cosine similarity, which yields a score in the range [0,1] indicating their semantic relatedness.

In this work, we utilize 14 SOTA pre-trained sentence transformer models, as summarized in Table~\ref{tab:pretrainedmodels}. These models are recognized as top performers in semantic similarity~\cite{muennighoff2022mteb,colangelo2025comparative, choi2021evaluation, pretrained}. All models are sourced from the Hugging Face Model Hub~\footnotemark[1], \footnotetext[1]{\url{Hugging Face: https://huggingface.co/}}which provides a comprehensive repository of transformer-based models for natural language processing tasks.
The selected models have been extensively benchmarked and fine-tuned for sentence-level tasks, such as semantic textual similarity, information retrieval, and clustering~\cite{muennighoff2022mteb,colangelo2025comparative, choi2021evaluation, pretrained}. Notably, models like MPNet, MiniLM, MSMARCO, RoBERTa, and T5 consistently achieve strong results across benchmarks~\cite{ muennighoff2022mteb, colangelo2025comparative}. Lightweight models, including PAlbert, have also demonstrated competitive performance in the evaluations~\cite{choi2021evaluation}.

\begin{table*}[htb]
\caption{Overview of SOTA Sentence Transformer Models per architecture and ordered by embedding dimensions.}
\label{tab:pretrainedmodels}
\centering
 \fontsize{8pt}{13pt}\selectfont
\begin{tabular}{llccc}
\hline  
\textbf{Acronym}&\textbf{Model} & \textbf{Architecture} & \textbf{Embedding Dimension}  & \textbf{Model Size} \\
\hline
PAlbert	&	paraphrase-albert-small-v2\footnotemark[2] 	&	 ALBERT 	&	768	&	 43 \, MB \\
PTinyBERT	&	paraphrase-TinyBERT-L6-v2\footnotemark[3] 	&	 TinyBERT 	&	768	& 240  MB \\
MDBERT	&	multi-qa-distilbert-cos-v1\footnotemark[4] 	&	 DistilBERT 	&	768	&	 250 MB \\
MSMBERT	&	msmarco-bert-base-dot-v5\footnotemark[5] 	&	 BERT 	&	768	&	 420 MB \\
DRoBERTa	&	all-distilroberta-v1\footnotemark[6] 	&	 DistilRoBERTa 	&	768	&	 290 MB \\
Roberta	&	all-roberta-large-v1\footnotemark[7] 	&	 RoBERTa 	&	1024	&	 1.36 GB \\
\hline
MiniLM6	&	all-MiniLM-L6-v2\footnotemark[8] 	&	 MiniLM 	&	384	&	 80 \, MB \\
MiniLM12	&	all-MiniLM-L12-v2\footnotemark[9] 	&	 MiniLM 	&	384	&	 120 MB \\
MMiniLM6	&	multi-qa-MiniLM-L6-cos-v1\footnotemark[10] 	&	 MiniLM 	&	384	&	 80 \, MB \\
PMiniLM6	&	paraphrase-MiniLM-L6-v2\footnotemark[11] 	&	 MiniLM 	&	384	&	 80 \, MB \\
PMiniLM12	&	paraphrase-multilingual-MiniLM-L12-v2\footnotemark[12] 	&	 MiniLM 	&	384	&	 420 MB \\
\hline
MPNet	&	all-mpnet-base-v2\footnotemark[13] 	&	 MPNet 	&	768	&	 420 MB \\
MMPNet	&	multi-qa-mpnet-base-dot-v1\footnotemark[14] 	&	 MPNet 	&	768	&	 420 MB \\
\hline
XXLT5	&	gtr-t5-xxl\footnotemark[15] 	&	 T5-XXL 	&	4096	&	 9.23 GB \\
\hline
\end{tabular}%
\end{table*}
\hspace{-4em}
\footnotetext[2]{\url{PAlbert:https://huggingface.co/sentence-transformers/paraphrase-albert-small-v2}}
\footnotetext[3]{\url{PTinyBERT: https://huggingface.co/sentence-transformers/paraphrase-TinyBERT-L6-v2}}
\footnotetext[4]{\url{MDBERT: https://huggingface.co/sentence-transformers/multi-qa-distilbert-cos-v1}}
\footnotetext[5]{\url{MSMBERT: https://huggingface.co/sentence-transformers/msmarco-bert-base-dot-v5}}
\footnotetext[6]{\url{DRoBERTa: https://huggingface.co/sentence-transformers/all-distilroberta-v1}}
\footnotetext[7]{\url{Roberta: https://huggingface.co/sentence-transformers/all-roberta-large-v1}}
\footnotetext[8]{\url{MiniLM6: https://huggingface.co/sentence-transformers/all-MiniLM-L6-v2}}
\footnotetext[9]{\url{MiniLM12: https://huggingface.co/sentence-transformers/all-MiniLM-L12-v2}}
\footnotetext[10]{\url{MMiniLM6: https://huggingface.co/sentence-transformers/multi-qa-MiniLM-L6-cos-v1}}
\footnotetext[11]{\url{PMiniLM6: https://huggingface.co/sentence-transformers/paraphrase-MiniLM-L6-v2}}
\footnotetext[12]{\href{PMiniLM12:https://huggingface.co/sentence-transformers/paraphrase-multilingual-MiniLM-L12-v2}{PMiniLM12: \texttt{https://huggingface.co/sentence-transformers/paraphrase-multilingual-}\\\hspace*{2em}\texttt{ MiniLM-L12-v2}}}
\footnotetext[13]{\url{MPNet: https://huggingface.co/sentence-transformers/all-mpnet-base-v2}}

\footnotetext[14]{\url{MMPNet: https://huggingface.co/sentence-transformers/multi-qa-mpnet-base-dot-v1}}
\footnotetext[15]{\url{XXLT5: https://huggingface.co/sentence-transformers/gtr-t5-xxl}}
A summary table listing these models along with their architecture, embedding size, and benchmark performance for semantic similarity is available in the official sentence transformers documentation~\cite{pretrained}, supporting the transparency and reproducibility of our model selection.
In addition, we selected models with different architectures and examined how these architectural differences affect performance in linking attack descriptions to CVEs.

\par\noindent \textbf{BERT.} 
BERT~\cite{reimers2019sentence} is a pre-trained transformer model designed for natural language understanding. Its key innovation lies in a deep bidirectional transformer encoder that captures context from both left and right of a token simultaneously, in contrast to traditional left-to-right or right-to-left models. This bidirectional context significantly improves performance in tasks such as classification, question answering, and semantic similarity. In this thesis, we utilize several BERT-based sentence transformers. Among them, \texttt{MSMBERT} is fine-tuned on the MS MARCO dataset for passage retrieval, making it highly effective for semantic search. By contrast, \texttt{PAlbert} and \texttt{PTinyBERT} represent lightweight BERT variants: 
ALBERT~\cite{lan2019albert} reduces BERT’s parameter count through cross-layer weight sharing and factorized embeddings, while TinyBERT~\cite{jiao2019tinybert} compresses BERT through multi-stage knowledge distillation~\cite{HintonEtAl2015}, yielding compact models optimized for paraphrase and similarity tasks.

\par\noindent \textbf{DistilBERT and DistilRoBERTa.}
Distillation-based models reduce the computational footprint of large transformers while retaining most of their accuracy. \texttt{MDBERT} is derived from DistilBERT~\cite{sanh2019distilbert}, a compressed version of BERT that reduces model size by 40\% and inference time by 60\% while maintaining 97\% of the original model’s accuracy. Similarly, \texttt{DRoBERTa} is based on DistilRoBERTa, which distills the RoBERTa architecture. Both models are well-suited for real-time applications where computational efficiency is critical. RoBERTa itself~\cite{liu2019roberta} improves upon BERT by removing the next-sentence prediction objective, using dynamic masking, and training on much larger corpora, which enhances generalization and robustness.

\par\noindent \textbf{MiniLM.} 
MiniLM~\cite{wang2020minilm} is a line of highly efficient transformer encoders designed to compress the representational power of large-scale models into compact architectures. Instead of distilling outputs directly~\cite{HintonEtAl2015}, MiniLM transfers the self-attention distributions from the teacher model (BERT) to the student, capturing relational information between queries, keys, and values. As a result, MiniLM achieves strong accuracy with far fewer parameters. In this thesis, we evaluate several MiniLM-based checkpoints, including \texttt{MMiniLM6}, \texttt{MiniLM6}, \texttt{MiniLM12}, \texttt{PMiniLM6}, and \texttt{PMiniLM12}. These models vary in depth, 6 or 12 transformer layers, and training objective, multi-qa retrieval vs. paraphrase STS. Notably, the multilingual variant \texttt{PMiniLM12} supports more than 50 languages, extending the applicability of our framework to multilingual attack descriptions.


\par\noindent \textbf{MPNet.}
MPNet~\cite{song2020mpnet} builds on BERT and XLNet by combining masked language modeling with permutation language modeling, enabling the model to capture bidirectional context while also modeling token dependencies more effectively. This hybrid approach allows MPNet to produce richer contextual embeddings, especially for long or complex sentences. In this thesis, we evaluate two different models: \texttt{MPNet} and \texttt{MMPNet}, extend BERT by integrating XLNet’s permutation-based training, making it suitable for contextual understanding and retrieval accuracy.

\par\noindent\textbf{T5.}
T5~\cite{raffel2020exploring} reformulates all NLP tasks into a text-to-text framework, allowing the same model to handle classification, summarization, translation, and retrieval by conditioning on task-specific prefixes. Within the sentence-transformer family, the T5 models use only the encoder component for embedding tasks. The \texttt{XXLT5} variant included in our evaluation represents the largest model in our set, with over 9 billion parameters and 4096-dimensional embeddings.

In this thesis, we employ pre-trained transformer models to generate embeddings for attack and vulnerability descriptions. As illustrated in Figure~\ref{fig:embedding-pipeline}, and then apply a similarity layer based on cosine similarity to compare the embeddings to determine whether an attack is linked to a vulnerability, and predictions are accepted when similarity exceeds a threshold (see Section~\ref{sec:TuningEmbedding}).

\chapter{Related Work}
\label{ch:lr}
\huge{I}\normalsize{n this chapter, we present a comprehensive overview of the existing research relevant to this PhD project. 
Research in this domain spans multiple approaches, from classical NLP methods to advanced transformer-based models.
In this chapter, we structure this discussion into three thematic areas: (1) Traditional NLP methods for cybersecurity text analysis, (2) Vulnerability-to-Attack mapping, and (3) Attack-to-Vulnerability mapping
Together, these sections establish the foundation for this thesis and motivate the need for novel approaches that can bridge the gaps in automated attack-to-vulnerability mapping.}

\section{Traditional NLP Methods for Cybersecurity Text Analysis}
\label{sec:classicalNLP}
The earliest studies in cybersecurity text analysis relied on statistical NLP and classical machine learning techniques. These approaches represented text as vectors and compared documents by measuring similarity or applying classification algorithms. Common methods included TF–IDF and LSI.
TF–IDF assigns weights to words based on their frequency in a document relative to their frequency across the corpus, thereby emphasizing terms that are particularly important in a given context. In contrast, LSI assigns text into a lower-dimensional semantic space, capturing latent structures across documents and revealing co-occurrence patterns between terms. 
A significant contribution in this area is the study by Rahman and Williams~\cite{rahman2022threat}, which compared multiple classical feature extraction methods for classifying attack TTPs from unstructured threat reports. They implemented five representative methods, including TF–IDF and LSI. Their evaluation showed that TF–IDF and LSI consistently outperformed the other methods. This work provides an important baseline for cybersecurity text analysis, demonstrating both the strengths and weaknesses of statistical methods.
Other research also applied TF–IDF and related techniques to establish mappings between structured cybersecurity repositories such as CAPEC and CVE~\cite{kanakogi2021tracing}. For instance, TF–IDF was shown to be effective in ranking the top-$N$ relevant CAPEC documents for a given vulnerability description. Our own early experiments in developing VULDAT~\cite{othman2023vuldat} also utilized TF–IDF and LSI as baselines, confirming their value as starting points while revealing their limitations in capturing the deeper semantics of attack or vulnerability descriptions.
In summary, classical NLP techniques demonstrated the feasibility of analyzing and classifying cybersecurity text using statistical features. Nonetheless, their emphasis on surface-level word matching restricted them from effectively modeling contextual semantics and specialized technical expressions.

\section{Vulnerability-to-Attack Mapping}
\label{sec:vul2attack}
A large portion of the literature has focused on vulnerability-to-attack mapping, where the starting point is a vulnerability that is linked to related attack information. 
These approaches typically enrich vulnerability reports with contextual attack tactics, techniques, or patterns. 
For instance, Kuppa et al.~\cite{kuppa2021linking} proposed a multi-head deep embedding model that linked CVEs to ATT\&CK techniques. 
Their approach combined regular expressions with cosine similarity on embeddings, but it only covered 17 ATT\&CK techniques and depended heavily on handcrafted patterns. 
Sun et al.~\cite{sun2021generating} employed BERT to enrich CVE descriptions by generating additional context, improving the utility of CVEs for downstream tasks. 
Similarly, Lakhdhar et al.~\cite{lakhdhar2021machine} explored a multi-label classification approach to automatically map CVEs to ATT\&CK tactics using machine learning classifiers.  
Another notable contribution is the Cve2att\&ck model~\cite{grigorescu2022cve2att}, which annotated CVE reports with ATT\&CK tactics using BERT-based embeddings. 
Although this model improved contextualization, it was limited to a dataset of 1,813 CVEs and 31 ATT\&CK techniques, restricting its generalizability. 
The CVE Transformer (CVET)~\cite{ampel2021linking} fine-tuned RoBERTa with a self-distillation mechanism to map CVEs to 10 ATT\&CK tactics. 
While CVET demonstrated strong performance, its focus on tactic-level mappings reduced granularity and practical utility.  
Graph-based efforts also contributed to vulnerability-to-attack mapping. 
The BRON framework~\cite{hemberg2021linking} aggregated multiple repositories (CWE, CVE, CAPEC, and ATT\&CK) into a relational graph, enabling relational path tracing. 
However, BRON struggled with integrating recent CVEs and lacked semantic embedding capabilities.
Overall, vulnerability-to-attack mapping research showed that transformers outperform classical NLP in enriching CVE descriptions. 
Nevertheless, these studies remained limited in scope, and focus on vulnerability to attack direction leave open the complementary challenge of inferring CVEs starting from attack
information.

\section{Attack-to-Vulnerability Mapping}
\label{sec:attack2Vul}
Attack-to-vulnerability mapping represents a research direction that begins with descriptions of adversarial behavior, such as tactics, techniques, and procedures, and seeks to identify the vulnerabilities that are related to attacks. In addition, this direction is directly relevant to this thesis, as it supports proactive defense by inferring which CVEs may be exploited in real-world attacks.
Early work primarily leveraged CAPEC attack patterns. 
By comparing CAPEC descriptions with CVE reports, researchers applied TF-IDF, Doc2Vec, and RoBERTa to establish mappings~\cite{kanakogi2021tracing}. 
RoBERTa consistently outperformed BERT, showing the importance of large-scale pre-training. 
However, these efforts were limited to CAPEC patterns and did not generalize across other attack types, such as Tactics, Techniques, or Procedures.
To expand coverage, Othman et al.~\cite{othman2024comparison} compared TF-IDF, LSI, BERT, MiniLM, and RoBERTa for CAPEC-to-CVE mapping, producing a dataset of 133 CAPEC patterns linked to 685 CVEs via CWEs. 
This work confirmed that transformers capture semantic relationships more effectively than statistical methods, though TF-IDF remained competitive for top-$N$ retrieval tasks.
A significant step forward was the introduction of VULDAT~\cite{othman2024cybersecurity}, which used MPNet embeddings to link ATT\&CK Technique descriptions with CVEs. 
This approach achieved an F$_1$-score of 85\%, demonstrating that semantic similarity at the Technique level is highly informative. 
However, VULDAT was limited to a single attack type and did not evaluate a broad set of transformer models.


\begin{table*}[htbp!]
    \centering
    \scriptsize
    \caption{Summary of Related Work in Vulnerability-Attack Mapping}
    \label{tab:related_work}
    \begin{tabular}{l@{\hskip 4pt}p{0.18\textwidth}@{\hskip 4pt}p{0.3\textwidth}@{\hskip 4pt}p{0.25\textwidth}}
    \hline
        \textbf{Study} & \textbf{Approach}& \textbf{Key Contribution} & \textbf{Limitations} \\
    \hline
    \textbf{Kuppa, A. et al.~\cite{hemberg2022sourcing}}&Multi-head Deep: Cosine similarity on embeddings.& Links CVEs to 17 ATT\&CK techniques using ATT\&CK and CVE datasets. & Limited scope, regex dependency. \\
    \textbf{Grigorescu, Octavian, et al.~\cite{grigorescu2022cve2att}} & Cve2att\&ck: BERT-based classification. &Annotates CVEs with tactics using 1,813 CVEs and 31 ATT\&CK techniques. & Small training set, narrow Technique coverage. \\
    \textbf{Ampel, Benjamin, et al.~\cite{ampel2021linking}} & CVET: RoBERTa fine-tuning. &Self-awareness distillation for Tactic mapping using 10 ATT\&CK tactics. & Only 10 tactics, no technique-level granularity. \\
    \textbf{Kanakogi, Kenta, et al..~\cite{kanakogi2021tracing}}&CAPEC-CVE: TF-IDF, RoBERTa, Doc2Vec. &Top-$N$ CAPEC matches for CVEs using CAPEC and CVE datasets. & Unidirectional (CVE$\to$CAPEC). \\
    \textbf{Othman et al.~\cite{othman2024comparison}} & CAPEC-CVE & Empirical comparison models (TF-IDF, LSI, BERT, MiniLM, RoBERTa)  and provides a mapping dataset linking 133 CAPEC attack patterns to 685 CVEs via CWEs.&  TF-IDF and LSI fail to capture contextual or semantic relationships in attack descriptions. \\
    \textbf{Hemberg, Erik, et al.~\cite{hemberg2021linking}} &  BRON: graph aggregation. &Bidirectional relational path tracing using CWE, CVE, CAPEC, and ATT\&CK datasets. & Fails on recent CVEs, no NLP integration. \\
   
\textbf{Othman et al.~\cite{othman2024cybersecurity}} & Technique-CVE. & Empirical comparison transformers models (BERT, MPNeT, MiniLM)  and provides a mapping dataset linking 100 attack techniques to 610 CVEs via CWEs and CAPEC.& Limited in its ability to compare all types of attack information (Tactic, Technique, Procedure, and Attack Pattern).\\
   
\textbf{Othman et al.~\cite{othmanVul}} & Att\&ck2Cve: Transformer-based approach for linking attacks to CVEs across four attack information types (Tactic, Technique, Procedure, Attack Pattern) using cosine similarity on embeddings.
&
Evaluation of 14 SOTA sentence transformers, accompanied by a comprehensive analysis that identifies the most effective attack information type (Technique) for vulnerability prediction; the MMPNet model achieves an F$_1$-score of 89\% and enables the identification of 275 validated missing CVE links, thereby enriching the MITRE repositories.
&
Provides broad coverage across attack types, but evaluation is restricted to MITRE repositories and does not yet extend to larger real-world datasets.
\\

    \hline
    \end{tabular}
\end{table*}

A comparison of representative prior works is summarized in Table~\ref{tab:related_work}. Most existing studies have focused on linking CVEs to a single type of attack information, such as Tactics or Techniques. For instance, CVET~\cite{ampel2021linking} achieved an F$_1$-score of 76.2\% when mapping CVEs to ATT\&CK Tactics, while Cve2att\&ck~\cite{grigorescu2022cve2att} reported an F$_1$-score of 47.8\% for CVE-to-Technique mappings, and Kanakogi et al.\cite{kanakogi2021tracing} obtained an F$_1$-score of 44.5\% when linking CVEs to CAPEC patterns. Our earlier work, VULDAT\cite{othman2024cybersecurity}, extended this line by achieving an F$_1$-score of 85\% for linking Technique descriptions to CVEs. Despite these contributions, the majority of prior studies concentrated on enriching CVE reports with attack information and were often restricted to small datasets, narrow subsets of attack types, or limited granularity. Few works systematically compared transformer architectures, and almost none addressed the prediction of CVEs directly from real-world, unstructured cyberattack news. 
This thesis addresses these gaps by:  
(1) Systematically evaluating 14 SOTA sentence transformers across four attack information types;  
(2) Developing novel annotated datasets linking ATT\&CK, CAPEC, CWE, and CVE;  
(3) Extending vulnerability detection to real-world cyberattack news;  
(4) Enriching MITRE repositories with 275 validated missing links.  
Consequently, the thesis presents a comprehensive and scalable framework that advances both academic understanding and practical tools for automated vulnerability detection.


\chapter{Methodology}
\label{ch:methods}

\huge{T}\normalsize{his chapter of this thesis consolidates the methodologies, datasets, and evaluation strategies employed throughout this research. By gathering and structuring each component, we highlight the underlying logic and processes that drive our approach to detecting known vulnerabilities from textual attack descriptions and real-world cyberattack news.}

The chapter begins by describing the data foundations, including the construction of annotated mapping datasets linking MITRE repositories (ATT\&CK, CAPEC, CWE, and CVE) and the curation of a real-world news dataset. It then introduces the preprocessing techniques applied to standardize unstructured attack text. The subsequent sections detail the sentence transformer models explored, ranging from lightweight architectures to SOTA large models, and explain the similarity computation and validation methods adopted. Finally, this chapter outlines the evaluation framework, encompassing both automated measures (Precision, Recall, F1, Jaccard Similarity, Mapping and Detection Accuracy) and manual validation of predicted links.

\begin{figure}[t!]
    \centering
    \includegraphics[width=\linewidth]{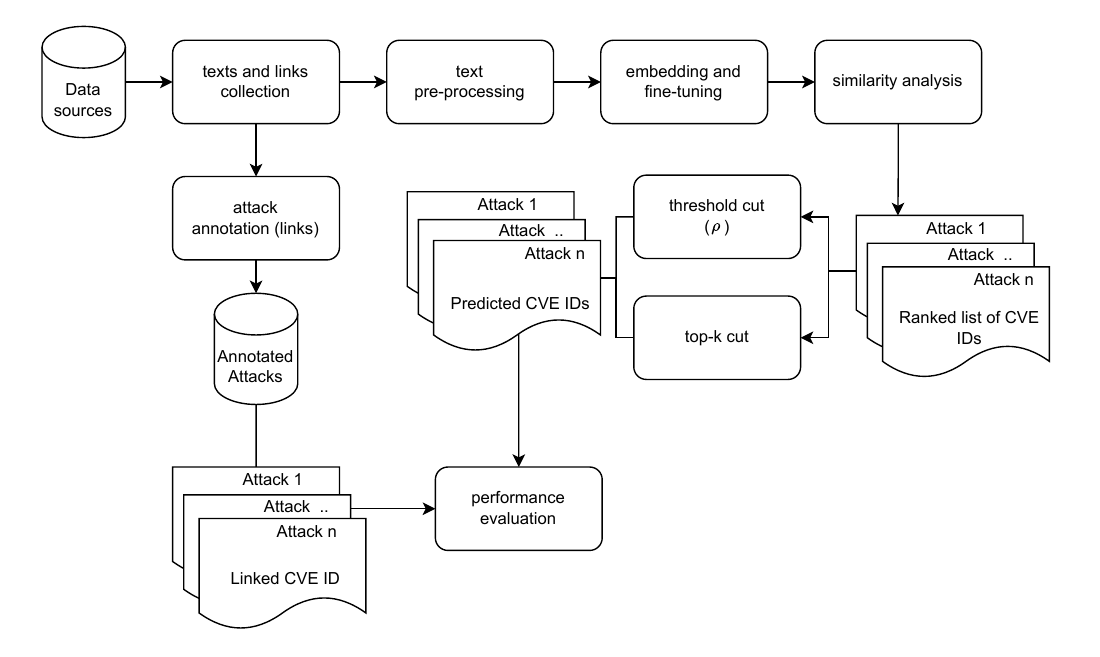}
    \caption{Overview of the methodology for automated vulnerability detection.}
    \label{fig:TopMethodology}
\end{figure}
\begin{figure}[t!]
    \centering
    \includegraphics[width=\linewidth]{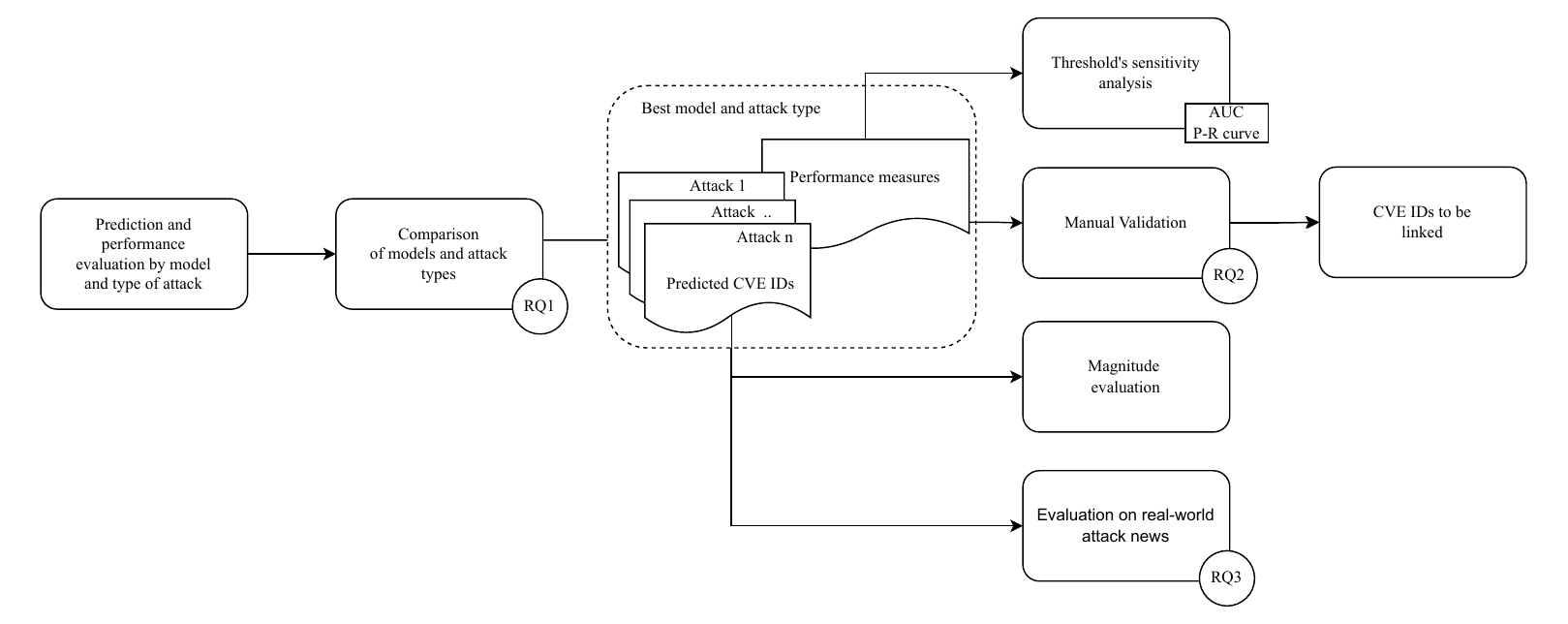}
    \caption{Methodology overview.}
    \label{fig:overviewMethodology}
\end{figure}

The methodology follows a structured but iterative process, where each phase builds on insights gained from earlier experiments and validations. Figure~\ref{fig:TopMethodology} illustrates the overall research flow, showing how data collection, preprocessing, model training, and evaluation are integrated into a cohesive framework. By combining these components, the chapter establishes the foundation for understanding how experiments were designed, how models were benchmarked, and how new attack–vulnerability links were discovered and validated. Together, these methodological steps form a rigorous research flow that underpins the contributions of this thesis toward automated, reliable, and scalable vulnerability detection.


As illustrated in Figure~\ref{fig:TopMethodology} and Figure~\ref{fig:overviewMethodology}, our methodology is designed to answer the three main research questions of this thesis:
\begin{itemize}
    \item \rqone
    \item \rqtwo
    \item \rqthree
\end{itemize}




To address these questions, the methodology integrates multiple phases as illustrated in Figure~\ref{fig:overviewMethodology}:

Firstly, for each selected sentence transformer model and type of attack text (Tactic, Technique, Procedure, or Attack Pattern), we perform a \textit{vulnerability prediction and performance analysis} (Section~\ref{sec:Prediction}), to generate ranked lists of potential CVE matches and quantify model effectiveness.
Secondly, we conduct a \textit{comparative evaluation} of models and attack types (Section~\ref{sec:Comparison}), identifying the model–attack-type combination that achieves the highest prediction accuracy. With the best-performing combination, we extend the analysis by applying a \textit{sensitivity analysis of the similarity threshold} (Section~\ref{sec:SensitivityAnalysis}), to identify the threshold that best balances precision and recall. We also perform a \textit{magnitude evaluation} (Section~\ref{sec:Magnitude}), measuring how many vulnerabilities are correctly predicted per attack text. Together, these analyses provide a deeper assessment of model performance and address the effectiveness of different models and attack types (RQ1).
Thirdly, we carry out a \textit{manual validation} of predicted vulnerabilities (Section~\ref{sec:Validation}). This step allows us to assess whether the approach can recommend new semantic links between attacks and vulnerabilities that are not explicitly documented in the MITRE repositories, thus answering R2.
Finally, we evaluate the generalizability of the approach by applying the best-performing transformer model to a curated dataset of real-world attack news reports (Section~\ref{sec:Newsevaluation}). This phase tests how well the methodology extends beyond structured repositories to unstructured, noisy data, thereby addressing RQ3.
Through this structured methodology, we ensure that each research phase contributes directly to answering the thesis research questions. By combining dataset construction, text preprocessing, transformer-based embedding, similarity analysis, and both automated and manual evaluation, our methodology provides a comprehensive framework for investigating vulnerability detection from both curated repositories and real-world data sources.

\section{Vulnerability prediction}\label{sec:Prediction}
\begin{figure}[t!]
    \centering
    \includegraphics[width=\linewidth]{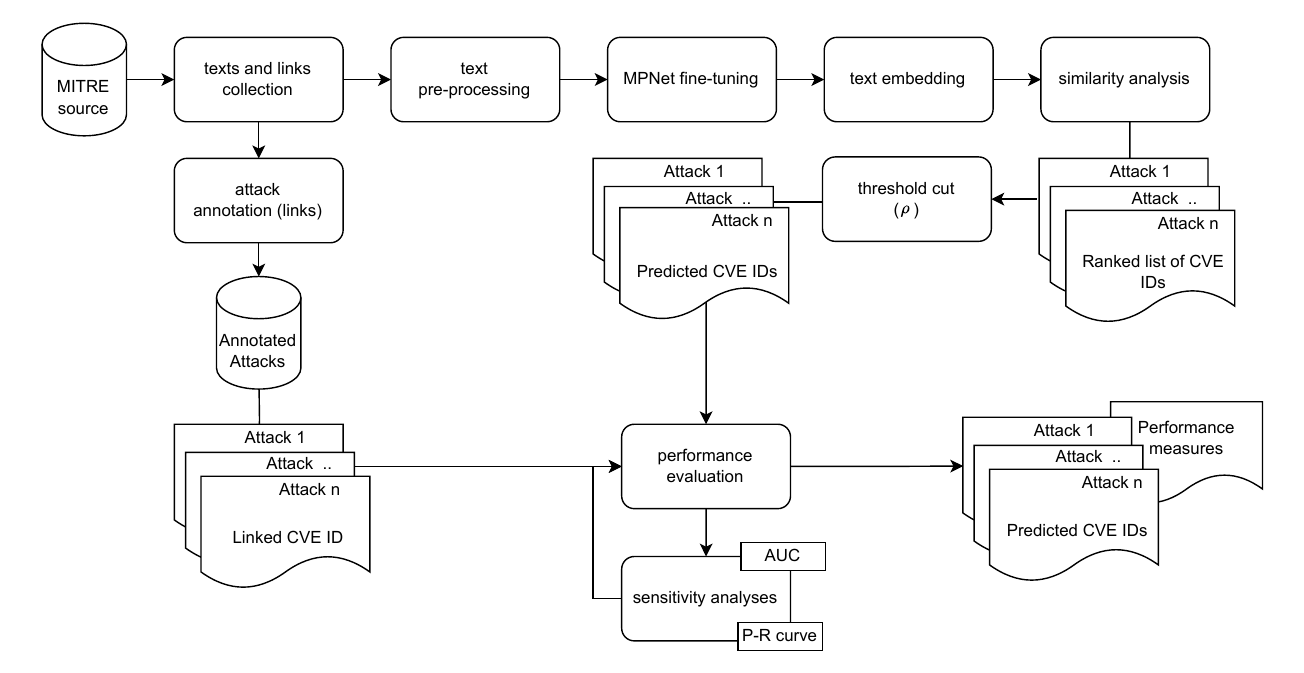}
    \caption{Prediction and performance evaluation by model and attack type.}
    \label{fig:MitreMethodology}
\end{figure}

Figure~\ref{fig:MitreMethodology} shows how we use a sentence transformer to predict vulnerabilities from an attack text of a given type. We collect all attack descriptions related to that type, along with all vulnerability descriptions and their IDs (\textit{texts and links collection}).
For each attack ID, we associate the CVE IDs eventually found in the attack's description page (links) (\textit{attacks annotation}). 
Then, we pre-process all texts  (\textit{text pre-processing}). After fine-tuning the model, we embed all texts with the selected model  (\textit{text embedding}). 
The embedding of each attack text is compared for similarity with the embeddings of all vulnerabilities (\textit{text similarity analysis}).
Vulnerability embeddings that have a similarity score greater than the threshold with a given attack embedding are
labeled as predicted vulnerabilities for that attack.
In this way, we obtain a mapping between the attack IDs and  CVE IDs. 
At this point, we evaluate the performance of the sentence transformer by comparing the sets of predicted and linked CVE IDs. 
In the next step, we compare the performance by varying the model and attack type. In the following, we describe this process in more detail.

\subsection{Texts and links collection}
\label{sec:DataCollection}

The initial phase of the methodology consists of the systematic collection and structuring of attack descriptions and vulnerability records from the MITRE repositories. This step is crucial because it provides a transparent and reproducible foundation for evaluating our automated linking approach. By extracting both textual content and explicit cross-references from the repositories, we generate a dataset that serves as the ground truth for training and assessing automated linking approaches.
The study considers four primary repositories maintained by MITRE, namely ATT\&CK, CAPEC, CWE, and CVE. Each of these repositories captures a different dimension of cybersecurity knowledge. The ATT\&CK repository provides adversarial behaviors represented through Tactics, Techniques, and Procedures (TTPs). CAPEC catalogues common attack patterns that describe general strategies used to exploit weaknesses. CWE enumerates recurring weaknesses in software or system design that may give rise to vulnerabilities. Finally, the CVE database contains standardized identifiers and descriptions of publicly disclosed vulnerabilities affecting a wide range of systems and applications. 
Collecting data from these repositories allows us to study the relationships among them and, where possible, automate the discovery of missing links.

From these repositories, we collect all available textual descriptions of the attacks (covering Tactics, Techniques, Procedures, and Attack Patterns) as well as the descriptions of vulnerabilities from CVE issues. In addition to descriptive texts, the process includes extracting explicit links, that is, links provided by MITRE between different entities. For example, CWE entries frequently include curated lists of CVEs that exemplify how a particular weakness can manifest in practice. These explicit links constitute reliable evidence of attack–weakness–vulnerability relations and are therefore incorporated into the dataset as part of the annotation process.
To illustrate, consider the weakness entry CWE-770: Allocation of Resources Without Limits or Throttling. On its official MITRE page, CWE-770 lists several CVE identifiers that correspond to real-world vulnerabilities caused by this weakness. For instance, CVE-2008-1700 and CVE-2020-7218 are explicitly cited as examples. In our dataset, CWE-770 is therefore mapped to these CVEs, thereby establishing a verified connection between the weakness and its manifestations. 
However, it is important to note that MITRE emphasizes that such lists are curated examples and not exhaustive.
As noted in the CWE pages, the explicit links to CVEs is  \textit{``a curated list of examples for users to understand the variety of ways in which this weakness can be introduced. It is not a complete list of all CVEs that are related to this CWE entry."} Thus, some existing links may not be reported in the page. 

\begin{figure*}[hbt!]
\centerline{\includegraphics[width=\columnwidth]{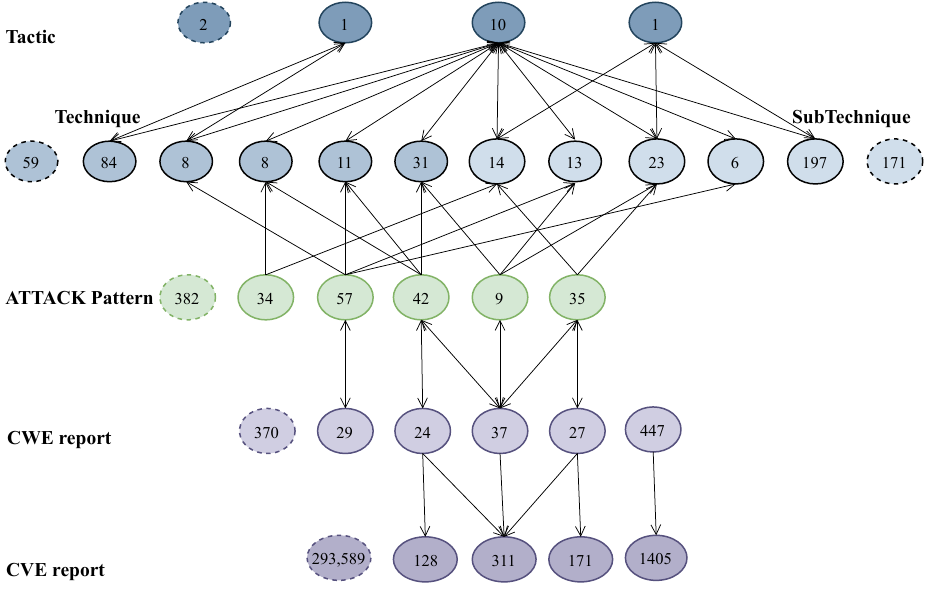}}
\caption{Graph representation of explicit links between attack types, weaknesses, and vulnerabilities in MITRE repositories, showing the density and direction of existing connections. Nodes with many connections represent “Super Entries,” while isolated nodes indicate “floating entries.”}
\label{fig:graphM}
\end{figure*}

To visualize the results of this extraction process, we construct a graph representation, shown in Figure~\ref{fig:graphM}. Each node in the graph represents a repository entity (e.g., an attack type, a weakness, or a vulnerability), and edges correspond to explicit links discovered in the repository pages. The size and density of the nodes indicate the number of links. For instance, we observe that certain nodes are well connected, forming what we call “Super Entries”. These represent mature knowledge points in the repositories, where many attacks, weaknesses, or vulnerabilities are explicitly interconnected. In contrast, a large number of nodes remain isolated, which we refer to as “floating entries”. Most CVE reports fall into this latter category, highlighting the substantial portion of vulnerabilities that lack explicit links to attacks or weaknesses.
In addition, the directions of the graph edges indicate where the links have been found. 
For instance, there are 447 CWE web pages that include links to 1405 CVEs. As Procedures are only listed in the Techniques' pages,  there are also no explicit links between Patterns or Tactics and Procedures. Therefore,
we omitted the Procedures from the graph representation in Figure~\ref{fig:graphM}, though they remain relevant in later prediction experiments.

\begin{table*}[htb!]
\caption{Number of CVEs and CWEs linked and not linked to attacks.}
\label{tab:golddataset}
\centering
\bgroup
\begin{tabular}{llccccc}
\hline
 & &\textbf{Tactic} &\textbf{Technique} &\textbf{Procedure} & \textbf{Attack Pattern} \\
\hline
\textbf{CWE} & linked& 117&79& 117& 117  \\
 & not linked &818&856&818&818\\
\hline
\textbf{CVEs} &linked&610&610& 610& 610 \\
&not linked &294994&295165&294994 &294994 \\
\hline
\end{tabular}%
\egroup
\end{table*}

Table~\ref{tab:golddataset} reports the number of CVEs and CWEs that are linked or not linked to the different types of attacks. The results indicate that a small fraction of vulnerabilities are explicitly linked to attacks, the overwhelming majority are not. For example, out of nearly 295,610 CVEs, only 610 are explicitly linked to Techniques, while the remaining 294,994 have no such connection. A similar pattern emerges for CWEs, where only 79 are linked to Techniques, compared to 856 that are not. The same pattern is observed across other attack types.
Moreover, the data presented in Table~\ref{tab:golddataset} highlights the incompleteness of current repositories. The fact that most vulnerabilities remain unlinked suggests two things: first, there is a significant knowledge gap for practitioners seeking to connect observed attacks to concrete vulnerabilities; and second, there is a strong need for automated approaches that can infer the missing links. Our methodology directly addresses this challenge by treating the explicitly annotated links as ground truth for training and evaluation, while using sentence transformers to predict and recommend additional, undocumented connections.
In summary, the texts and links collection phase establishes the empirical foundation of this research by systematically integrating data from ATT\&CK, CAPEC, CWE, and CVE repositories. The resulting dataset captures both the descriptive content of attacks and vulnerabilities and the explicit links that connect them, while these explicit links provide reliable ground truth for evaluation. However, the quantitative analysis reveals that the majority of vulnerabilities remain unlinked, confirming the fragmented nature of current repositories. This imbalance highlights the necessity of automated methods capable of inferring additional, semantically valid connections. Consequently, the collected dataset not only provides the benchmark required for evaluating model performance but also defines the context in which automated link discovery can contribute to the enrichment and completeness of existing cybersecurity knowledge bases.

\subsection{Attacks' annotation}
\label{sec:AttacksAnnotation}

After collecting attack descriptions, vulnerability reports, and explicit cross-references from the MITRE repositories, the next step is the annotation of attacks with vulnerabilities. The objective of this step is to establish a traceable and reproducible mapping between each attack entity and the CVEs to which it is related. This annotated mapping forms the ground truth used to train and evaluate the automated prediction models presented in later stages of the methodology.

We utilize explicit links to establish connections between attacks and a defined set of vulnerabilities.
As illustrated in Figure~\ref{fig:graphM}, there are no explicit links between Tactics, Techniques, or Procedures and CWEs or CVEs.  
We then exploit the explicit links between CWEs and attack patterns to annotate attacks with the CVEs to which they are linked. 
This step creates a traceable mapping between attacks and vulnerabilities, forming the ground truth for model evaluation.
For instance, given a Technique, we retain in CAPEC all attack patterns that mention it. For each Attack Pattern, we then collect all the CWEs whose explicit link mention it, and from those CWEs, we retrieve the explicit links to CVEs.


\begin{table*}[htb]
\caption{Attacks linked and not linked to CVE reports.}
\label{tab:golddataset2}
\centering
\bgroup
\def\arraystretch{1}
\setlength{\tabcolsep}{4pt}
\begin{tabular}{lccccccc}
\hline  
& \textbf{Tactic} &\textbf{Technique} &\textbf{Procedure} & \textbf{Attack Pattern} \\
\hline
 \textbf{Linked}&11 & 100  & 721 &86\\
\textbf{Not linked}&3&525&88&473\\
\hline
\textbf{Total}&14&625&809&559\\
\hline
\end{tabular}%
\egroup
\end{table*} 

 \begin{table*}[htbp!]%
\centering
\caption{An example of a Technique and its chain to one of the linked CVEs.}%
\label{tab:cveDataset}
\begin{small}
\begin{tabular*}{\textwidth}{@{\extracolsep{\fill}}p{3.5cm}p{3.5cm}p{3.5cm}p{3.5cm}}%
\hline

\textbf{Technique-T1574.007:} Adversaries may execute their own malicious payloads by hijacking environment variables used to load libraries. Adversaries may place a program in an earlier entry in the list of directories stored in the PATH environment variable, which Windows will then execute when it searches sequentially through that PATH listing in search of the binary that was called from a script or the command line.

& \textbf{CAPEC-38:} This Pattern of attack sees an adversary load a malicious resource into a program's standard path so that when a known command is executed then the system instead executes the malicious component. The adversary can either modify the search path a program uses, like a PATH variable or classpath, or they can manipulate resources on the path to point to their malicious components. 
If one of these libraries and/or references is controllable by the attacker then application controls can be circumvented by the attacker.
&\textbf{CWE-427:} The product searches for critical resources using an externally-supplied search path that can point to resources that are not under the product's direct control.
& \textbf{CVE-2022-4826:} The Simple Tooltips WordPress plugin before 2.1.4 does not validate and escape some of its shortcode attributes before outputting them back in a page/post where the shortcode is embed, which could allow users with the contributor role and above to perform Stored Cross-Site Scripting attacks.
\\\hline
\end{tabular*}
\end{small}
\end{table*}

This procedure results in a structured mapping where each attack entity is annotated with zero, one, or several CVEs. Table~\ref{tab:golddataset2} summarizes the number of attacks of each type that are linked to at least one CVE versus those that remain unlinked. 
The results show marked differences across attack types: \emph{Tactics} exhibit 11 linked out of 14 total entries (78.6\%); \emph{Techniques} show 100 linked out of 625 (16.0\%); \emph{Procedures} present 721 linked out of 809 (89.1\%); and \emph{Attack Patterns} display 86 linked out of 559 (15.4\%). Consequently, Procedures achieve the highest coverage, followed by Tactics, whereas Techniques and Attack Patterns have substantially lower coverage in terms of explicit or derived links to CVEs.
Table~\ref{tab:cveDataset} illustrates an example of such links in the case of the T1574.007: DLL Search Order Hijacking. This Technique describes adversaries executing their own malicious payloads by hijacking environment variables used to load libraries. It can be associated with the attack pattern CAPEC-38: Load Malicious Resource, which models scenarios where adversaries insert malicious resources into a program’s execution path. CAPEC-38, in turn, references the weakness CWE-427: Uncontrolled Search Path Element, which documents cases where software searches for critical resources using externally supplied paths. Finally, CWE-427 is explicitly linked to several CVEs, among them CVE-2022-4826, a vulnerability in the WordPress Simple Tooltips plugin that allows stored cross-site scripting through improper validation of shortcode attributes. Following this chain, we annotate Technique T1574.007 with CVE-2022-4826, establishing a traceable mapping from the abstract adversarial behavior to a concrete vulnerability instance. The table includes only one representative instance among all those that can be linked to Technique T1574.007 using our procedure.
Finally, this annotation procedure is applied systematically across all attack entities. For each attack $a$, we construct the set of CVEs linked to it, denoted as: \[ \mathcal{M}(a) = \{c \in \mathcal{C} \; | \; \exists \, (a \rightarrow c)\}
\]

where $\mathcal{C}$ is the set of all CVE identifiers, and $a \rightarrow c$ denotes the existence of a link between attack $a$ and CVE $c$ established through the annotation procedure. The resulting annotated dataset, which contains the complete mapping between attacks and CVEs, is included in our replication package~\cite{VULDATDataSet} and serves as the ground truth for evaluating the models in Section~\ref{sec:PerformanceEvaluation}.

\subsection{Text pre-processing}\label{sec:Preprocessing}

Once the attacks have been annotated with vulnerabilities, the next step is to pre-process the textual descriptions of both attacks and CVEs. Pre-processing is necessary to remove irrelevant information and noise while retaining the semantic content that is essential for embedding and similarity analysis. The objective of this phase is to standardize the text so that the models can focus on meaningful linguistic features without being distracted by superficial inconsistencies.
We begin by cleaning the text to eliminate elements that do not contribute to semantic meaning. Specifically, citations, URLs, HTML tags, and general non-alphanumeric characters are removed. This process ensures that the models are not distracted by extraneous tokens such as hyperlinks or reference markers, which are frequent in attack descriptions but provide little linguistic value.
Unlike traditional NLP pipelines, we do not apply stemming, lemmatization, or stop-word removal. Prior work has shown that such techniques are beneficial for models based on bag-of-words or statistical representations. However, transformer-based models rely on subword tokenization and contextual embeddings, which are designed to benefit from the full grammatical and lexical structure of the text~\cite{okonkwo2023leveraging, siino2024text}.
Stop words such as in, on, or by, although frequent, contribute to the syntactic flow of sentences and often enhance the contextual representation captured by attention mechanisms. By preserving these words, we allow the model to retain the full sentence structure, which is particularly important for distinguishing relationships expressed in natural language. For instance, the difference between “executed \textbf{by} the adversary” and “executed \textbf{on} the adversary” is semantically significant but would be lost if prepositions were removed.
Similarly, stemming and lemmatization are not applied in order to maintain morphological variants. Words such as execute, executing, and execution carry subtle distinctions in meaning that transformers are capable of modeling. Reducing them to a single stem (e.g., execut) would result in the loss of valuable information about action, process, or outcome. Retaining these distinctions enables the embedding models to capture nuanced semantic relations between attack descriptions and CVE reports.
Although aggressive preprocessing is avoided, lightweight cleaning steps are applied to ensure textual consistency. Redundant whitespace is collapsed, punctuation with no syntactic function is removed, and formatting inconsistencies are corrected. 
However, sentence delimiters such as dots are preserved, as they provide and support contextual embedding, for example, when indicating software versions.
Our pre-processing strategy adopts a minimalistic yet targeted approach. By removing only clearly irrelevant elements while preserving grammatical and lexical features, we balance noise reduction with semantic preservation. This design enables sentence transformer models to fully exploit contextual dependencies and produce embeddings that better capture the relationships between attack descriptions and CVE reports. In contrast, traditional NLP methods such as TF-IDF and LSI, we applied a full preprocessing pipeline to reduce dimensionality and noise, since these methods operate on surface-level term frequencies without considering meaning or context. As a result, they are highly sensitive to lexical variability and irrelevant tokens, requiring aggressive cleaning to produce reliable similarity measures. 

\subsection{Model fine-tuning and text embedding }\label{sec:TuningEmbedding}

Following the pre-processing of attack and vulnerability descriptions, the next step is to represent the cleaned texts in a numerical format that models can process. To this aim, we employed 14 pre-trained sentence transformer models (see Section~\ref{sec:transformer}), which are specifically designed to generate high-quality sentence embeddings. These embeddings are fixed-size vector representations of input texts that capture their semantic content in a shared vector space, thereby enabling the comparison between attack descriptions and vulnerability reports.
To generate these embeddings, each of the selected models relies on a multi-layer transformer architecture to process input text. Tokens are first generated from the raw text using subword tokenization. These tokens are then passed through multiple transformer encoder layers, which employ self-attention mechanisms and feed-forward networks to capture contextual relationships both within and across sentences. Finally, the token-level embeddings are condensed into a fixed-length sentence embedding through a pooling strategy, such as mean pooling. The resulting embeddings allow semantically similar sentences to be placed close together in the vector space, even if they do not share identical wording. Since the 14 models selected for this study vary considerably in architecture and capacity, with embedding sizes ranging from 384 to 4096 dimensions. This diversity enables us to assess the trade-offs between lightweight architectures (e.g., MiniLM, DistilBERT, TinyBERT), which are computationally efficient, and larger architectures (e.g., RoBERTa, T5-XXL), which are designed to capture more nuanced semantic aspects. 
To adapt the models to the task of linking attack descriptions with CVE reports, we applied a fine-tuning procedure. The dataset was split into three subsets with fixed ratios of 80\% for training, 10\% for validation, and 10\% for testing. This fixed split was chosen to ensure fair comparability across models, to avoid variability caused by random re-sampling, and to guarantee reproducibility of results~\cite{HinidumaEtAl2025}. Before the split, dataset statistics, including the number of linked and unlinked samples for each attack type, were computed and are reported in Table~\ref{tab:golddataset2}.
All models were fine-tuned using the \textit{CosineSimilarityLoss} function, which is particularly suited to embedding-based tasks where the goal is to maximize the similarity of semantically related pairs while minimizing the similarity of unrelated ones. Training was conducted for four epochs with 100 warm-up steps, and intermediate evaluation was performed every 500 steps. These hyperparameters follow established practices in the sentence-transformer literature and training examples~\cite{hyperparams}. The validation set was used to monitor intermediate performance and adjust training dynamics, while the held-out test set was used for final reporting of performance metrics.
In summary, this stage enabled us to generate semantically rich embeddings from both attack and vulnerability descriptions, fine-tuned consistently across diverse transformer architectures, and prepared for the similarity analysis described in Section~\ref{sec:Similarity}.

\subsection{Similarity analysis}\label{sec:Similarity}

After generating embeddings for both attack and vulnerability descriptions, the next step is to measure their semantic closeness. Following established practices in embedding-based natural language processing~\cite{reimers2019sentence, muennighoff2022mteb}, we adopt \textit{cosine similarity} as the primary metric for comparing sentence embeddings. This choice is motivated by its widespread use and proven effectiveness across diverse tasks, including semantic search, clustering, and information retrieval. For example, Reimers and Gurevych~\cite{reimers2019sentence} demonstrated that cosine similarity in Sentence-BERT significantly improved performance on semantic textual similarity benchmarks, while Muennighoff et al.~\cite{muennighoff2022mteb} established it as the standard metric in large-scale evaluations of over 50 transformer-based models.

Cosine similarity evaluates the angular relationship between two vectors rather than their magnitude. This property makes it particularly well-suited for sentence embeddings, where semantic meaning is primarily encoded in vector orientation. Formally, given an attack embedding $\vec{p} = (p_1, p_2, \ldots, p_n)$ and a vulnerability embedding $\vec{q} = (q_1, q_2, \ldots, q_n)$, cosine similarity is defined as:
\begin{equation}
\label{eqSim}
Sim(\vec{p}, \vec{q}) = \frac{\vec{p} \cdot \vec{q}}{|\vec{p}| \cdot |\vec{q}|} = \frac{\sum_{i=1}^{n} p_i q_i}{\sqrt{\sum_{i=1}^{n} p_i^2} \cdot \sqrt{\sum_{i=1}^{n} q_i^2}}
\end{equation}
The similarity score theoretically ranges from $-1$ to $1$, where $1$ denotes maximum similarity, $0$ indicates no correlation, and $-1$ indicates complete opposition. Since embeddings are normalized and the focus of this study is on semantically related pairs, the practical range of interest is $[0,1]$. For consistency with the thresholding procedure introduced later, we express these values on a 0–100 scale in our experiments.

Using this metric, all CVE embeddings are ranked according to their similarity with a given attack embedding. A threshold $\rho$ is then applied to cut the ranked list at a chosen similarity level, retaining only the most relevant predictions. Formally, for an attack $a$, the set of predicted CVEs is defined as:

$$\mathcal{L}_{\rho}(a) = \{ c \in C \,:\, Sim(\vec{a} \cdot \vec{c}) > \rho\}
$$

where $C$ is the set of all CVE identifiers. The set $\mathcal{L}_{\rho}(a)$ therefore represents the \textit{predicted vulnerabilities} for attack $a$ at threshold $\rho$.
Based on this definition, we can frame vulnerability detection as a classification problem. A model is said to predict at least one vulnerability for an attack if $\mathcal{L}{\rho}(a)$ is non-empty; conversely, if $\mathcal{L}{\rho}(a) = \emptyset$, the model predicts no vulnerabilities for that attack. This formulation enables systematic evaluation of different models, as described in the following sections.
\subsection{Performance evaluation}
\label{sec:PerformanceEvaluation}

To assess the effectiveness of our approach, we evaluate model predictions on the classification problem defined in Section~\ref{sec:Similarity}.
A perfect model should satisfy the following condition for any attack ID $a$: either the intersection between the predicted and the linked CVE IDs is non-empty,
$$\mathcal{L}_{\rho}(a) \cap \mathcal{M}(a) \neq \emptyset,
$$
or both sets are empty:
$$\mathcal{L}_{\rho}(a) = \emptyset \, \land \, \mathcal{M}(a) = \emptyset.
$$  
To operationalize this classification, we define positives and negatives, predicted positives and negatives, and the standard four outcomes of binary classification: true positives (TP), false positives (FP), false negatives (FN), and true negatives (TN). These sets, summarized in Table~\ref{tab:postivesNegatives}, provide the foundation for computing performance metrics.
\begin{table}[htbp!]
\caption{Classification in the attacks' set $\mathcal{A}$.}
\label{tab:postivesNegatives}
\begin{center}
\begin{tabular}{ll}
\hline
\hline
\textbf{Type}  & \textbf{Description}  \\
\hline
Positives & $\{ a \in \mathcal{A} \, : \exists \, {c} \in \mathcal{M}(a) \}$\\
Negatives &  $\{ a \in \mathcal{A} \, : \nexists \, {c} \in \mathcal{M}(a) \}$\\
Predicted Positives (PP) & $\{ a \in \mathcal{A} \, : \exists \, {c} \in \mathcal{L}_{\rho}(a) \}$\\
Predicted Negatives (PN) &  $\{ a \in \mathcal{A} \, : \nexists \, {c} \in \mathcal{L}_{\rho}(a) \}$\\
True Positives  (TP)& $\{ a \in \mathcal{A} \, :  \mathcal{L}_{\rho}(a) \cap \mathcal{M}(a) \not = \emptyset   \} $\\
False Positives  (FP) & $\{ a \in \mathcal{A} \, :\left( \mathcal{L}_{\rho}(a) \not = \emptyset   \, \land \, \mathcal{M}(a)  = \emptyset  \right)  \}$ \\
False Negatives (FN)  & $\{ a \in \mathcal{A} \, : \left( \mathcal{L}_{\rho}(a)  = \emptyset    \, \land \,  \mathcal{M}(a)  \not = \emptyset   \right)\}$ \\
True Negatives (TN)  & $\{ a \in \mathcal{A} \, : \left( \mathcal{L}_{\rho}(a)   = \emptyset   \,\land \,  \mathcal{M}(a)  = \emptyset   \right)\}$ \\

\hline
\end{tabular}
\end{center}
\end{table}

Based on these definitions, we adopt three widely used metrics to quantify performance: Precision, Recall, and their harmonic mean, the F1-score. Precision measures the proportion of correctly predicted CVEs among all predicted CVEs, Recall captures the proportion of correctly predicted CVEs among all ground-truth CVEs, and F1 provides a balanced measure that accounts for both Precision and Recall. These metrics are formally defined as:

\begin{equation}
\textit{Precision} = \frac{\textit{TP}}{\textit{TP} + \textit{FP}}
\end{equation}
\begin{equation}
\textit{Recall} = \frac{\textit{TP}}{\textit{TP} + \textit{FN}}
\end{equation}
\begin{equation}
\textit{F1} = 2 \times \frac{\textit{Precision} \times \textit{Recall}}{\textit{Precision} + \textit{Recall}}
\end{equation}

The F1-score in particular provides a concise indicator of how well a model balances correctness precision with completeness recall: the higher the F1-score, the better the model’s overall effectiveness. In addition, we fine-tune each of the pre-trained models
on our dataset. To this aim, we applied the same 80/10/10 train, validation, and test split introduced in Section~\ref{sec:TuningEmbedding}. 
For each attack type, we evaluate the performance of each model in the testing set using Precision,
Recall, and F1-Score. The experiments were executed on a six-node GPU cluster, where each node was equipped with an NVIDIA A100 (80 GB), 192 GB of RAM, and an Intel Xeon 4208 (16-core) CPU. Model training and evaluation were implemented using PyTorch 2.8.0 and Hugging Face Transformers 4.55.2. In summary, this evaluation framework enables a systematic comparison of models and attack types, ensuring that results are both reliable and replicable. Precision, Recall, and F-1 provide the quantitative basis for determining which sentence transformer models and attack type descriptions are most effective in predicting vulnerabilities, as discussed in the following sections.

\section{Comparison of models and attack types}\label{sec:Comparison}

To address RQ1, we conduct a comparative evaluation of all 14 pre-trained sentence transformer models introduced in Section~\ref{sec:transformer}, across the four types of attack descriptions: Tactic, Technique, Procedure, and Attack Pattern. For each model–attack-type pair, we compute Precision, Recall, and F1 as defined in Section~\ref{sec:PerformanceEvaluation}. Additionally, this systematic evaluation offers a comprehensive understanding of how various transformer architectures perform when applied to different types of attack information.
Finally, the results of this step allow us to identify which combinations of models and attack types are most effective for vulnerability prediction. To ensure consistency, we use the maximum value of F1 to determine the best model(s) and the attack type(s) on which the model(s) predict best.
In this way, we derive pairs $(\textit{attack type}, \textit{model})$ that represent the most effective configurations. The outcome of this comparison highlights the best-performing model and the attack description type on which it achieves its strongest results, thus providing the foundation for the subsequent sensitivity analysis of the similarity threshold.

\section{Sensitivity analysis}
\label{sec:SensitivityAnalysis}

After identifying the best-performing model–attack-type pair, we analyze the role of the similarity threshold $\rho$ in determining whether a CVE is predicted for a given attack. In classification tasks, a threshold defines the decision boundary for class membership. While a default threshold of 0.5 is common, prior work has shown that this value is often suboptimal and should instead be adapted to the dataset in order to optimize performance~\cite{lobo2022cost, sheng2006thresholding}. The threshold directly governs the trade-off between Precision and Recall: higher values of $\rho$ reduce false positives but may miss valid links, whereas lower values improve coverage but risk introducing spurious predictions~\cite{davis2006relationship}.

To avoid arbitrary selection, we adopt two complementary strategies. First, following established practices in embedding-based classification, we perform a sensitivity analysis in which $\rho$ is systematically varied across its range. For each value of $\rho$, we compute Precision, Recall, and F1, and then summarize the results using a Receiver Operating Characteristic (ROC) curve. 
We employ the ROC curve to plot the True Positive Rate (TPR)
\begin{equation}
TPR=\frac{TP}{TP+FN}
\end{equation}
vs. the False Positive Rate (FPR)
\begin{equation}
FPR =\frac{FP}{TN+FP}
\end{equation}
across a range of threshold values from 1 to 100. The Area Under the Curve (AUC) provides an aggregate measure of discriminative ability, while the optimal threshold is selected as the point that minimizes the Euclidean distance between the ROC curve and the ideal classification point $(0,1)$. This ensures that the chosen $\rho$ balances detection with error minimization.
Second, we complement this analysis with a Precision–Recall (PR) curve~\cite{davis2006relationship}. In this approach, we plot Precision and Recall across different $\rho$ values and select the threshold at which the two measures intersect, known as the Equal Error Rate (EER). This point provides a principled operating threshold where false positives and false negatives are balanced.
In summary, threshold sensitivity analysis combines ROC-based, which emphasizes the balance between true and false detections, with PR-based, which emphasizes the trade-off between Precision and Recall. Together, these strategies ensure that the threshold used in our experiments is systematically derived and empirically grounded, rather than arbitrarily fixed.

\begin{figure}[htb]
\centering
\small
\includegraphics[width=\columnwidth]{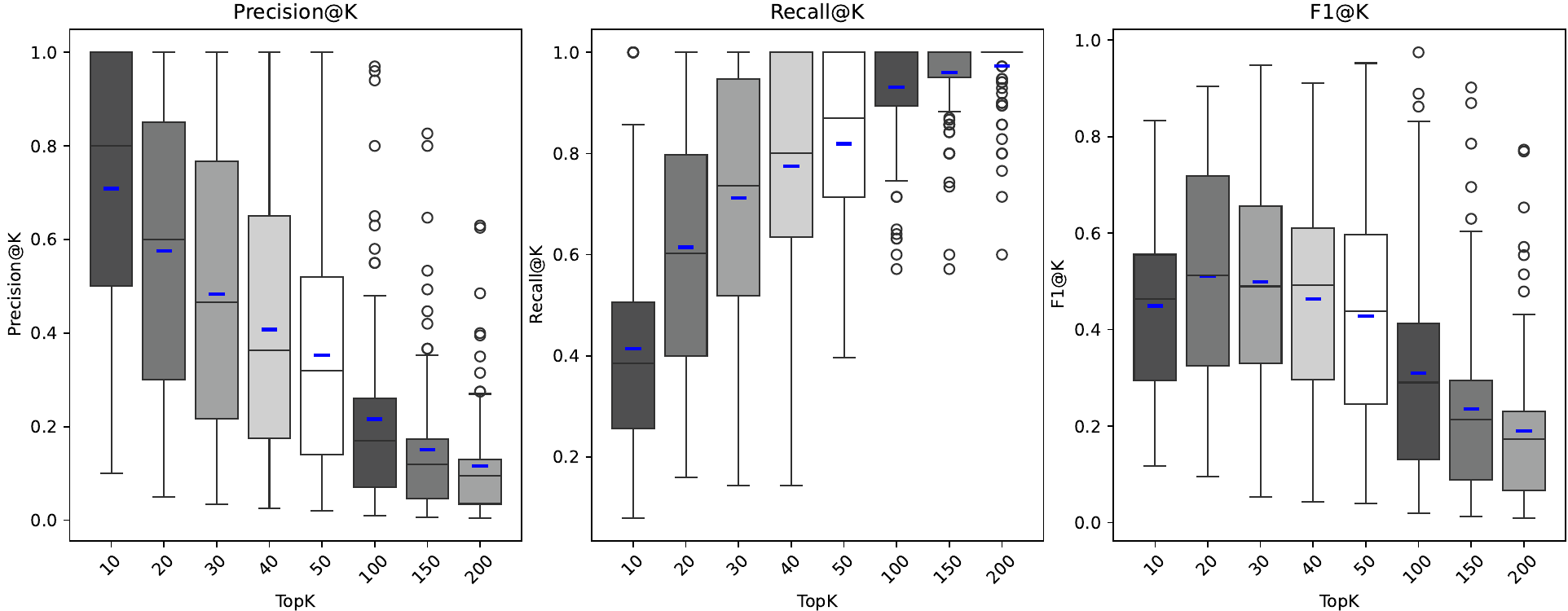}
\caption{Performance measures of the approach across different K values in top-K predictions.}
\label{fig:sensitivityAnalysisK}
\end{figure}

\noindent
\paragraph{Top-K sensitivity analysis.}\label{sec:TopKAnalysis}
In addition to the similarity threshold $\rho$, the value of K, i.e, the number of top-ranked CVE predictions retained for each attack description, has a direct impact on model performance. Smaller values of K tend to reduce false positives but may increase false negatives by discarding relevant predictions, while larger values of K increase coverage but risk introducing irrelevant vulnerabilities. To identify an optimal balance, we performed a sensitivity analysis on a balanced dataset derived from MITRE ATT\&CK, composed of 50 positive and 50 negative technique descriptions. For each value of K, we computed Precision and Recall and visualized their distribution using boxplots. As shown in Figure~\ref{fig:sensitivityAnalysisK}, Precision is higher for small K values, while Recall increases as K grows; at $k=20$, the average Precision and Recall converge, resulting in the highest F1 score. Based on this analysis, $k=20$ was selected as the most balanced operating point, guiding the number of CVE predictions considered for each attack description in the subsequent experiments.  

\section{Magnitude evaluation}\label{sec:Magnitude}
While threshold sensitivity analysis determines the decision boundary for classification, it does not fully capture how closely the predicted vulnerabilities match the actual ground truth. To this aim, we evaluate the magnitude of overlap between predicted and actual CVE sets. Thus, we adopt three complementary set-based metrics: the Jaccard Similarity Index, Mapping Accuracy, and Detection Accuracy. Together, these measures provide a more fine-grained assessment of how well the model identifies and maps vulnerabilities.

First, the Jaccard Similarity Index evaluates the similarity between the predicted set of CVEs ($\mathcal{L}_{a}$) and the ground-truth set ($\mathcal{M}_{a}$). It is defined as the ratio of the size of the intersection to the size of the union of the two sets:

\begin{equation}
\mathrm{Jaccard \, Similarity }= \frac{{|\mathcal{L}_{a} \cap \mathcal{M}_{a}|}}{|\mathcal{L}_{a} \cup \mathcal{M}_{a}|}
\end{equation}
\par\noindent

This measure ranges from 0 to 1, where 0 indicates no overlap and 1 indicates perfect agreement between the predicted and actual sets. Unlike Precision or Recall alone, the Jaccard Similarity accounts for both false positives and false negatives in a single measure of set overlap.

Second, Mapping Accuracy focuses on the extent to which the actual ground-truth CVEs are successfully recovered. It is computed as the ratio of the intersection of the predicted set ($\mathcal{L}_{a}$) and actual set ($\mathcal{M}_{a}$) to the size of the ground-truth set:

\begin{equation}
\mathrm{Mapping \, Accuracy} = \frac{|\mathcal{L}_{a} \cap \mathcal{M}_{a}|}{|\mathcal{M}_{a}|}.
\end{equation}

This measure can be interpreted as a recall-oriented metric, showing the fraction of relevant CVEs correctly mapped by the model.

Third, Detection Accuracy measures the portion of linked CVE IDs in the set of predicted CVE IDs and is the ratio of the intersection of the predicted set ($\mathcal{L}_{a}$) and the actual set ($\mathcal{M}_{a}$) to the size of the predicted set ($\mathcal{L}_{a}$):

\begin{equation}
\mathrm{Detection \, Accuracy} = \frac{|\mathcal{L}_{a} \cap \mathcal{M}_{a}|}{|\mathcal{L}_{a}|}.
\end{equation}

This metric has a precision-oriented interpretation, as it highlights the fraction of model predictions that correspond to true links.
In summary, the combination of Jaccard Similarity, Mapping Accuracy, and Detection Accuracy provides a detailed picture of how well the predicted sets align with the ground truth, complementing the earlier classification-based evaluation. While these measures quantify the overlap numerically, they do not verify the semantic correctness of predictions beyond the existing repository links. To address this limitation, the next section presents a manual validation process, in which predicted attack–CVE links are inspected to assess their practical relevance and to identify new, previously undocumented connections.

\section{Manual validation}\label{sec:Validation}

Manual validation process designed to address RQ2 by examining whether transformer-based models can identify vulnerabilities that are not explicitly documented in the MITRE repositories. For this purpose, we use the best-performing model–attack-type pair identified in Section~\ref{sec:Comparison}.
The focus of this validation lies on CVE identifiers predicted by the model but not explicitly linked to the corresponding attacks in the repositories. These predictions are categorized as \textit{False Positives} in the automated evaluation. However, this set is heterogeneous: it may contain truly incorrect predictions, but it may also include valid links that are missing from the MITRE repositories due to their incomplete coverage (see Section~\ref{sec:AttacksAnnotation}). Distinguishing between these two cases requires expert inspection.
To conduct this validation, we implemented a structured and iterative review process. First, all predicted but unlinked CVEs were extracted for inspection. Each CVE description was then compared with the corresponding attack description to assess their semantic relationship. Second, a CVE was marked as “linked” if we judged that the vulnerability description clearly matched the behavior or characteristic described in the attack entry. If we initially disagreed, the case was re-examined in a second round of discussion. After two iterations, only those cases on which consensus was reached were retained as valid links.
This procedure serves two complementary purposes. From a methodological perspective, it refines the classification of False Positives by separating true errors from repository omissions. From a practical perspective, it highlights undocumented but meaningful attack–vulnerability links that could enrich existing cybersecurity knowledge bases.
In summary, the manual validation complements the automated evaluation by assessing the semantic correctness of predictions beyond the explicit repository annotations. This step ensures that undocumented but meaningful attack–vulnerability links are not discarded as mere errors, but rather considered as potential contributions to the enrichment of MITRE’s knowledge bases.

\section{Evaluation on real-world data}
\label{sec:Newsevaluation}

\begin{figure}[t!]
    \centering
    \includegraphics[width=\linewidth]{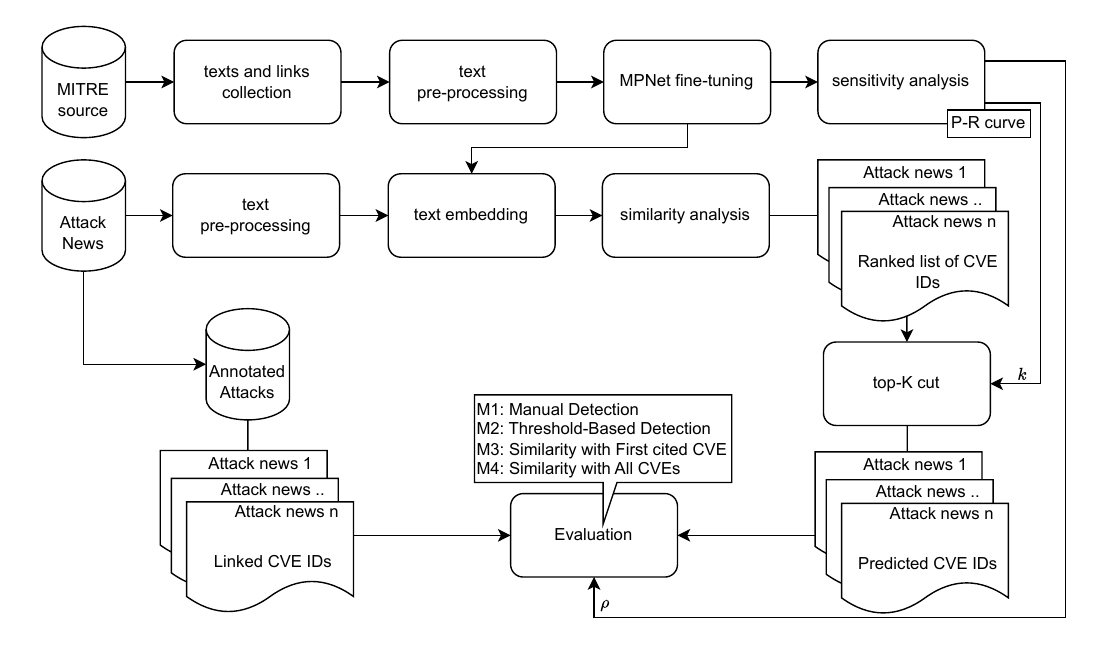}
    \caption{News methodology overview.}
    \label{fig:NewsMethodology}
\end{figure}
To address RQ3, we extend our methodology to evaluate the ability of the best-performing sentence transformer model to generalize from curated repositories to real-world cyberattack news. As illustrated in Figure~\ref{fig:NewsMethodology}, the process follows a top-K prediction framework in which each news description is embedded and compared against all CVE embeddings, generating a ranked list of candidate vulnerabilities. 
Let $a$ denote an attack news description and let $C = {c_1, c_2, \ldots, c_n}$ represent the set of vulnerability descriptions. For any pair $(a, c)$, $\text{sim}(a, c)$ denotes the cosine similarity between their embeddings, and $\rho$ is the fixed similarity threshold determined during the sensitivity analysis (Section~\ref{sec:SensitivityAnalysis}).
The resulting top-K predictions are then validated through four complementary strategies, each targeting a different dimension of evaluation. Manual validation (M1) provides expert-driven confirmation of semantic correctness, while threshold-based validation (M2) applies the similarity threshold $\rho$ to filter predictions. In addition, predictions are compared with explicit references in the news text, either by using the first CVE mentioned as the primary reference (M3) or by aggregating all mentioned CVEs into a single embedding to assess broader semantic coverage (M4). Together, these methods ensure that model predictions are evaluated not only in terms of quantitative similarity but also in terms of contextual and practical relevance, thereby providing a rigorous framework for assessing generalization to real-world data.

\paragraph{M1: Manual validation.}
In this method, model predictions are validated through manual validation. For each attack description, the top-k predicted CVEs are reviewed to determine whether their textual descriptions are related to the attack. A CVE is considered valid if the evaluation confirms a meaningful match.

\[
\text{Predicted}_{\text{M1}}(a) = 
\{c_i \in C \mid c_i \text{ is judged related to } a \text{ by manual inspection}\}
\]

\paragraph{M2: Threshold-based validation.}
This method retains only those CVEs from the top-k predictions whose similarity score to the attack description exceeds the threshold $\rho$. Rather than adopting the default classification threshold of 0.5, we employ threshold sensitivity analysis~\cite{sheng2006thresholding} and Precision–Recall analysis~\cite{davis2006relationship} to select an empirically grounded value of $\rho$, ensuring balanced performance between precision and recall.
\[
\text{Predicted}_{\text{M2}}(a) = 
\{c \in \text{top-}k(C) \mid \text{sim}(a, c) \geq \rho\}
\]

\paragraph{M3: Validation with the first CVE reference.}
When news articles reference a specific CVE, the first-mentioned CVE is often central to the report. In this method, we embed the description of this CVE, denoted as $c_1^{\text{attack}}$, and compare it to each predicted CVE from the model output. A prediction is validated if its similarity to $c_1^{\text{attack}}$ exceeds the threshold $\rho$.

\[
\text{Predicted}_{\text{M3}}(a) = 
\{c \in C \mid \text{sim}(c_1^{\text{attack}}, c) \geq \rho\}
\]

\paragraph{M4: Validation with all CVE references.}
Some attack news reports reference multiple CVEs. To capture this, we concatenate all referenced CVE descriptions, denoted as $C^{\prime} \subseteq C$, into a single aggregated text, which is then embedded as $\text{concat}(C^{\prime})^{\text{attack}}$. Each predicted CVE is validated if its similarity score with this aggregated embedding exceeds $\rho$.

\[
\text{Predicted}_{\text{M4}}(a) = \{c \in \text{top-}k(C) \mid \text{sim}(\text{concat}(C^{\prime})^{\text{attack}}, c) \geq \rho\}
\]

These four methods provide a comprehensive evaluation framework for real-world news data. Manual validation (M1) ensures semantic correctness, while our oracle automated methods (M2–M4) offer systematic and reproducible ways of aligning predictions with known or referenced vulnerabilities. By combining manual and automated approaches, we achieve a balanced evaluation that captures both the quantitative accuracy and the practical relevance of the model’s predictions, thereby ensuring a rigorous assessment of its generalization capability to real-world attack scenarios.






\chapter{Results}
\label{ch:results}
\huge{I}\normalsize{n this chapter, we present the experimental results of our empirical evaluation with respect to the research questions stated in Sec.~\ref{sec:intro:researchgoals}.}
The chapter is divided into three sections according to the research goals: \textbf{RG1}: Identify effective sentence transformer models and attack information types
for vulnerability prediction, \textbf{RG2}: Recommend semantic CVE links that are not explicitly listed in MITRE
repositories, and \textbf{RG3}: Evaluate model generalization to unseen real-world data, as shown in Table~\ref{tab:mappingTable}.

\begin{table}[htbp]
\centering
\caption{Mapping of the thesis papers and chapters to the research goals and questions}
\label{tab:papers-rg-rq}
\renewcommand{\arraystretch}{1.3}
\setlength{\tabcolsep}{8pt}

\begin{tabular}{>{\raggedright\arraybackslash}p{3.2cm}|ccc}
\hline
\multirow{2}{*}{\parbox[c]{3.2cm}{\centering\textbf{Papers / Chapters}}}
  & \textbf{RG1} & \textbf{RG2} & \textbf{RG3} \\
\cline{2-4}
  & RQ1 & RQ2 & RQ3 \\
\hline
Paper A   & \checkmark & \checkmark &   \\
Paper B   & \checkmark &            &   \\
Paper C   & \checkmark &            &   \\
Paper D   &            & \checkmark & \checkmark \\
Paper E   &            & \checkmark & \checkmark \\
Chapter F &            &            & \checkmark \\
\hline
\end{tabular}
\label{tab:mappingTable}
\end{table}

\section{RG1: Identify effective sentence transformer models and attack information types for vulnerability prediction}

This section addresses \textbf{RG1}, which aims to determine which sentence transformer models and which types of attack descriptions are most effective for predicting known vulnerabilities. To achieve this goal, we derived the following research question:

\begin{quote}
\rqone
\end{quote}

To answer RQ1, we performed a comparative evaluation of 14 SOTA sentence transformer models in Table~\ref{tab:ATT2V}, each tested on the four types of attack descriptions provided by MITRE repositories as described in Section~\ref{sec:PerformanceEvaluation}. 
We evaluated the models across four categories of attack information extracted from MITRE repositories: (1) Tactic, (2) Technique, (3) Procedure, and (4) Attack Pattern (see Section~\ref{sec:PerformanceEvaluation}).
Evaluating models across these categories enables us to assess not only model performance but also the informativeness of the attack descriptions themselves.

The comparative analysis in Table~\ref{tab:ATT2V} shows that \textbf{MMPNet} consistently achieves the highest performance overall, obtaining an F1-score of 89.0 for \textit{Techniques} and 72.4 for \textit{Attack Patterns}. This demonstrates its ability to capture both detailed and generalized attack semantics. The \textbf{XXLT5} model, a large-scale T5 variant designed for dense retrieval, performs best on \textit{Tactics} with an F1-score of 88.0, and is remarkable for achieving recall values of 100\% across several categories, although its precision drops in fine-grained contexts such as \textit{Procedures} (Precision = 48.6). Other strong performers include \textbf{MPNet}, which achieved F1 = 80.1 on Techniques and 60.9 on Procedures, showing balanced precision and recall, and \textbf{MSMBERT}, which outperformed all other models on Procedures with an F1 of 66.7.  
\begin{table*}[htbp!]
    \centering
    \fontsize{8pt}{13pt}\selectfont
    \caption{Performance metrics for all types of attacks and models.}
    \begin{tabular}{l@{\hskip 1pt}|c@{\hskip 2pt}c@{\hskip 2pt}c|c@{\hskip 2pt}c@{\hskip 2pt}c|c@{\hskip 2pt}c@{\hskip 2pt}c|c@{\hskip 2pt}c@{\hskip 2pt}c}
    \hline
         &  \multicolumn{3}{c|}{\textbf{Tactic}} & \multicolumn{3}{c|}{\textbf{Technique}} & \multicolumn{3}{c|}{\textbf{Procedure}} & \multicolumn{3}{c}{\textbf{ Pattern}} \\
         \hline 
    \textbf{Acronym} & \textbf{Precision} & \textbf{Recall} & \textbf{F1-Score} & \textbf{Precision} & \textbf{Recall} & \textbf{F1-Score} & \textbf{Precision} & \textbf{Recall} & \textbf{F1-Score} & \textbf{Precision} & \textbf{Recall} & \textbf{F1-Score} \\
    \hline
PAlbert	&	50	&	66.7	&	57.1	&	96.8	&	22.6	&	36.6	&	100	&	6.9	&	12.8	&	47.8	&	50	&	 48.9 \\	
PTinyBERT	&	100	&	18.2	&	30.8	&	100	&	22.6	&	36.8	&	92.1	&	19.1	&	31.6	&	53.3	&	46.5	&	 49.7 \\	  
MDBERT	&	75	&	27.3	&	40	&	100	&	15	&	26.1	&	70.8	&	15.2	&	24.1	&	37.5	&	3.5	&	 6.4 \\	
MSMBERT	&	50	&	100	&	66.7	&	66.5	&	100	&	79.9	&	50	&	100	&	66.7	&	48.6	&	100	&	 65.4 \\	
DRoBERTa	&	66.7	&	18.2	&	28.6	&	79.7	&	41.4	&	54.5	&	77.8	&	8	&	14.4	&	52.1	&	43	&	 47.1 \\	
Roberta	&	75	&	27.3	&	40	&	84.8	&	66.9	&	74.8	&	83.3	&	17	&	28.3	&	52.3	&	53.5	&	 52.9 \\	  \hline
MiniLM6	&	100	&	18.2	&	30.8	&	86.7	&	29.3	&	43.8	&	100	&	1.1	&	2.2	&	48.6	&	20.9	&	 29.3 \\	
MiniLM12	&	100	&	33.3	&	50	&	94.5	&	51.9	&	66.9	&	100	&	3.4	&	6.5	&	52.8	&	32.6	&	 40.3 \\	
MMiniLM6	&	100	&	9.1	&	16.7	&	95.2	&	15	&	26	&	4.4	&	60	&	3.5	&	50	&	10.5	&	 17.3 \\	
PMiniLM6	&	40	&	66.7	&	50	&	93.3	&	31.6	&	47.2	&	86.9	&	2.8	&	5.4	&	52	&	75.6	&	 61.6 \\	
PMiniLM12	&	40	&	66.7	&	50	&	93.7	&	44.4	&	60.2	&	76.9	&	22.7	&	35	&	57.2	&	87.3	&	 69.1 \\	  \hline
MPNet	&	100	&	9.1	&	16.7	&	77.1	&	83.5	&	80.1	&	97.5	&	44.3	&	60.9	&	52.9	&	53.5	&	 53.2 \\	
MMPNet	&	100	&	33.3	&	50	&	84	&	94.7	&	 \textbf{89.0} 	&	89.7	&	29.5	&	44.4	&	71.6	&	73.3	&	 \textbf{72.4}  \\	  \hline
XXLT5	&	78.6	&	100	&	 \textbf{88.0} 	&	66.5	&	100	&	79.9	&	48.6	&	100	&	65.4	&	50.1	&	100	&	 66.7 \\			    \hline
    \end{tabular}
    \label{tab:ATT2V}
\end{table*}
In contrast, lightweight models such as MiniLM and TinyBERT consistently underperform. For example, DRoBERTa and PTinyBERT achieved F1 scores below 40 in multiple categories, reflecting their limited capacity to capture the complex semantic relationships required for effective vulnerability prediction. These results indicate that higher-capacity architectures with richer embedding spaces (e.g., MMPNet, XXLT5) offer significant advantages over compressed, distilled models when applied to cyberattack–vulnerability linking tasks.
When comparing the four types of attack descriptions, \textbf{Techniques} emerge as the most informative input type. Most models achieved F1-scores above 60\% in this category, with MMPNet reaching 89.0, XXLT5 achieving 79.9, and RoBERTa reaching 74.8. This confirms that Techniques, which describe specific adversarial methods, provide rich semantic signals that are most easily mapped to CVE descriptions.  
The second most effective category is \textbf{Attack Patterns}, where multiple models achieved F1 scores above 65 (e.g., MMPNet: 72.4, PMiniLM12: 69.1, MSMBERT: 65.4). This reflects the fact that attack patterns encode recurring exploitation strategies that generalize well across different vulnerabilities.  
In contrast, \textbf{Procedures} performed poorly for most models, often with F1-scores below 50. This weakness can be attributed to the highly specific nature of procedure descriptions, which often reference particular products, versions, or configurations. Only a few models managed reasonable performance, with MSMBERT achieving the highest score (F1 = 66.7), followed by MPNet (60.9). Finally, \textbf{Tactics}, being highly abstract and goal-oriented, are less discriminative overall. However, large-scale models such as XXLT5 are able to exploit their representational capacity to achieve strong results (F1 = 88.0) by capturing high-level semantic regularities.  
\medskip

\begin{figure*}[htb!]
    \centering
    \includegraphics[width=0.9\textwidth]{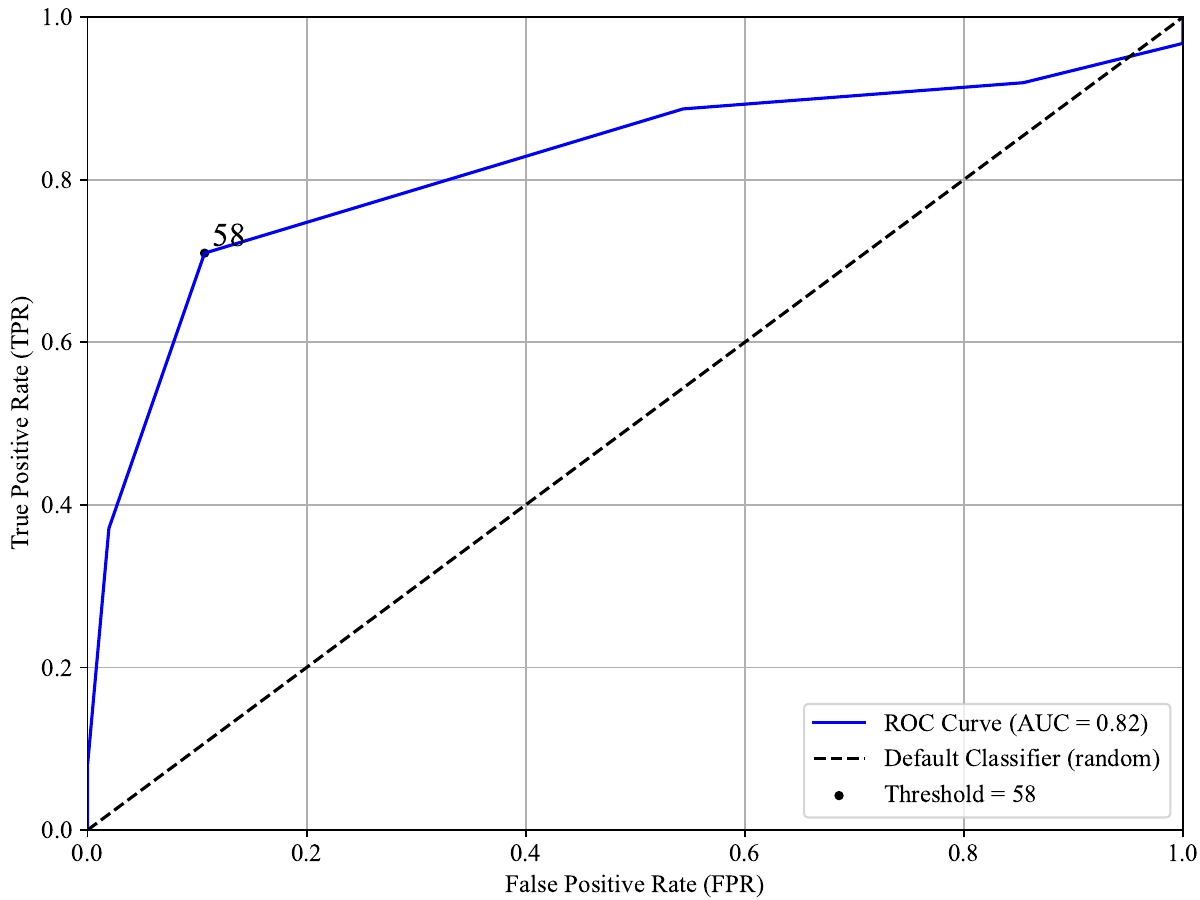}
    \caption{ROC analysis of the pair (MMPNet, Technique). The AUC = 0.82 and the optimal threshold occurs at $\rho = 58\%$.}
    \label{fig:AUCsinsitivity}
\end{figure*}

\begin{figure*}[htb!]
    \centering
    \includegraphics[width=\textwidth]{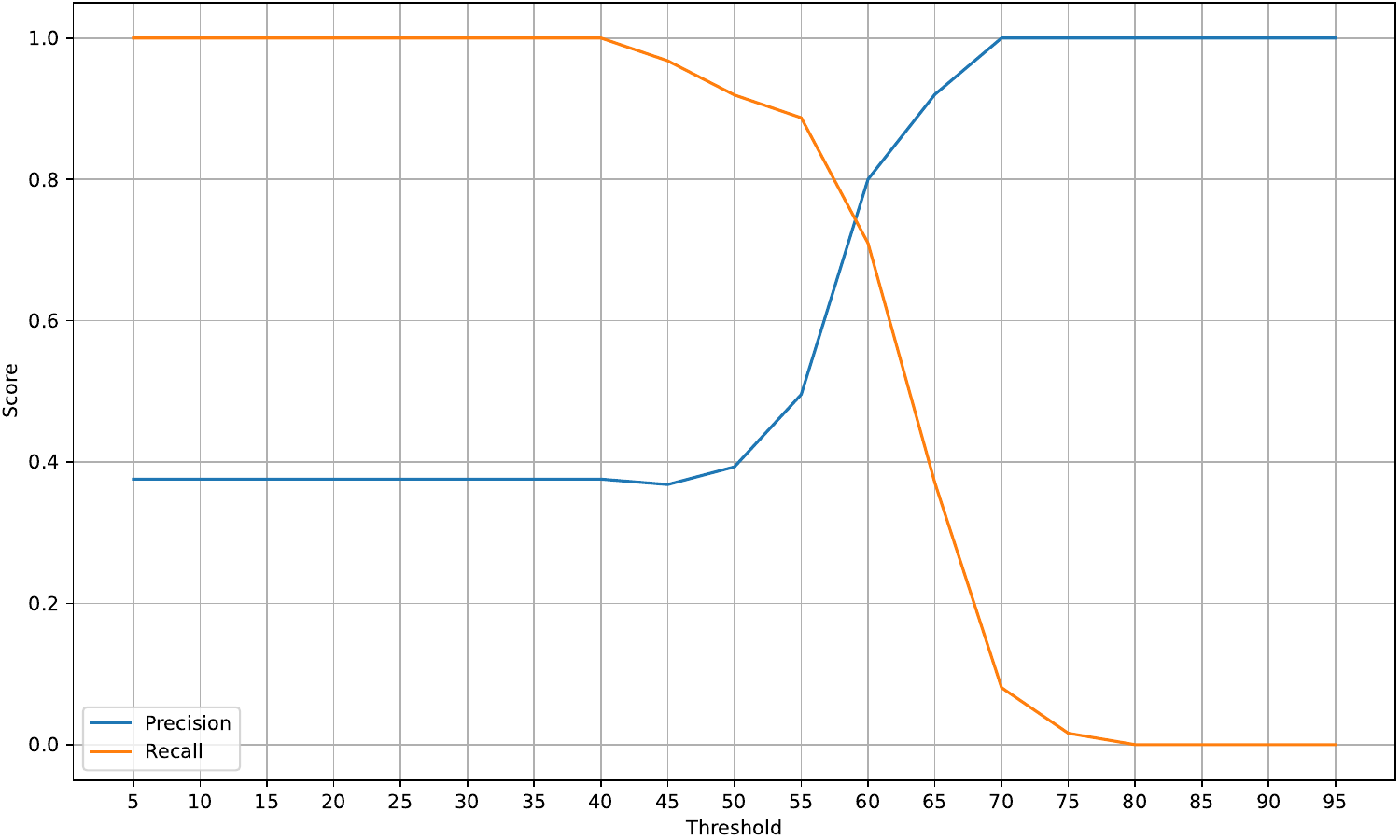}
    \caption{Precision–Recall analysis of the pair (MMPNet, Technique). The intersection at $\rho = 58\%$ indicates the Equal Error Rate point where precision and recall are balanced.}
    \label{fig:sensitivityAnalysis}
\end{figure*}

\paragraph{\textbf{Sensitivity analysis.}}  
As described in Section~\ref{sec:SensitivityAnalysis}, we determined the optimal threshold $\rho$ using both ROC- and PR-based analyses. Here we present the empirical results of applying this methodology to the best-performing pair, namely (MMPNet, Technique).  
Figure~\ref{fig:AUCsinsitivity} shows the ROC curve obtained from a balanced dataset of 50 positive and 50 negative examples. The resulting AUC of 0.82 confirms the strong discriminative ability of the model. The optimal threshold selected by minimizing the Euclidean distance to the ideal point $(0,1)$ was $\rho = 58\%$, which is close to our initial heuristic choice of $\rho = 60\%$.  
Complementary evidence is provided by the Precision–Recall analysis (Figure~\ref{fig:sensitivityAnalysis}). The precision and recall curves intersect at $\rho = 58\%$, corresponding to the EER, where false positives and false negatives are balanced.  
Together, these results validate the robustness of our chosen threshold. Both ROC and PR perspectives converge on $\rho = 58\%$ as the optimal operating point, while our use of $\rho = 60\%$ in the experiments remains consistent with this outcome. These findings further highlight that threshold choice can be adapted depending on practical priorities: analysts may prefer a lower threshold to maximize recall (reducing missed vulnerabilities) or a higher threshold to maximize precision (reducing false alarms).  

The superior performance of \textbf{MMPNet} can be explained by its underlying architecture, which leverages permuted language modeling to capture bidirectional dependencies and richer contextual semantics than other models. Together with MSMBERT, MMPNet is among the only models trained using a dot-product similarity objective, which accounts for both vector length and orientation. This choice enhances the model’s ability to discriminate subtle contextual differences in embeddings. Indeed, while MMPNet dominates for \textit{Techniques} and \textit{Patterns}, MSMBERT achieves the best performance on \textit{Procedures}, demonstrating that the loss function plays a role in aligning embeddings for specific attack types. In contrast, lightweight models trained primarily with cosine objectives (e.g., MiniLM, TinyBERT) fail to capture such nuances and exhibit much lower F1 scores.  
The evaluation shows that \textbf{Technique} is the most informative attack description type for vulnerability prediction, and transformer models with higher capacity or more effective training objectives consistently outperform lighter architectures. Large-scale models such as XXLT5 excel at abstract categories like Tactics, while specialized models such as MMPNet achieve SOTA results on the most informative category, Techniques. 

\begin{tcolorbox}
\textbf{RQ1 Summary:} Sentence transformers are effective in predicting vulnerabilities from attack descriptions. Among the evaluated models, MMPNet achieved the best overall results, with F1 = 89.0 on Techniques and 72.4 on Attack Patterns. XXLT5 excelled on Tactics with F1 = 88.0, while MSMBERT was the best model for Procedures (F1 = 66.7). Techniques were the most informative input type, followed by Attack Patterns, whereas Procedures proved the least informative. High-capacity models consistently outperformed lightweight variants, and dot-product–based training objectives (MMPNet, MSMBERT) offered further advantages.  These findings validate the effectiveness of transformer-based methods for vulnerability prediction and highlight the importance of selecting both the right model and the right type of attack description.
\end{tcolorbox}

\section{RG2: Recommend semantic CVE links that are not explicitly listed in MITRE repositories}

This section addresses \textbf{RG2}, which aims to evaluate whether sentence transformer models can (1) reproduce existing CVE links from MITRE repositories and (2) recommend additional links that are currently missing but semantically valid. To achieve this goal, we derived the following research question:

\begin{quote}
\rqtwo
\end{quote}

\begin{figure}[htb!]
\centering
\includegraphics[width=\columnwidth]{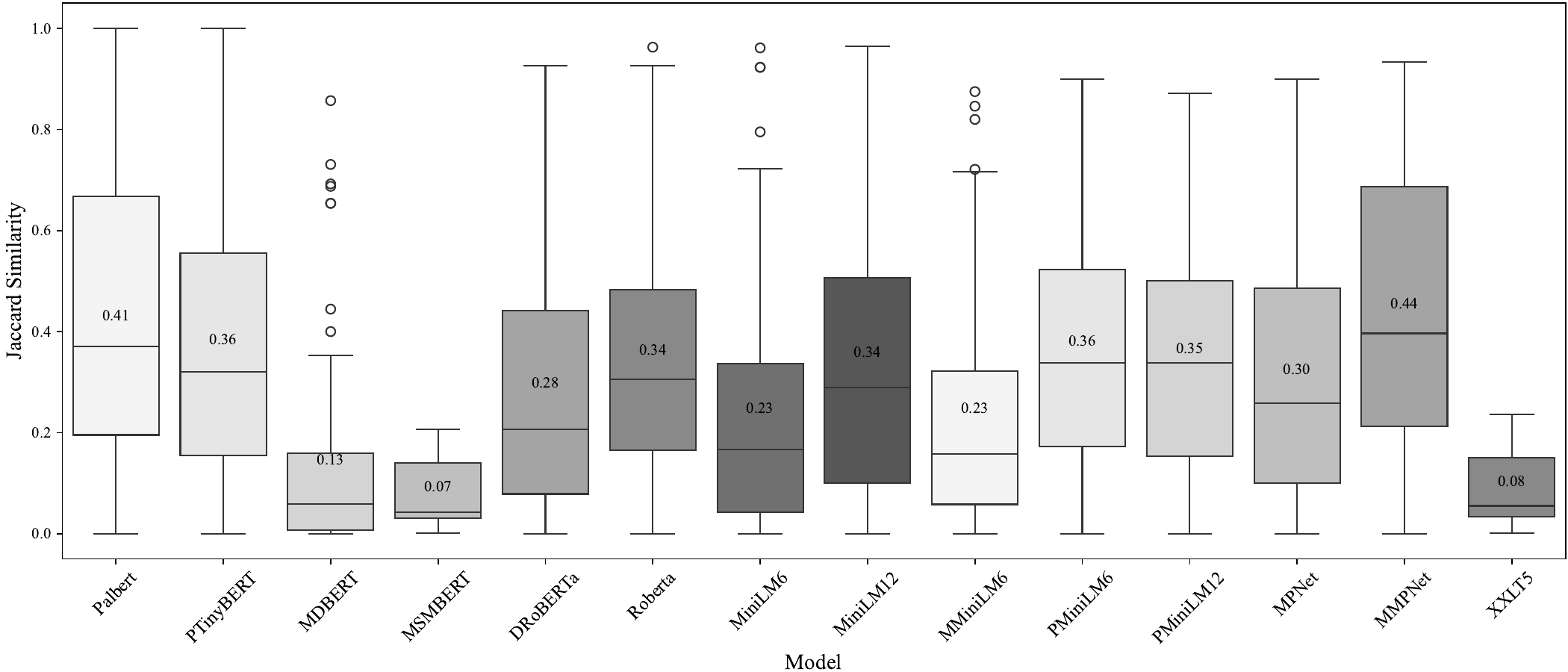}
\caption{Jaccard similarity across SOTA models.}\label{fig:allmodelsJaccard}
\end{figure}

After answering RQ1, we established that the most effective configuration for vulnerability prediction is the combination of the MMPNet model with \textit{Technique} descriptions. Building on this result, we use this best-performing setup as the foundation for RQ2. In particular, we investigate whether MMPNet applied to Techniques can not only reproduce the existing CVE links documented in MITRE repositories but also recommend new, previously undocumented links that extend and enrich these repositories. 
Thus, to answer RQ2, we first conducted a comparative evaluation of SOTA transformer models using Jaccard Similarity as the primary metric (Figure~\ref{fig:allmodelsJaccard}). The results indicate that \textbf{MMPNet} achieved the highest Jaccard Similarity (mean = 0.44), demonstrating superior capability in identifying linked CVE IDs from attack descriptions. After establishing that MMPNet combined with \textit{Technique} descriptions represents the most effective configuration, we selected this pair for deeper evaluation with three complementary metrics: Jaccard Similarity, Mapping Accuracy, and Detection Accuracy (Section~\ref{sec:Magnitude}). This second step allowed us to go beyond model-to-model comparison and quantify in detail how well the chosen configuration is able to reproduce known links and uncover potentially missing ones.  
\begin{figure}[htb!]
\centering
\includegraphics[width=\columnwidth]{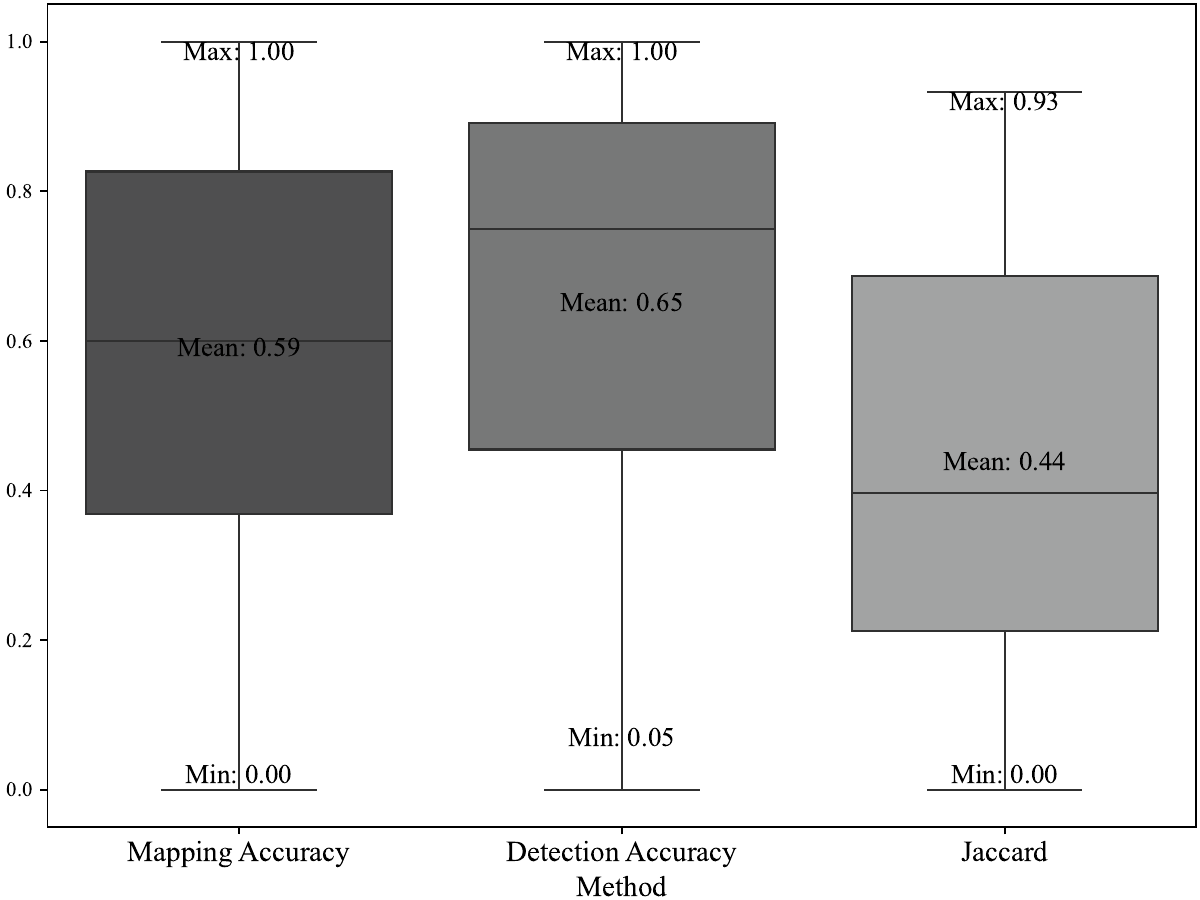}
\caption{Performance of the MMPNet-based approach across attack techniques description.}\label{fig:jaccardplot}
\end{figure}

Figure~\ref{fig:jaccardplot} shows the distribution of these measures across all ATT\&CK Techniques. On average, the Mapping Accuracy reached 56\%, the Detection Accuracy 61\%, and the Jaccard Similarity 40\%. These values demonstrate that MMPNet is able to reproduce a significant proportion of the CVE links documented in MITRE repositories. For more than 50\% of the Techniques, the model predicted CVE IDs that were over 70\% linked (i.e., median Detection Accuracy above 70\%). A representative example is Technique T1539 “Steal Web Session Cookie,” where the model achieved a Jaccard Similarity of 82\%. Specifically, $\mathcal{L}(T1539)$ contained 150 CVEs, $\mathcal{M}(T1539)$ contained 125, and the model retrieved all but one existing linked CVE IDs, while also predicting a number of additional ones.

\begin{table*}[htb]
\centering
\footnotesize
\caption{Summary of our approach validation process for new candidates' links.}
\label{tab:validationResult}
\setlength{\tabcolsep}{3pt} 
\begin{tabular}{@{}lcc|cc@{}} 
\hline
&\textbf{Retrieved}&&\textbf{Validated}&\\
\hline
&\textbf{Total Predicted}&\textbf{False positives}&\textbf{False recommended}& \textbf{Recommended to be linked} \\
\hline
\textbf{Number of linked CVE reports} &  2230 &434& 159& 275
 (12.3\%)\\
\textbf{Number of techniques} & 201 & 74 & 54 & 60 (29.8\%)\\
 \hline
\end{tabular}
\end{table*}

These additional predictions may represent false positives, but they may also reflect valid attack–vulnerability links that are missing from the MITRE repositories. To investigate this further, we manually inspected the model’s recommendations. Table~\ref{tab:validationResult} summarizes the results of this validation. Out of 2230 linked CVE reports across 201 Techniques, the model predicted 434 additional CVE IDs not listed in the repositories. Manual inspection revealed that 159 were false positives, but 275 were validated as legitimate links across 60 Techniques (29.8\%). This corresponds to 12.3\% of all CVE reports in our evaluation.  

An illustrative example is Technique T1039 “Data from Network Shared Drive.” Here, MMPNet predicted CVE-2005-1205, CVE-2009-3107, and CVE-2020-3452, all of which involve unauthorized access or data exposure through shared network resources. These vulnerabilities were missing from the ATT\&CK repositories at the time of annotation but were later independently added, thereby reinforcing the validity of our model’s predictions. Similar cases were observed for several other Techniques, confirming that our approach is able not only to replicate existing knowledge but also to uncover previously undocumented links. In addition, we are currently preparing a formal submission of the 275 validated links to the MITRE board in order to support the enrichment of ATT\&CK repositories. These findings demonstrate the value of automated approaches in enhancing the coverage and completeness of curated security knowledge bases.

\begin{tcolorbox}
\textbf{RQ2 Summary.}  
Our evaluation shows that transformer-based models can both reproduce existing attack–CVE links and recommend missing ones. Using MMPNet with Technique descriptions, we achieved Mapping Accuracy of 56\%, Detection Accuracy of 61\%, and Jaccard Similarity of 40\% across ATT\&CK Techniques. Beyond reproducing known links, our manual validation confirmed 275 previously undocumented CVE links across 60 Techniques (about 12\% of all CVE reports), many of which were later incorporated into MITRE repositories. This demonstrates that our approach can significantly contribute to repository enrichment by systematically identifying missing semantic links.
\end{tcolorbox}

\section{RG3: Evaluate model generalization to unseen real-world data}

This section addresses \textbf{RG3}, which aims to assess whether the proposed approach can generalize from structured repositories to unstructured sources such as cybersecurity news. After identifying the best-performing configuration in RQ1 (MMPNet with \textit{Technique} descriptions) and validating its ability to reproduce and enrich repository links in RQ2, we now investigate how well this setup performs on real-world data. Cybersecurity news reports differ substantially from structured repositories: they are noisy, informal, and highly diverse in language. To evaluate robustness in this setting, we derived the following research question:

\begin{quote}
\rqthree
\end{quote}

To answer RQ3, we evaluated the MMPNet–Technique configuration on a dataset of 100 real-world cybersecurity news reports. For each report, the model generated a ranked list of top-K candidate CVEs, where we set $k=20$ following established practices in information retrieval as shown in Section~\ref{sec:TopKAnalysis}. The evaluation combined two complementary strategies: (1) manual validation of the predictions against the content of the news reports, and (2) automated oracle-based validation using similarity-based heuristics.  

\medskip
\noindent\textbf{Manual validation.}  
We manually validated 2000 predicted CVEs by comparing their descriptions to the corresponding news reports (M1: Manual Detection). Out of these, 1418 predictions (70\%) were confirmed as relevant, while 582 (30\%) were not, as illustrated in Table~\ref{tab:cveComparison}. This result demonstrates that the model is able to retrieve vulnerabilities that align semantically with real-world attack narratives in a majority of cases. 
\begin{table*}[htb]
\centering
\scriptsize
\caption{Validation outcomes for each method (M1–M4).}
\label{tab:cveComparison}
\begin{tabular}{lc|cc}
\hline
 & \textbf{Retrieved} & \multicolumn{2}{c}{\textbf{Validated}} \\ 
 & \textbf{Total} & \textbf{Relevant} & \textbf{Not Relevant} \\ 
\hline
M1: Manual Detection & 2000 & 1418 (70\%) & 582 (30\%) \\ 
M2: Threshold-Based Detection & 2000 & 1625 (81\%) & 375 (19\%) \\ 
M3: Similarity with First CVE & 2000 & 1607 (80\%) & 393 (20\%) \\ 
M4: Similarity with All CVEs & 2000 & 1575 (78\%) & 425 (22\%) \\ 
\hline
\end{tabular}
\end{table*}
Additionally, we analyzed the overlap between predicted and explicitly mentioned CVE IDs in the reports. As shown in Table~\ref{tab:attackNewsCVE}, in 57\% of the news reports, at least one predicted CVE matched a CVE identifier explicitly mentioned in the report. In 40\% of the cases, the model suggested CVEs not explicitly listed but potentially relevant, while 3\% of the reports did not mention any CVE identifiers at all. 
These results suggest that the model is not only capable of recovering directly referenced vulnerabilities but also of inferring plausible vulnerabilities when explicit identifiers are absent. This highlights its practical utility for analysts, who often face incomplete or noisy reporting in real-world intelligence feeds.


\begin{table*}[htb]
\centering
\caption{Number of news reports where the model’s predicted CVEs matched at least one CVE mentioned in the report.}
\label{tab:attackNewsCVE}
\setlength{\tabcolsep}{5pt} 
\renewcommand{\arraystretch}{1} 
\resizebox{\textwidth}{!}{ 
\begin{tabular}{lcccc}
\hline
 & \multicolumn{4}{c}{\textbf{Retrieved}} \\ 
\cline{2-5}
\textbf{} & \textbf{Total} & \textbf{Attack News Containing Matching CVE-ID} & \textbf{Attack News No Matching CVE-ID} & \textbf{Attack News Without CVE-ID Link} \\ 
\hline
\textbf{Number of Attack News} & 100 & 57 (57\%) & 40 (40\%) & 3 (3\%) \\ 
\hline
\end{tabular}
}
\end{table*}

\medskip
\noindent\textbf{Oracle-based validation.}  
Manual validation is accurate but resource-intensive. To scale evaluation, we introduced three automated oracle methods: (M2) threshold-based detection, (M3) similarity with the first-mentioned CVE, and (M4) similarity with all CVEs mentioned in the report. Each oracle uses a calibrated threshold ($\rho = 0.58$, determined in Section~\ref{sec:SensitivityAnalysis}) to filter predictions.  

As summarized in Table~\ref{tab:cveComparison}, the oracle methods achieved comparable precision: 81\% for M2, 80\% for M3, and 78\% for M4. These results confirm that predictions retained above the similarity threshold are contextually relevant to the attack narratives. Moreover, M3 and M4 highlight the model’s ability to align predictions with the primary and secondary vulnerabilities described in the reports, capturing both single-CVE and multi-CVE contexts.  

The convergence of manual and oracle validation provides strong evidence that the model generalizes effectively to real-world, unstructured data. While manual validation confirmed that 70\% of predictions were relevant, the oracle methods consistently reported precision above 78\%. The overlap analysis further showed that more than half of the predictions directly matched explicitly mentioned CVEs, while in the remaining cases the model proposed plausible but undocumented vulnerabilities. This dual capability, of recovering known links and suggesting implied ones, demonstrates the potential of transformer-based approaches for supporting vulnerability detection in dynamic, heterogeneous intelligence sources.  

It is worth noting that performance is sensitive to the choice of top-K. In our study, $k=20$ balanced precision and recall, but different values may be more appropriate depending on the use case. A larger K increases coverage but introduces noise, while a smaller K improves precision at the expense of recall. Exploring the optimal K across different datasets of cyber threat reports is a promising direction for future work.  

\begin{tcolorbox}
\textbf{RQ3 Summary.}  
Our evaluation shows that the MMPNet–Technique configuration generalizes effectively to unstructured cybersecurity news. Manual validation confirmed that 70\% of 2000 predicted CVEs were contextually relevant, while oracle-based validation methods reported precision between 78–81\%. In 57\% of news reports, the model’s predictions matched at least one explicitly mentioned CVE, and in many of the remaining cases it suggested plausible vulnerabilities not explicitly referenced. These results highlight the model’s robustness in handling noisy, heterogeneous data sources and its potential to support analysts by identifying both documented and implied vulnerabilities in real-world threat reports.
\end{tcolorbox}


\chapter{Threats to validity}
\label{ch:threats}
\huge{I}\normalsize{n this chapter, we discuss potential threats to the validity of our study, structured according to the common categories of construct, internal, and external validity~\cite{sjoberg2022construct,wohlin2012experimentation,Shadish2002expdesign}. While each individual study presented its own validity considerations, here we provide a consolidated discussion across all our experiments.}

\section{Construct validity}
Construct validity refers to how well the operational measures used in our studies represent the underlying conceptual constructs~\cite{devellis2021scale}. Several issues arise in this regard.  
First, the choice of the cosine similarity threshold ($\rho$) constitutes a major source of bias. The threshold directly determines which CVEs are included in the list of recommended vulnerabilities for each attack description. If $\rho$ is set too high, potentially relevant CVEs may be excluded, while a low threshold may include CVEs that are not actually related to the underlying vulnerability. We empirically selected $\rho = 58$ by balancing false positives and false negatives through sensitivity analyses. While this value proved robust across experiments, a single global threshold may not optimally capture the operating characteristics of all models or attack types. 
We also considered an alternative truncation strategy by selecting the top-$k$ most similar CVEs for each attack description, which provided a complementary perspective on prediction quality. Nevertheless, other adaptive or model-specific thresholding techniques could be explored in future work to further improve the accuracy of vulnerability recommendations.
Second, transformer models are limited by the maximum input length they can process. Although all attack and vulnerability descriptions in our datasets fall well below the 384-token limit, this constraint could affect scalability if longer, more detailed reports are considered.  
Third, certain evaluation assumptions may not fully capture the intended construct. For example, in news-based evaluations we assumed that the first or aggregated CVEs mentioned in an article reflect ground truth, but reports may occasionally cite irrelevant or incomplete CVEs. Similarly, in repository-based evaluations, the absence of explicit links in MITRE datasets does not imply non-existence, since repositories are known to be incomplete and evolve over time. These factors could bias the measurement of semantic similarity and link prediction.
Finally, reported or predicted vulnerabilities may not correspond to the actual exploitable weaknesses in the systems under study. Textual descriptions alone cannot confirm whether a referenced CVE truly manifests in a given software environment. Future work should therefore aim to reproduce and execute these vulnerabilities in controlled settings to validate their real-world exploitability and ensure that the predicted links represent genuine vulnerabilities.

\section{Internal validity}
Internal validity concerns whether the observed results can be reliably attributed to our approach rather than confounding factors~\cite{petrivc2024validation}.  
One threat stems from the variability and noise in textual data. Attack and vulnerability descriptions often include highly technical terms (e.g., version numbers, library names, or acronyms) that may not contribute to semantic similarity but can influence model embeddings. We mitigated this by applying consistent preprocessing (removal of URLs, citations, and noise) across all datasets.  
Another threat is dataset selection. Repository-based studies rely on snapshots of MITRE ATT\&CK, CAPEC, CWE, and CVE, which are manually curated and updated periodically. While we documented the dataset version used, future updates may alter the ground truth links. In the news-based study, we used 100 articles from SecurityWeek. Although credible, this single source may not capture the full spectrum of reporting styles and detail levels, which could limit robustness.  
Finally, design choices such as fixing the top-$k$ parameter at $k=20$ ensured consistency across experiments, but may not be equally optimal for all cases, potentially underestimating or overestimating performance. Our sensitivity analysis suggests that $k=20$ is a reasonable compromise, but further experiments with dynamic $k$ values are warranted.

\section{External validity}
External validity relates to the generalizability of our findings beyond the studied datasets and settings~\cite{pan2023fine}.  
First, the scale of our ground truth was limited. In repository-based experiments, links were constructed from 100 ATT\&CK Techniques to 610 CVEs, a subset of the more than 201 techniques and 295,000 CVEs available. While suitable for exploratory analysis, this sample may not fully represent the breadth of adversarial behaviors or vulnerabilities.  
Second, both repositories and news sources are dynamic. The ATT\&CK and CVE datasets are updated continuously, meaning that models trained on static snapshots may degrade over time. Similarly, new vulnerabilities and attack techniques may not align with patterns learned from historical data.  
Third, generalizability across sources and languages remains a challenge. Our news-based evaluation was conducted exclusively on English articles from a single outlet. Other sources such as advisories, blogs, social media posts, or non-English reports may employ different terminology, style, or structure. Likewise, our main evaluations relied heavily on the MPNet model; although this proved effective, results may differ for other transformers or domain-adapted variants.  
To mitigate these threats, we plan to extend evaluations to larger and more diverse datasets, integrate additional threat intelligence sources such as ExploitDB, and test multilingual or domain-specific sentence transformers. These steps would increase confidence in the transferability of our findings to real-world, heterogeneous cybersecurity contexts.

Overall, while our approach demonstrates strong empirical performance and practical potential, its validity is subject to limitations stemming from threshold choices, dataset coverage, repository evolution, and generalizability to unseen sources. We addressed these threats through sensitivity analyses, careful preprocessing, manual validation, and documentation of dataset snapshots. Nonetheless, future research should expand the empirical basis of our studies, refine parameter selection strategies, and test broader contexts to strengthen both internal and external validity.

\chapter{Conclusion and Future work}
\label{ch:conclusion}
\huge{T}\normalsize{his thesis presented a comprehensive study on predicting software vulnerabilities from textual descriptions of adversary behaviors using sentence transformer models. By evaluating fourteen state-of-the-art transformer architectures across four attack description types (\textit{Tactic}, \textit{Technique}, \textit{Procedure}, and \textit{Attack Pattern}), the study demonstrated that transformer-based approaches can effectively automate the linking of attacks to vulnerabilities. The results confirmed that \textit{Technique} descriptions provide the most informative and discriminative signal, with the \texttt{multi-qa-mpnet-base-dot-v1 (MMPNet)} model achieving an F1-score of 89.0. This superior performance is attributed to its hybrid pre-training strategy, which combines masked and permuted language modeling and enables the model to capture long-range contextual dependencies and nuanced semantic relations in technical text. Importantly, manual validation identified 275 previously undocumented Technique--CVE links, highlighting the incompleteness of existing repositories and the need for automation. These findings provide empirical evidence that automated semantic linking not only enhances detection accuracy but also supports proactive and timely mitigation. By coupling comparative model evidence with a curated set of newly validated links, this work contributes both a practical pathway for decision-makers and a solid research foundation for advancing scalable, transparent, and operationally relevant attack--vulnerability linking.}

This research has two main implications. \textbf{For practitioners}, the findings offer actionable guidance for vulnerability management and threat intelligence operations: (i) prioritize \textit{Technique} descriptions when predicting candidate CVEs, as they yield the most accurate and informative results; (ii) operationalize high-performing models such as \texttt{MMPNet} to pre-populate candidate CVE sets for observed techniques, thereby reducing analyst workload and accelerating triage; and (iii) integrate the 275 validated new links into existing vulnerability management systems and CTI platforms to enrich lookups, shorten patching lead times, and strengthen defensive posture. Collectively, these practices support earlier mitigation, more consistent triage, and greater organizational resilience against evolving cyber threats.

\textbf{For researchers}, the comparative evaluation across fourteen transformer models and four attack information types establishes a replicable baseline for automated attack–to–vulnerability linking. Building upon this foundation, several promising directions emerge: (i) \textit{Model evolution}, assessing newer architectures and domain-specific or multilingual variants to improve robustness across heterogeneous cyber threat datasets; (ii) \textit{Data breadth}, extending beyond ATT\&CK–CVE linkages to other sources such as CWE, CAPEC, ExploitDB, and incident reports to evaluate generalization under distributional shift; (iii) \textit{Timeliness}, enabling near real-time ingestion of attack reports and news to recommend relevant CVEs as incidents unfold, thereby improving situational awareness and response speed; and (iv) \textit{Transparency}, incorporating continuous data streams from multiple intelligence feeds, retraining models to adapt to emerging vulnerabilities, and developing interpretability and visualization tools for model outputs. Together, these directions lay the groundwork for scalable, adaptive, and explainable vulnerability–attack linking systems that can advance both academic research and practical cybersecurity operations.

Building on the current findings, several promising research directions can further advance this work toward a comprehensive and automated framework for vulnerability detection and threat intelligence integration. Future efforts will extend the current pipeline beyond CVE prediction to infer the corresponding CWEs and the MITRE ATT\&CK techniques that are likely to exploit them. This will be complemented by static code analysis to locate and classify weaknesses at the source-code level, enabling a deeper understanding of how vulnerabilities manifest within software components. By integrating adversary emulation frameworks such as MITRE CALDERA~\cite{CALDERA}, the identified weaknesses can be reproduced and tested in controlled environments to verify exploitability and prioritize high-risk cases for remediation. 
In addition, forthcoming research will focus on constructing diverse datasets of vulnerable code functions drawn from repositories such as the MITRE family (CVE, CWE), DiverseVul, and GitHub. These datasets will support experiments aimed at identifying and classifying vulnerabilities across multiple programming languages and mapping each function to its corresponding CWE identifier and ATT\&CK technique. The resulting framework will provide an end-to-end mapping that traces an observed attack through its predicted CVE, associated CWE, and corresponding exploitation technique, thereby explaining “what was attacked, what was vulnerable, why the weakness occurred, and how it can be exploited.” 
Furthermore, the robustness and generalizability of the proposed methodology will be evaluated across heterogeneous datasets, including ExploitDB~\cite{exploitdb}, incident databases, and large-scale vulnerability repositories, to assess its adaptability and uncover previously undocumented semantic relationships between attacks and weaknesses. 
Finally, collaboration with the MITRE Corporation will be pursued to integrate the newly discovered attack–vulnerability–weakness mappings into the ATT\&CK, CVE, and CWE repositories, ensuring that the knowledge produced by this research directly contributes to advancing global cybersecurity intelligence and practice.

\printbibliography[title={References}] 
\appendix

\myPaper{Paper A \\ From attack descriptions to vulnerabilities: A sentence transformer-based
approach}
\begin{pubinfo}{
\renewcommand{\arraystretch}{1.3}
\begin{tabularx}{\textwidth}{>{\bfseries}l X}
Current status:      & Published \\
Journal:             & Journal of Systems and Software (Elsevier, Q1) \\
Date of acceptance:  & 12 September 2025 \\
Full citation:       & Refat Othman, Diaeddin Rimawi, Bruno Rossi, Barbara Russo (2025). 
From Attack Descriptions to Vulnerabilities: A Sentence Transformer-Based Approach. 
\textit{Journal of Systems and Software}, 200, 112615. DOI: \url{https://doi.org/10.1016/j.jss.2025.112615} \\
Journal description: & The Journal of Systems and Software (JSS) is a Q1 journal ranked 
\#3 worldwide in Google Scholar Software Systems venues. It publishes original, 
peer-reviewed research in software engineering and systems. \\
\end{tabularx}
}
\end{pubinfo}

\label{paper:A}
\includepdf[pages=-,pagecommand={\thispagestyle{plain}},fitpaper=true]{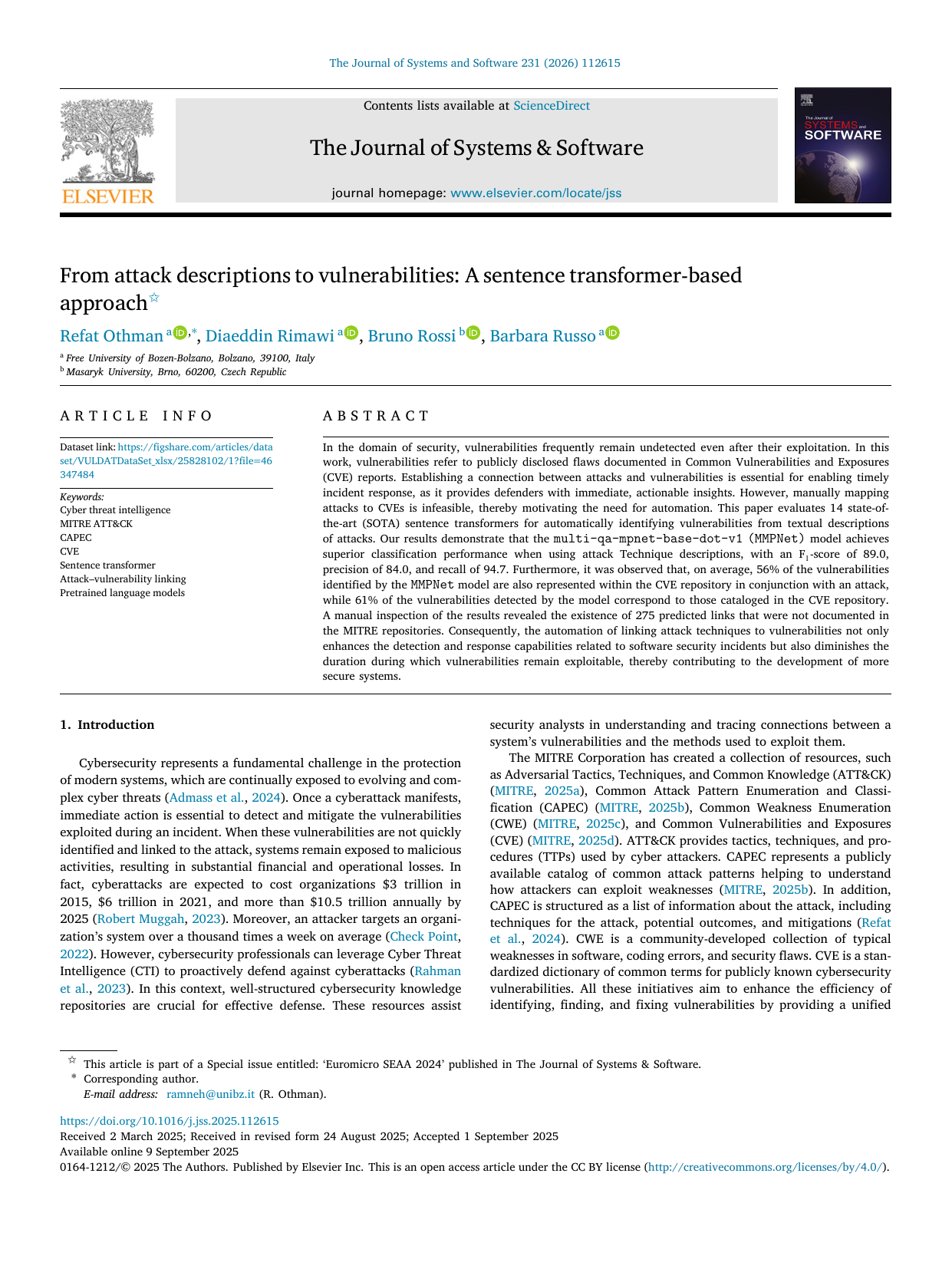}

\myPaper{Paper B \\ A Comparison of Vulnerability Feature Extraction
Methods from Textual Attack Patterns}
\begin{pubinfo}{
\renewcommand{\arraystretch}{1.3}
\begin{tabularx}{\textwidth}{>{\bfseries}l X}
Current status:      & Published \\
Conference:          & 2024 50th Euromicro Conference on Software Engineering and Advanced Applications (SEAA) \\
Date of acceptance:  & 28 August 2024 \\
Full citation:       & Refat Othman, Bruno Rossi, and Barbara Russo (2024). 
A Comparison of Vulnerability Feature Extraction Methods from Textual Attack Patterns. 
In: \textit{Proceedings of the 50th Euromicro Conference on Software Engineering and Advanced Applications (SEAA 2024)}, 
pp. 419--422. IEEE. DOI: \url{https://ieeexplore.ieee.org/document/10803510} \\
Conference description: & SEAA (Software Engineering and Advanced Applications) is a long-running IEEE/Euromicro conference, 
CORE B ranked, focusing on software engineering, systems, and applications. It provides a platform for 
presenting advanced methods, tools, and industrial experiences. \\
\end{tabularx}
}
\end{pubinfo}

\label{paper:B}
\includepdf[pages=-,pagecommand={\thispagestyle{plain}},fitpaper=true]{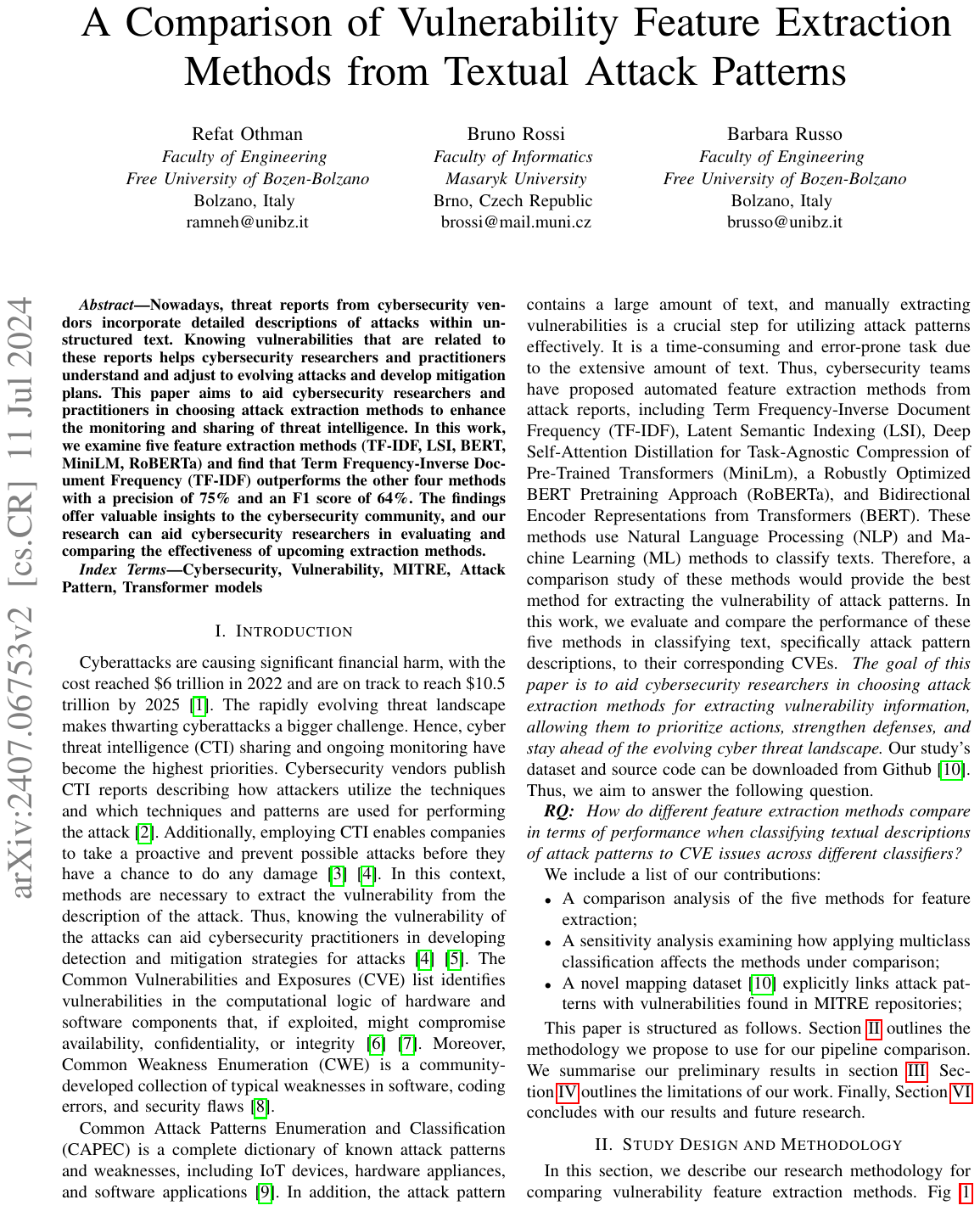}

\myPaper{Paper C \\ Cybersecurity Defenses: Exploration of CVE Types
through Attack Descriptions}
\begin{pubinfo}{
\renewcommand{\arraystretch}{1.3}
\begin{tabularx}{\textwidth}{>{\bfseries}l X}
Current status:      & Published \\
Conference:          & 2024 50th Euromicro Conference on Software Engineering and Advanced Applications (SEAA) \\
Date of acceptance:  & 28 August 2024 \\
Full citation:       & Refat Othman, Bruno Rossi, and Barbara Russo (2024). 
Cybersecurity Defenses: Exploration of CVE Types
through Attack Descriptions. 
In: \textit{Proceedings of the 50th Euromicro Conference on Software Engineering and Advanced Applications (SEAA 2024)}, 
pp. 415--418. IEEE. DOI: \url{https://ieeexplore.ieee.org/document/10803317} \\
Conference description: & SEAA (Software Engineering and Advanced Applications) is a long-running IEEE/Euromicro conference, 
CORE B ranked, focusing on software engineering, systems, and applications. It provides a platform for 
presenting advanced methods, tools, and industrial experiences. \\
\end{tabularx}
}
\end{pubinfo}

\label{paper:C}
\includepdf[pages=-,pagecommand={\thispagestyle{plain}},fitpaper=true]{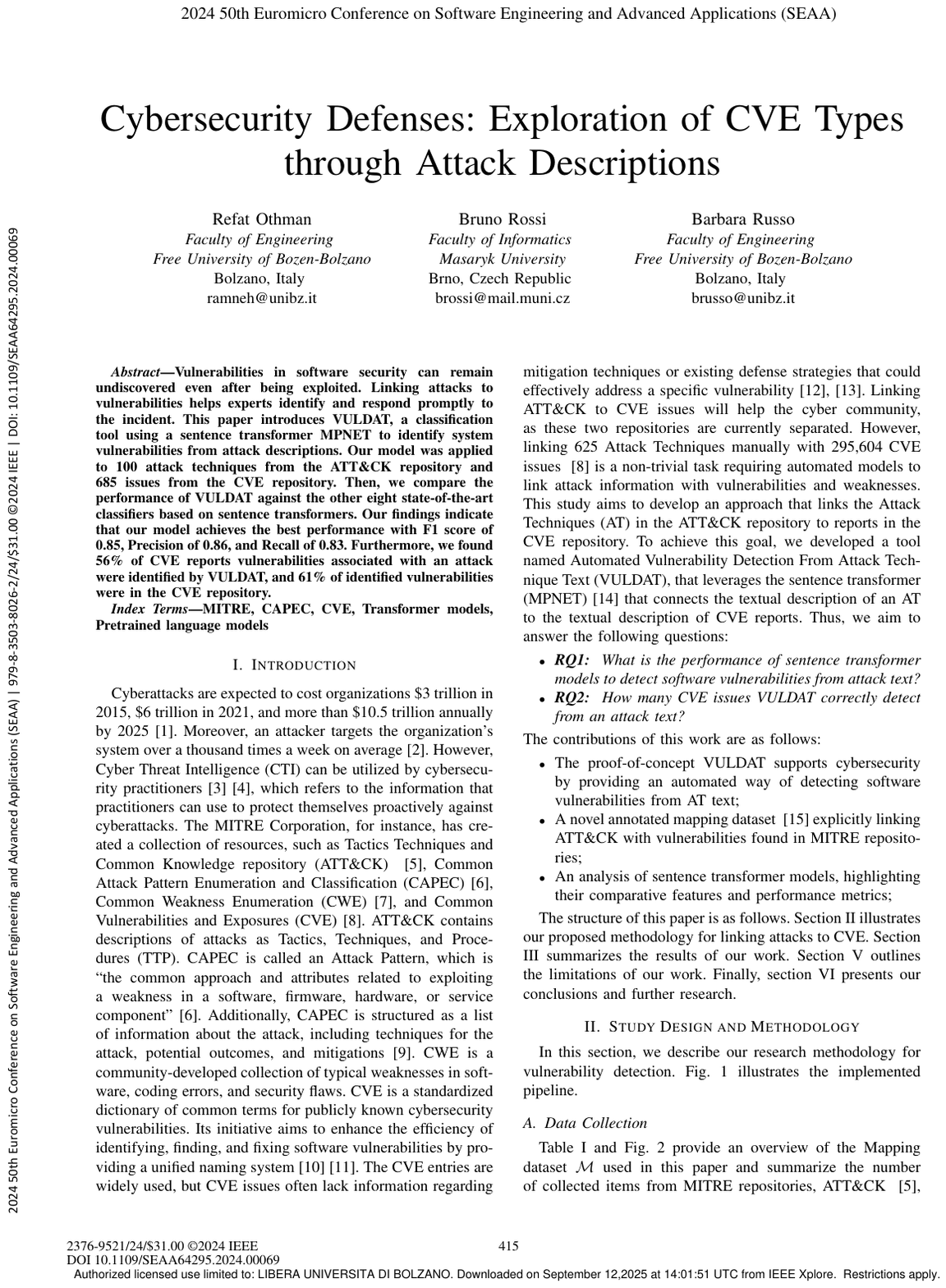}

\myPaper{Paper D \\ VULDAT: Automated Vulnerability
Detection from Cyberattack Text}
\begin{pubinfo}{
\renewcommand{\arraystretch}{1.3}
\begin{tabularx}{\textwidth}{>{\bfseries}l X}
Current status:      & Published \\
Conference:          & International Conference on Embedded Computer Systems: Architectures, Modeling, and Simulation (SAMOS XXIII) \\
Date of publication: & 2 July 2023 \\
Full citation:       & Refat Othman and Barbara Russo (2023). 
VULDAT: Automated Vulnerability Detection from Cyberattack Text. 
In: \textit{Proceedings of the 23rd International Conference on Embedded Computer Systems: Architectures, Modeling, and Simulation (SAMOS 2023)}. 
Springer. DOI: \url{https://link.springer.com/chapter/10.1007/978-3-031-46077-7_36} \\ 

Conference description: & SAMOS is a well-established international conference on embedded computer systems, 
architectures, modeling, and simulation, attracting both academic and industrial researchers. \\
\end{tabularx}
}
\end{pubinfo}

\label{paper:D}
\includepdf[pages=-,pagecommand={\thispagestyle{plain}},fitpaper=true]{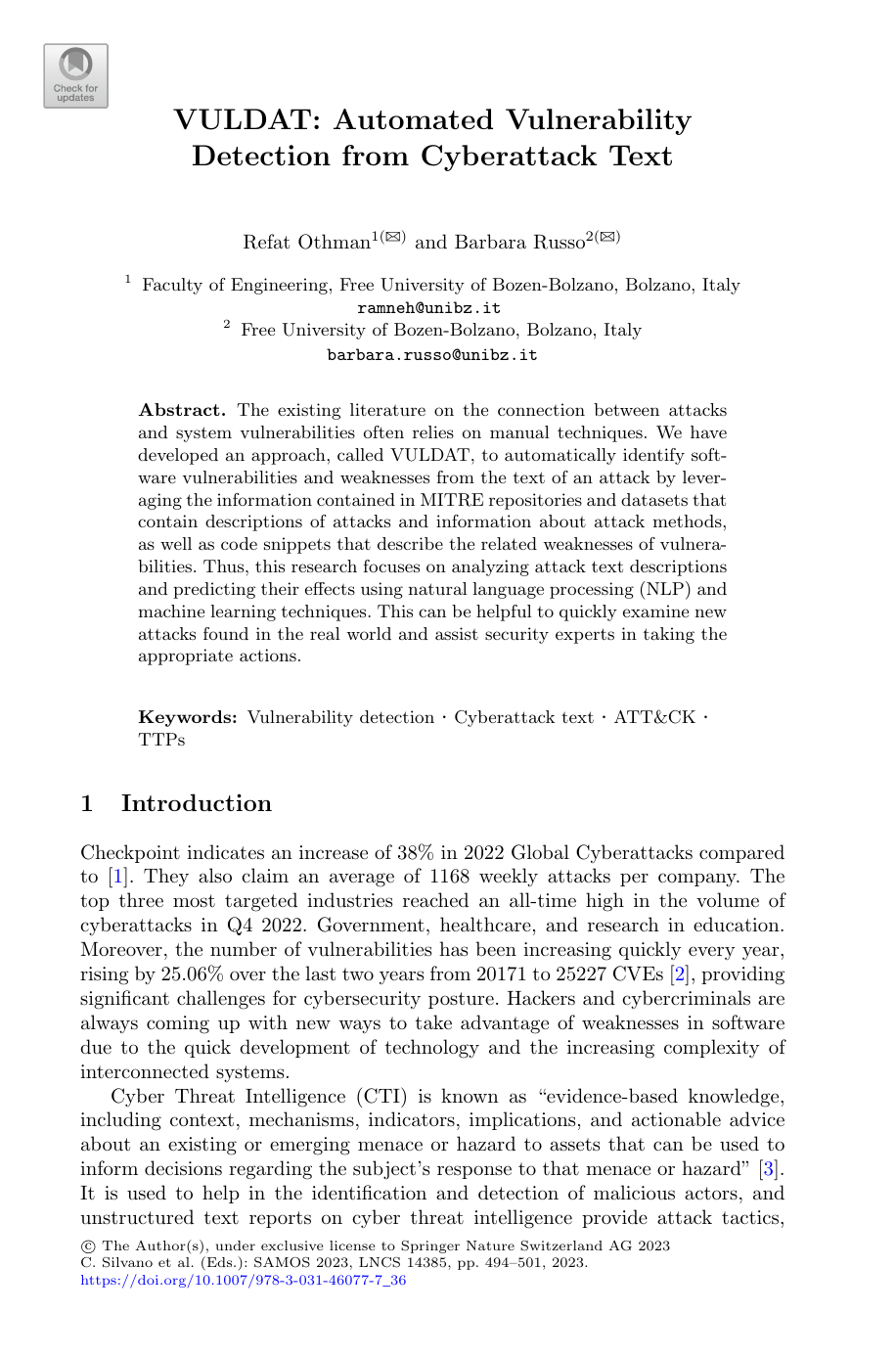}

\myPaper{Paper E \\ Vulnerability Detection for software-intensive system}

\begin{pubinfo}{
\renewcommand{\arraystretch}{1.3}
\begin{tabularx}{\textwidth}{>{\bfseries}l X}
Current status:      & Published \\
Conference:          & 28th International Conference on Evaluation and Assessment in Software Engineering (EASE 2024) \\
Date of publication: & 18 June 2024 \\
Full citation:       & Refat Tahseen Othman (2024). 
Vulnerability Detection for Software-Intensive System. 
In: \textit{Proceedings of the 28th International Conference on Evaluation and Assessment in Software Engineering (EASE 2024)}, 
pp. 510-515. ACM. DOI: \url{https://doi.org/10.1145/3661167.3661170} \\
Conference description: & EASE is a CORE A-ranked international conference focusing on empirical and evaluation methods 
in software engineering research and practice. \\
\end{tabularx}
}
\end{pubinfo}

\label{paper:E}
\includepdf[pages=-,pagecommand={\thispagestyle{plain}},fitpaper=true]{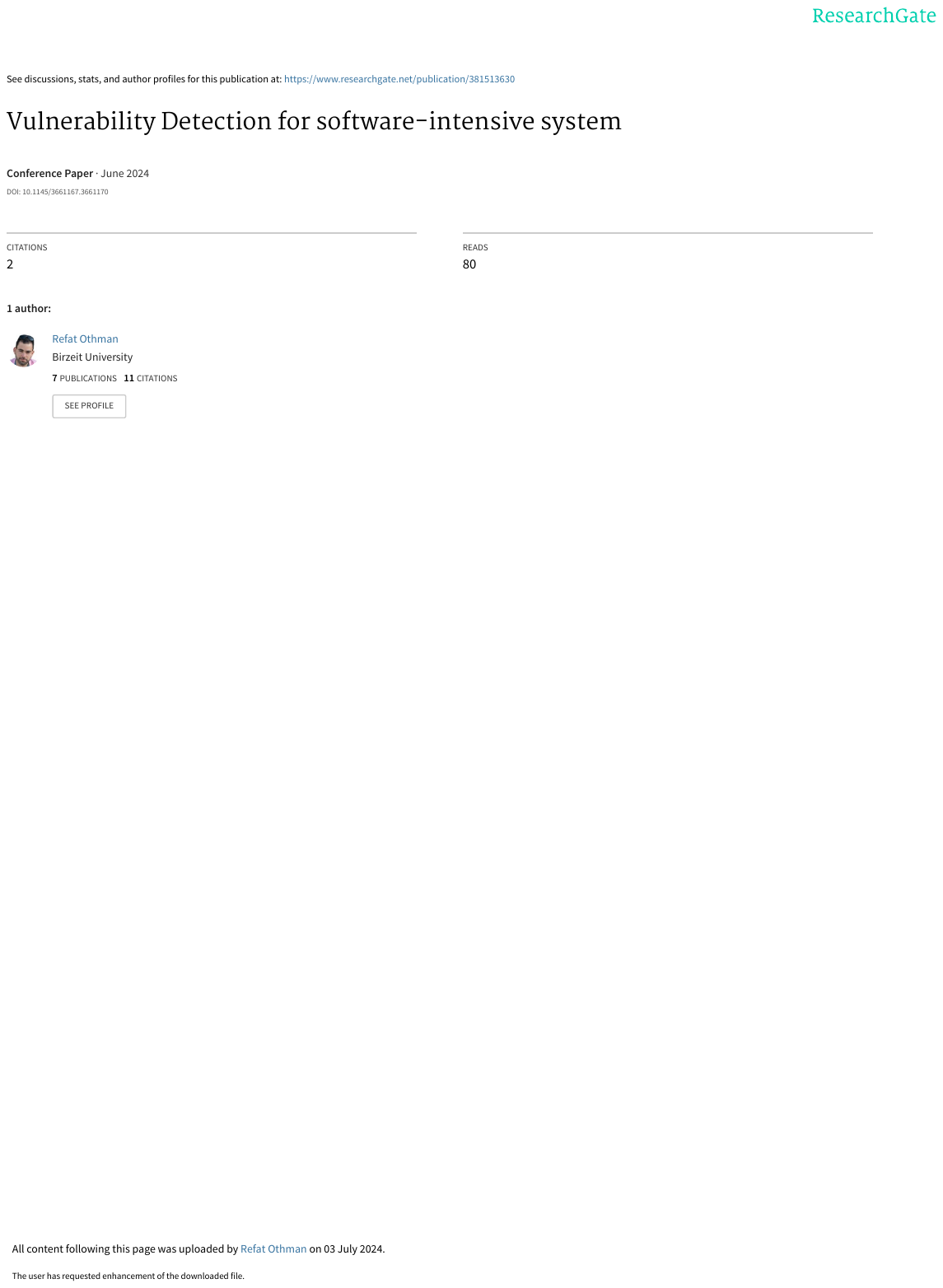}


\letterChapter[F]{Chapter F \\ Predicting known Vulnerabilities from Attack News}

\begin{pubinfo}{
\renewcommand{\arraystretch}{1.3}
\begin{tabularx}{\textwidth}{>{\bfseries}l X}
Current status:      & In preparation / Target journal submission \\
Target journal:      & Digital Threats: Research and Practice (ACM) \\
Planned submission:  & 2025 \\
Full citation (planned): & Refat Othman, Diaeddin Rimawi, Bruno Rossi, and Barbara Russo (2025). 
Predicting known Vulnerabilities from Attack News. 
Submitted to: \textit{ACM Digital Threats: Research and Practice}. \\
Journal description: & Digital Threats: Research and Practice (DTRAP) is an ACM journal focusing on 
cybersecurity threats, vulnerabilities, and defenses. It emphasizes bridging the gap between 
academic research and real-world practice. \\
\end{tabularx}
}
\end{pubinfo}

\label{Chapter:F}
\includepdf[pages=-,pagecommand={\thispagestyle{plain}},fitpaper=true]{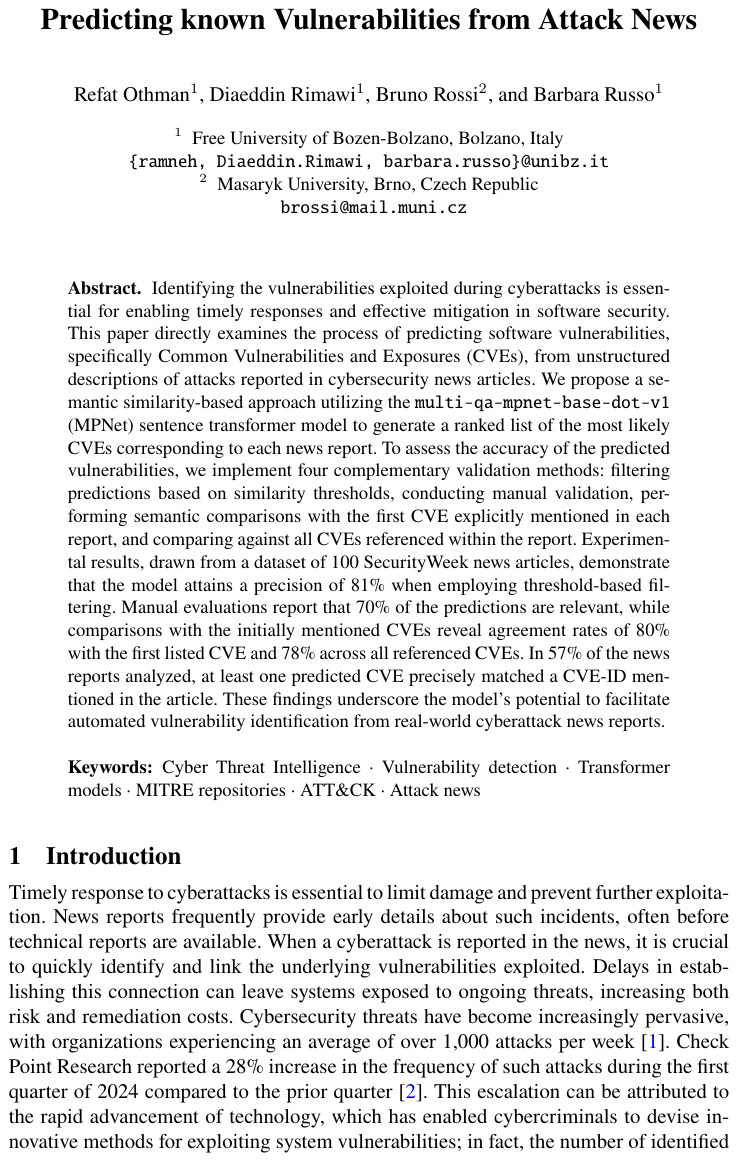}






\end{document}